\documentclass{article}

\usepackage{PRIMEarxiv}

\usepackage[utf8]{inputenc} 
\usepackage[T1]{fontenc}    
\usepackage{hyperref}       
\usepackage{url}            
\usepackage{booktabs}       
\usepackage{amsfonts}       
\usepackage{nicefrac}       
\usepackage{microtype}      
\usepackage{lipsum}
\usepackage{fancyhdr}       
\usepackage{graphicx}       
\graphicspath{{media/}}     
\usepackage{amsmath}
\usepackage{caption}
\usepackage{subcaption}
\usepackage{float}
\usepackage{xcolor}
\usepackage{natbib}
\usepackage{apalike}
\usepackage[export]{adjustbox}
\usepackage{multicol}
\usepackage{wrapfig}
\usepackage{rotating}
\usepackage{array}
\usepackage{tabularx}
\usepackage{natbib}

\title{Rotating neutron stars in the first order post-Newtonian approximation}

\author{
  A. Fotopoulos, V. Karageorgopoulos, and V. Geroyannis \\
  Department of Physics,  University of Patras, Patras, Greece 
                                                             \\
  \texttt{\{Athanasios Fotopoulos\}fotopoulos@upatras.gr}    \\
  \texttt{\{Vasileios Karageorgopoulos\}vkarageo@upatras.gr} \\
  \texttt{\{Vassilis Geroyannis\}vgeroyan@upatras.gr}}

\begin{document}
\maketitle

\begin{abstract}
We study models of uniformly and differentially rotating neutron stars in the framework of post-Newtonian approximation in general relativity as established by Chandrasechar. In particular, we adopt the polytropic equation of state in order to derive the appropriate hydrodynamic equations and a rotation law based on the generalized Clement's model. To compute equilibrium configurations at the mass-shedding limit, i.e. at critical angular velocity (equivalently, Keplerian angular velocity), we develop an iterative numerical method, belonging to the category of the well-known ``self-consistent field methods'', with two perturbation parameters: the ``rotation parameter'' $\bar{\upsilon}$ and the ``gravitation or relativity parameter'' $\bar{\sigma}$. These two parameters represent the effects of rotation and gravity on the configuration. We investigate the validity and the limits of our method by comparing our results with respective results of other computational methods and public domain codes. As it turns out, our method can derive satisfactory results for general-relativistic polytropic configurations at critical rotation. 

\end{abstract}



\section{Introduction}
\label{intro}
The theoretical framework of hydrodynamic equations in ``Post-Newtonian Approximation'' (PNA), concerning rotating neutron stars, has been established by \cite{C1965a, C1965b, C1965c,C1969}. This framework was applied by \cite{Kr1966, Kr1967a, Kr1967b} to the problem of uniformly rotating relativistic configurations with axial symmetry, under stationary conditions. Furthermore, based on the studies of Chandrasekhar and Krefetz, \cite{FA1971} developed a pertubation method for treating uniformly rotating relativistic polytropic configurations. \cite[Chapter~4]{2004Horedt} gives interesting details on the relativistic polytropes, emphasizing in Section~4.2 on the issue of their rotational distortion; and in Section~4.2.6 on the issue of rapidly rotating and fully relativistic polytropes. Recently, \cite{GK2014} revisited the problem of uniformly rotating relativistic polytropes by applying a non-iterative numerical method, so-called ``Complex-Plane Strategy'' (CPS). They further extended their research on critical uniform rotation of relativistic polytropes by developing an iterative numerical method, so-called  ``Hybrid Approximative Scheme'' (HAS)  \citep{GK2015}.

Differentially rotating relativistic polytropes have been studied by \cite{S1973}. He applied the post-Newtonian theory of first order to this  problem and described in detail an iterative method, belonging to the category of the well-known ``self-consistent field methods'', for computing equilibrium configurations. A numerical method based on the same theoretical framework has been also developed by \cite{liu2002postnewtonian}. To the extend of our knowledge, Seguin's method has not been implemented yet by other investigators. In the present study, we combine Seguin's method along with HAS and the so-called ``Complex Iterative Technique'' (CIT) \citep{G1991} for computing either rigidly or differentially critically rotating relativistic polytropes in the first-order post-Newtonian approximation. We then examine the extend of reliability and accuracy of our results in comparison with respective results of other computational methods and public domain codes.

\section{Equations of Hydrodynamics}
\label{sec2}
The components of the metric tensor, $g_{ij}$, in the framework of the second-order PNA are given by \citep{C1965a, S1973}
\begin{equation}
\begin{split}
    g_{00}& = 1 - \frac{2U}{c^2} + \frac{1}{c^4} \left( 2U^2-4\Phi \right) + \mathit{O} \left( \frac{1}{c^6} \right),\\
    g_{0\alpha}& = \frac{1}{c^3} \left(4U_{\alpha}-\frac{1}{2}\frac{\partial^2 \chi}{\partial t \, \partial x_{\alpha}}\right) + \mathit{O} \left( \frac{1}{c^5} \right),\\
    g_{\alpha\beta}& = - \left( 1 + \frac{2U}{c^2} \right) + \mathit{O} \left( \frac{1}{c^4} \right).
    \label{metric}
\end{split}    
\end{equation}
The stress-energy tensor, $T_{ij}$, is assumed in the form
\citep[Eq.~(16)]{S1973}
\begin{equation}
    T_{ij}=\left[\varrho(c^2+\Pi)+P\right]u_{i} \, u_{j}-P \, g_{ij},
\end{equation}
where $\varrho$ is the rest mass density, $\Pi$ the internal energy per unit rest mass, $u_{i}$ the four-velocity, $g_{ij}$ the metric tensor, $P$ the pressure and $U$ the gravitational potential. The so-called ``potentials'' $\chi, U, U_{\alpha}, \mathrm{and} \; \Phi$ are defined as \citep[Eqs.~(3, 41, 44, 45)]{C1965a}
\begin{align}
    \nabla^2 \chi &= - 2U,\\
    \nabla^2 U &= - 4 \pi G \varrho, \label{pot-U}\\
    \nabla^2 U_{\alpha} &= - 4 \pi G \varrho \upsilon_{\alpha}, \label{pot-Ua} \\
    \nabla^2 \Phi &= - 4 \pi G \varrho \phi. \label{pot-Phi}
\end{align}
Under these assumptions, the equations of motion are written as \citep[Eq.~(17)]{S1973}
\begin{equation}
\begin{split}
    \frac{\partial}{\partial t}(\sigma \upsilon_{\alpha}) + \frac{\partial}{\partial x_{\mu}}(\sigma \upsilon_{\alpha} \upsilon_{\mu}) + \frac{\partial}{\partial x_{\alpha}} \left[ \left( 1 + \frac{2U}{c^2} \right) P \right] - \varrho \frac{\partial U}{\partial x_{\alpha}} + \frac{4}{c^2} \varrho \frac{d}{dt} \left( \upsilon_{\alpha}U  - U_{\alpha} \right) 
    \\
     + \frac{4}{c^2} \varrho \upsilon_{\mu} \frac{\partial}{\partial x_{\alpha}} U_{\mu}
    + \frac{\varrho}{2c^2} \frac{\partial}{\partial t} \left( U_{\alpha} - U_{\mu;\alpha \mu} \right) - \frac{2}{c^2} \varrho \left( \phi \frac{\partial U}{\partial x_{\alpha}} + \frac{\partial \Phi}{\partial x_{\alpha}} \right) = 0,
    \label{EoM}
\end{split}    
\end{equation}
and the equation of continuity as \citep[Eq.~(21)]{S1973}
\begin{equation}
    \frac{\partial}{\partial t} \varrho^* + \frac{\partial}{\partial x_{\alpha}} \left( \varrho^*\upsilon_{\alpha} \right) = 0,
    \label{EoCont}
\end{equation}
where \cite[Eqs.~(22-25)]{S1973}
\begin{gather}
    \phi \equiv \upsilon^2 + U +\frac{\Pi}{2} + \frac{3P}{2\varrho}, \label{phi} \\
    \varrho^* \equiv \varrho \left[ 1+ \frac{1}{c^2}\left(\frac{\upsilon^2}{2} + 3U \right) \right],
    \label{rho-star}\\
    \sigma \equiv \varrho \left[ 1 + \frac{1}{c^2} \left( \upsilon^2 + 2U + \Pi +\frac{P}{\varrho} \right) \right], \label{sigma}\\
    U_{\mu ; \alpha \mu} \equiv G \int \frac{\varrho(\mathbf{x'}) \upsilon_{\mu}(\mathbf{x'})(x_\alpha - x_\alpha ')(x_\mu - x_\mu ')}{\lvert \mathbf{x - x'}\rvert^3} d^3x',
\end{gather}
and $\varrho^{*}$ being the ``conserved density''.
Eqs.~\eqref{EoM} and \eqref{EoCont} become under stationary conditions \cite[Eq.~(37)]{S1973}
\begin{equation}
\begin{split}
    \frac{\partial}{\partial x_{\mu}}(\sigma \upsilon_{\alpha} \upsilon_{\mu}) + \frac{\partial}{\partial x_{\alpha}} \left[ \left( 1 + \frac{2U}{c^2} \right) P \right] - \varrho \frac{\partial U}{\partial x_{\alpha}} + \frac{4}{c^2} \varrho \upsilon_{\mu} \frac{\partial}{\partial x_{\alpha}} U_{\mu} 
    - \frac{2}{c^2} \varrho \left( \phi \frac{\partial U}{\partial x_{\alpha}} + \frac{\partial \Phi}{\partial x_{\alpha}} \right)& 
    \\ 
    + \frac{4}{c^2}\varrho \upsilon_{\mu} \frac{\partial}{\partial x_{\mu}} \left( \upsilon_{\alpha}U-U_{\alpha}\right)& = 0,
    \label{EoM_static}
\end{split}    
\end{equation}
and \cite[Eq.~(2)]{C1965b}
\begin{equation}
    \frac{\partial}{\partial x_{\alpha}} \left( \varrho^*\upsilon_{\alpha} \right) = 0.
    \label{EoCont_static}
\end{equation}
Assuming axial symmetry around the axis of rotation, the only nonzero component of the velocity, in cylindrical coordinates is the one in $\phi$-direction, with $\upsilon^2=\Tilde{\omega}^2\Omega^2$. 

With use of cylindrical coordinates, some terms of the above equation can be written as \cite[Eqs.~(39-41)]{S1973}
\begin{gather}
    \hat{x}_{\alpha} \frac{\partial}{\partial x_{\mu}}(\sigma \upsilon_{\alpha} \upsilon_{\mu}) = -\sigma \Tilde{\omega} \Omega^{*2} \hat{\Tilde{\omega}}, 
    \label{subt1}\\
    \hat{x}_{\alpha} \upsilon_{\mu} \frac{\partial}{\partial x_{\mu}} \left( \upsilon_{\alpha}U\right) = - \Tilde{\omega} \Omega^{*2} U \hat{\Tilde{\omega}}, \label{subt2}\\ 
    \frac{1}{c^2}\hat{x}_{\alpha} \left[ \upsilon_{\mu} \frac{\partial U_{\mu}}{\partial x_{\alpha}} - \upsilon_{\mu} \frac{\partial U_{\alpha}}{\partial x_{\mu}} \right] = \frac{1}{c^2} \nabla \left( \Tilde{\omega} \Omega U_{\phi} \right) - \frac{\Tilde{\omega}}{c^2} U_{\phi} \frac{d\Omega}{d\Tilde{\omega}} \hat{\Tilde{\omega}}, \label{subt3}
\end{gather}
where $\Omega^{*}=d\phi/dt$ is an expression for the angular velocity \cite[Eq.~(33)]{S1973}. In Seguin's study, the angular velocity depends on both coordinates $\Tilde{\omega}$ and $z$ \cite[Eq.~(34)]{S1973},
\begin{equation}
    \Omega^{*2}(\Tilde{\omega},z)=\Omega^{2}(\Tilde{\omega})+\frac{1}{c^2} h^{2}(\Tilde{\omega},z),
    \label{Omega-star}
\end{equation}
where $h^{2}(\Tilde{\omega},z)$ is a function to be determined.
Since the required accuracy is of order $1/c^2$, $\Omega^{*}$ can be replaced by $\Omega$ whenever it is involved in product(s) with the term $1/c^2$. 

For simple adiabatic conditions, we can write \cite[Eq.~(15)]{Kr1966}
\begin{equation}
    \frac{\partial P}{\partial x_{\alpha}}=\varrho\frac{\partial}{\partial x_{\alpha}}\left(\Pi+\frac{P}{\varrho}\right). \label{kr-15}
\end{equation}
After having defined the quantities \cite[Eqs.~(43-45)]{S1973}
\begin{gather}
    \nabla B\equiv \Tilde{\omega}\Omega^2\hat{\Tilde{\omega}}, \label{pot-B} \\ 
    H\equiv U+B, \label{H} \\ 
    \nabla W \equiv \Tilde{\omega}^2 \Omega^2 \nabla B = \Tilde{\omega}^3\Omega^4\hat{\Tilde{\omega}}, \label{pot-W}
\end{gather}
and according to details given in Appendix~\ref{app1}, we can write Eq.~\eqref{EoM_static} in the form 
\begin{multline}
    \frac{1}{\varrho}\nabla{P}-\nabla{U}-\nabla{B}-\frac{1}{c^2}\left[ \nabla{(2\Tilde{\omega}^2\Omega^2U)}+\nabla{(2\Phi)}
    -\nabla{\left(4\Tilde{\omega}\Omega U_{\phi}\right)} +\nabla{W}+(n+1)\frac{P}{\varrho}\nabla{\left(\Pi+\frac{P}{\varrho}\right)}\right] 
    \\ +\frac{1}{c^2}\Tilde{\omega}\left[4\frac{d\Omega}{d\Tilde{\omega}}\left(\Tilde{\omega}\Omega U-U_{\phi}\right)-h^{2}(\Tilde{\omega},z)\right]\hat{\Tilde{\omega}}=0,
    \label{seg-eq-42c}
\end{multline}
where, to deduce this equation, we use the following relations in the post-Newtonian terms
\begin{gather}
    \Pi+(P/\varrho)-U-\frac{1}{2}\upsilon^2=\delta=\mathrm{constant}, \label{Pi_p_U_u} \\
    \Pi+(P/\varrho)-U-B-\mathcal{O}\left(1/c^2\right)=\delta=\mathrm{constant}. \label{Pi_p_U_B} 
\end{gather}
Eq.~\eqref{Pi_p_U_u} \citep[Eq.~(6)]{Kr1967a} is a Newtonian expression, while  Eq.~\eqref{Pi_p_U_B} results from Eq.~\eqref{seg-eq-42c} by using Eq.~\eqref{kr-15}. As our analysis is limited to a first-order approximation, both equations can be used only when applied on post-Newtonian terms.

\subsection{Equation of state}
In this study, we adopt the well-known polytropic ``Equation of State'' (EOS)
\begin{equation}
    P = K \, \varrho^{\Gamma},
    \label{pol-eq}
\end{equation}
where $K$ is the polytropic constant, $\Gamma$ the adiabatic index defined by $\Gamma=1+1/n$, and $n$ the polytropic index. Under simple adiabatic conditions, the adiabatic index $\Gamma$ interrelates the internal energy density and the pressure with the form \citep{F1966, C1965b, FA1971}
\begin{equation}
    \Pi=\frac{1}{\Gamma-1} \, \frac{P}{\varrho} \quad \rightarrow \quad \varrho \, \Pi = n \, P.
    \label{np}
\end{equation}
In view of Eqs.~\eqref{kr-15} and \eqref{np}, Eq.~\eqref{seg-eq-42c} becomes 
\begin{equation}
\begin{split}
    \nabla{\left(\Pi+\frac{P}{\varrho}\right)} = \nabla{U}&+\nabla{B}+\frac{1}{c^2}\bigg[ \nabla{(2\Tilde{\omega}^2\Omega^2U)}+\nabla{(2\Phi)}
    -\nabla{\left(4\Tilde{\omega}\Omega U_{\phi}\right)} +\nabla{W} 
    \\
    &+\left(\Pi+\frac{P}{\varrho}\right)\nabla{\left(\Pi+\frac{P}{\varrho}\right)}\bigg] 
    +\frac{1}{c^2}\Tilde{\omega}\left[4\frac{d\Omega}{d\Tilde{\omega}}\left(\Tilde{\omega}\Omega U-U_{\phi}\right)-h^{2}(\Tilde{\omega},z)\right].
    \label{kr-seg2}
\end{split}
\end{equation}
Next, in view of Eqs.~\eqref{H} and \eqref{Pi_p_U_B}, the above equation takes the form (for details, see Appendix~\ref{app1})
\begin{equation}
\begin{split}
    \nabla{\left(\Pi+\frac{P}{\varrho}\right)} = \nabla{H}+\frac{1}{c^2}\bigg[ \nabla{(2\Tilde{\omega}^2\Omega^2U)}+\nabla{(2\Phi)}
    -\nabla{\left(4\Tilde{\omega}\Omega U_{\phi}\right)} +\nabla{W} + \nabla{\left(\frac{(H+\delta)^2}{2}\right)}\bigg] 
    \\
    +\frac{1}{c^2}\Tilde{\omega}\left[4\frac{d\Omega}{d\Tilde{\omega}}\left(\Tilde{\omega}\Omega U-U_{\phi}\right)-h^{2}(\Tilde{\omega},z)\right]\hat{\Tilde{\omega}},
    \label{kr-seg3}
\end{split}
\end{equation}
where $\delta$ is the constant which results by solving Eq~\eqref{Pi_p_U_u} at the center of the star.
However, Eq.~\eqref{kr-seg3} has a solution only if \cite[Eq.~(62)]{S1973}
\begin{equation}
    \frac{\partial}{\partial z}\left[4\frac{d\Omega}{d\Tilde{\omega}}\left(\Tilde{\omega}\Omega U-U_{\phi}\right)-h^{2}(\Tilde{\omega},z)\right]=0,
\end{equation}
which means in turn that $h^{2}(\Tilde{\omega},z)$ must be of the form \cite[Eq.~(63)]{S1973}
\begin{equation}
    h^{2}(\Tilde{\omega},z) = 4\frac{d\Omega}{d\Tilde{\omega}}\left(\Tilde{\omega}\Omega U-U_{\phi}\right)+\beta(\Tilde{\omega}),
    \label{h^2}
\end{equation}
where $\beta$ is any function of $\Tilde{\omega}$. Since $\beta(\Tilde{\omega})$ is an arbitrary function, it can be assigned equal to zero, $\beta(\Tilde{\omega})=0$. 

Substituting Eq.~\eqref{h^2} in Eq.~\eqref{kr-seg3}, the last term vanishes, and the result reads
\begin{equation}
\begin{split}
    \nabla{\left(\Pi+\frac{P}{\varrho}\right)}=\nabla{H}+\frac{1}{c^2}\left[ \nabla{(2\Tilde{\omega}^2\Omega^2U)}+\nabla{(2\Phi)}
    -\nabla{\left(\Tilde{4\omega}\Omega U_{\phi}\right)} +\nabla{W}+\nabla{\left(\frac{(H+\delta)^2}{2}\right)}\right],
    \label{kr-seg4}
\end{split}
\end{equation}
taking finally the form
\begin{equation}
\begin{split}
   \nabla{\left(\Pi+\frac{P}{\varrho}\right)}=\nabla{\mathcal{U}},
   \label{EoM_to_be_integrated}
\end{split}
\end{equation}
where the quantity
\begin{equation}
    \mathcal{U}= H+\frac{1}{c^2}\left[ 2\Phi+W+2\Tilde{\omega}^2\Omega^2U-4\Tilde{\omega}\Omega U_{\phi}+\frac{(H+\delta)^2}{2} \right]
    \label{Ueff}
\end{equation}
is the so-called ``efficient potential''.

In view of the above relations, Eq.~\eqref{EoM_to_be_integrated} can be directly integrated to give \cite[Eq.~(66)]{S1973}
\begin{equation}
\begin{split}
    K(n+1)\varrho^{1/n}=H +\frac{1}{c^2}\left[2\Phi+W+2\Tilde{\omega}^2\Omega^2 U
    -4\Tilde{\omega}\Omega U_{\phi}+\frac{(H+\delta)^2}{2}\right]+D,
    \label{seg-64}
\end{split}    
\end{equation}
where $D$ denotes the ``constant of integration'' and is left to be computed by the numerical method.

The analysis so far has been accomplished  in terms of cylindrical coordinates, $\Tilde{\omega}, \, z, \, \phi$. Due to axial symmetry, however, the involved quantities are independent of the coordinate $\phi$; apparently, they can be also expressed in terms of the spherical coordinates $r$ and $\theta$, with $\Tilde{\omega} = r \, \mathrm{sin}(\theta)$ and $z = r \, \mathrm{cos}(\theta)$. 

\subsection{Potentials}

Eqs.~\eqref{pot-B},\eqref{pot-W},\eqref{pot-U},\eqref{pot-Ua} and \eqref{pot-Phi} can be easily expressed in integral form as \cite[Eqs.~(72-74,76); the last one results in the same way as the other]{S1973}
\begin{align}
    B(\Tilde{\omega}) &= \int_0^{\Tilde{\omega}'}\Tilde{\omega}\Omega^2(\Tilde{\omega})d\Tilde{\omega} \label{pot-B-int}\\
    W(\Tilde{\omega}) &= \int_0^{\Tilde{\omega}'}\Tilde{\omega}^3\Omega^4(\Tilde{\omega})d\Tilde{\omega} \label{pot-W-int}\\
    U(\mathbf{x'}) &= G\int \frac{d^3\mathbf{x'}}{|\mathbf{x}-\mathbf{x'}|} \varrho(\mathbf{x'}) \label{pot-U-int}\\
    \Phi(\mathbf{x'}) &= G\int \frac{d^3\mathbf{x'}}{|\mathbf{x}-\mathbf{x'}|} \varrho(\mathbf{x'})\phi(\mathbf{x'}) \label{pot-Phi-int}\\
    U_{\alpha}(\mathbf{x'}) &= G\int \frac{d^3\mathbf{x'}}{|\mathbf{x}-\mathbf{x'}|} \varrho(\mathbf{x'}) \upsilon_{\alpha}(\mathbf{x'}) \label{pot-Ua-int}
\end{align}
Since the only non zero component of the velocity is the one in $\phi$-direction ($\upsilon_{\phi}^2=\Tilde{\omega}^2\Omega^2$), the only component of the potential $U_{\alpha}$ is $U_{\phi}$, with \cite[Eq.~(84a)]{S1973}
\begin{equation}
    U_{\phi}(\mathbf{x'}) = G\int \frac{d^3\mathbf{x'}}{|\mathbf{x}-\mathbf{x'}|} \varrho(\mathbf{x'}) \upsilon_{\phi}(\mathbf{x'}) \label{pot-Uphi-int}
\end{equation}
Another useful potential is $U^*$, defined as \cite[Eq.~(29)]{S1973}
\begin{equation}
    \nabla^2 U^* = - 4 \pi G \varrho^* \label{pot-U*},
\end{equation}
involved in the calculation of the energy per unit coordinate volume. This potential can also be expressed in integral form \cite[Eq.~(75)]{S1973},
\begin{equation}
    U^{*}(\mathbf{x'}) = G\int \frac{d^3\mathbf{x'}}{|\mathbf{x}-\mathbf{x'}|} \varrho^{*}(\mathbf{x'}) .\label{pot-U*-int}
\end{equation}

For a given rotation law, the first two potentials, $B$ and $W$, can be easily calculated. The other can be integrated by using an appropriate numerical method (e.g. that used by \cite{H1986}).

\subsection{Mass and binding energy}

In order to determine the rest mass of the configuration, one must take into account the difference between the proper and coordinate volume. Given that the space part of the metric \eqref{metric} is
\begin{equation}
    g_{\alpha\beta}  = - \left( 1 + \frac{2U}{c^2} \right),
\end{equation}
the relation between an element of proper volume, $dV$, and an element of unit coordinate volume, $d^3x$, is
\begin{equation}
    dV = \sqrt{det|g_{\alpha\beta}|} d^3x = \left( 1 + \frac{2U}{c^2} \right)^{3/2} d^3x \simeq \left( 1+\frac{3U}{c^2} \right) d^3x.
\end{equation}
Given that $\varrho$ refers to the density of rest mass per unit proper volume, the rest mass of the star is given by \cite[Eq.~(96)]{S1973}
\begin{equation}
    M_{0} = \int \varrho dV = \int \varrho \left( 1+\frac{3U}{c^2} \right) d^3x.
    \label{rest_mass}
\end{equation}
The total energy per unit coordinate volume of the configuration is given from the expression \cite[Eq.~(28)]{S1973}
\begin{gather}
\begin{split}
    \mathfrak{E}=\left(\sigma - \frac{1}{2} \varrho^{*}\right) +& \varrho^{*}\Pi - \frac{1}{2}\varrho^{*}U^{*} 
    \\ 
    +& \frac{1}{c^2} \varrho \left( -\frac{1}{8}\upsilon^{4} + \frac{1}{2}U^{2}
     - U\Pi - \frac{1}{2}\upsilon^{2}\Pi +\frac{5}{2}\upsilon^{2}U - \frac{7}{4}\upsilon_{\alpha}U_{\alpha} - \frac{1}{4}\upsilon_{\alpha}U_{\mu;\alpha\mu} 
     \right).
     \label{goth_E_Chandra}
\end{split}
\end{gather}
The potential $U^{*}$ can be expressed as $U^{*}=U+\mathcal{O}(c^{-2})$. To the required order of approximation, we can say that $U^{*}=U$ on the post-Newtonian terms. Using this relation along with the adopted polytropic EOS (Eqs.~\eqref{pol-eq},  \eqref{np}) the above equation can be written as \cite[Eq.~(97)]{S1973} 
\begin{gather}
\begin{split}
    \mathfrak{E}=\frac{1}{2} \Tilde{\omega}^{2}\Omega^{2}\varrho +& Kn\varrho^{1+1/n} - \frac{1}{2}U^{*}\varrho 
    \\
    +& \frac{1}{c^2} \varrho \left[ \frac{5}{8}\Tilde{\omega}^{4}\Omega^{4}
    + \frac{11}{4}\Tilde{\omega}^{2}\Omega^{2}U + K(n+1)\Tilde{\omega}^{2}\Omega^{2}\varrho^{1/n} 
    + 2KnU\varrho^{1/n} - U^2 - 2\Tilde{\omega}\Omega U_{\phi} \right].
    \label{goth_E_Seguin}
\end{split}
\end{gather}
The total mass energy of the star is then given by \cite[Eq.~(98)]{S1973}
\begin{equation}
    Mc^{2} = \int \left[ \varrho c^{2}\left( 1+\frac{3U}{c^2} \right) + \mathfrak{E} \right] d^3x = M_{0}c^2 + \int \mathfrak{E} \, d^3x,
\end{equation}
and the gravitational mass, accordingly, by 
\begin{equation}
    M = M_{0} + \frac{1}{c^2} \int \mathfrak{E} \, d^3x = M_{0} - E_b,
    \label{grav_mass}
\end{equation}
where $E_{b}$ is the binding energy of the star, defined as the difference between the rest mass energy and the total energy \cite[Eq.~(99)]{S1973}, 
\begin{equation}
    E_b = - \int \mathfrak{E} \, d^3x.
    \label{bind_nrg}
\end{equation}

\subsection{Angular momentum, kinetic energy and gravitational potential energy}
The expression of the angular momentum per unit coordinate volume around the $z$ axis is given by \cite[Eq.~89]{S1973}
\begin{equation}
    \mathcal{J} = \sigma\Tilde{\omega}^{2}\Omega^{*} + \frac{\varrho}{c^2} \left( 4\Tilde{\omega}^{2}\Omega U - 4\Tilde{\omega}U_{\phi} \right).
    \label{J/d3x}
\end{equation}
The angular momentum is, then, calculated on the proper volume by integration of \eqref{J/d3x}, as
\begin{equation}
    J = \int \mathcal{J} \, dV.
    \label{J-dV}
\end{equation}
Consequently, the rotational kinetic energy can be calculated from the expression
\begin{equation}
    T = \frac{1}{2}\int \Omega\mathcal{J} \, dV .
    \label{T-dV}
\end{equation}
Finally, the gravitational potential energy can be calculated as the integral of the product of the gravitational mass density and the corresponding potential. The gravitational mass density can be expressed through Eq.~\eqref{grav_mass} as
\begin{equation}
    \varrho_{g}=\varrho \left[ 1+\frac{1}{c^2} \left( 3U -\frac{1}{2} \Tilde{\omega}^{2}\Omega^{2} - Kn\varrho^{1/n} + \frac{1}{2}U^{*} \right) \right],
    \label{rho-grav}
\end{equation}
the corresponding potential is, then, defined as
\begin{equation}
    \nabla^2 U_{g} = - 4 \pi G \varrho_{g},
\end{equation}
and the gravitational potential energy can be computed as 
\begin{equation}
    W = - \frac{1}{2} \int \varrho_{g}U_{g} \, dV.
    \label{W-dV}
\end{equation}

\subsection{Rotation and relativity parameters} \label{upsilon_sigma}

Before we go any further, we have to introduce two perturbation parameters that play important role in our method. The first is the ``rotation parameter'' $\bar{\upsilon}$, representing the effects of rotation, and the second is the ``gravitation or relativity parameter'' $\bar{\sigma}$, representing the post-Newtonian effects of relativity. These parameters are defined as \citep{FA1971, GK2014}
\begin{gather}
    \bar{\upsilon}=\frac{\Omega_{c}^2}{2\pi G\varrho_{c}}, \label{upsilon-par} \\ \bar{\sigma}=\frac{1}{c^2}\frac{P_{c}}{\varrho_{c}}. \label{sigma-par}
\end{gather}
For a spheroidal rotating configuration, there is a value of the angular velocity, $\Omega$, for which the mass shedding from the equator begins to occur. The critical value of the angular velocity, just below that causing mass shedding, is denoted by $\Omega_{\mathrm{crit}}$; the corresponding value of $\bar{\upsilon}$ is denoted by $\bar{\upsilon}_{\mathrm{max}}$. In our study, we are interested in configurations at critical rotation, i.e. configurations with  $\bar{\upsilon}=\bar{\upsilon}_{max}$.

With use of the polytropic EOS (Eq.~\eqref{pol-eq}), Eq.~\eqref{sigma-par} takes the form
\begin{equation}
    \bar{\sigma}=\frac{1}{c^2}K{\varrho_{c}}^{1/n}.
    \label{sigma-par-eos}
\end{equation}

\begin{wraptable}{r}{5cm}
\centering
\vspace{-3mm}
\begin{tabular}{cc} 
\vspace{-1mm}
n & $\bar{\sigma}_{max}$ \vspace{1mm}  \\
\hline 
& \vspace{-2mm}\\
0.5  & 8.27448$ \times 10^{-1}$ \\
1.0  & 3.19773$ \times 10^{-1}$ \\
1.5  & 1.50569$ \times 10^{-1}$ \\
2.0  & 7.10464$ \times 10^{-2}$ \\
2.5  & 2.68066$ \times 10^{-2}$ \\
2.9  & 4.41591$ \times 10^{-3}$ \\ 
\vspace{-3mm} \\
\hline
\end{tabular}
\captionsetup{width=0.2\textwidth}
\caption{Values of $\bar{\sigma}_\mathrm{max}$ for several values of $n$.}
\label{table:sigma}
\end{wraptable}

This relation shows that the relativistic parameter is directly related to the central rest-mass density and, accordingly, to the central mass-energy density. For a given value of the polytropic constant, $K$, higher values of $\bar{\sigma}$ signify higher values of the central mass-energy density, and thus a more compact configuration. On the other hand, the mass of the configuration does not follow the same behaviour. As $\bar{\sigma}$ increases, the mass of the configuration gets increasing until a specific value of $\bar{\sigma}$, after which it gets decreasing. The value of $\bar{\sigma}$ for which the configuration obtains its maximum mass is denoted by $\bar{\sigma}_\mathrm{max}$. As can be seen from Eq~\eqref{sigma-par-eos}, $\bar{\sigma}$ depends also on the polytropic index; in particular, it increases as the polytropic index decreases. The values of $\bar{\sigma}_\mathrm{max}$ are quoted in Table \ref{table:sigma} for several polytropic indices (see e.g. \cite{GK2015}). 
\\

\section{Units}
The systems of units used in this study (either explicitly or implicitly) are the ``cgs units'' (cgs), the ``gravitational, or $c=G=1$ units'' (gu), the ``polytropic units related to the gravitational units'' (pu), and the ``classical polytropic units'' (cpu). To quote the measure of a physical quantity in a system of units other than that in which it has been determined, we multiply the measure of this quantity by a coefficient, so-called ``conversion coefficient''. For instance, if a quantity $q$ has measure $q_\mathrm{cpu}$ in cpu, then the respective measure $q_\mathrm{cgs}$ of $q$ in cgs is given by  
\begin{equation}
    q_\mathrm{cgs}=[Q]_\mathrm{cpu} \times q_\mathrm{cpu},
\label{cgs-cpu}
\end{equation}
where $[Q]_\mathrm{cpu}$ is the conversion coefficient for this quantity. Inversely, for converting measures from cgs to cpu, the conversion coefficient is $1/[Q]_\mathrm{cpu}$,
\begin{equation}
    q_\mathrm{cpu}=(1/[Q]_\mathrm{cpu}) \times q_\mathrm{cgs}.
\label{cpu-cgs}
\end{equation}
Likewise, one can convert values from gu to cgs and from pu to gu by the relations
\begin{gather}
    q_\mathrm{cgs}=[Q]_\mathrm{gu} \times q_\mathrm{gu}, \label{cgs-gu} \\
    q_\mathrm{gu}=[Q]_\mathrm{pu} \times q_\mathrm{pu}. \label{gu-pu}
\end{gather}
Further details on cpu are given in \cite{GTV79}, on pu in \cite{CST94}, and on gu and pu in \cite{GS11}.

Table \ref{tab:unts} shows conversion coefficients for several  physical quantities. The second column gives the conversion coefficients from cpu to cgs, the third column those from gu to cgs, and the fourth column those from pu to gu. Apparently, as defined here, the conversion coefficient for a physical quantity does coincide with the unit of this quantity in the system in which it is determined. For instance, the coefficient $[D]_\mathrm{cpu} = \varrho_\mathrm{c}$ coincides with the unit of density in cpu; and the coefficient $[D]_\mathrm{gu} = c^2/G$ coincides with the unit of density in gu. 

\begin{table}[ht]
\begin{center}
\renewcommand*{\arraystretch}{1.3}
\caption{Conversion coefficients from cpu to cgs, from gu to cgs, and from pu to gu, for several physical quantities. 
\\}
\vspace{3mm}
\label{tab:unts}
\begin{tabular}{l|lll} 
\hline
\\ [-2.0ex]
Physical quantity  & \multicolumn{3}{c}{Conversion coefficients: symbols and definitions}          \\
\\ [-2.0ex]
\hline
\\ [-1.5ex]
Polytropic constant  & $[K]_\mathrm{cpu}$ = $K_\mathrm{cgs}$
& $[K]_\mathrm{gu}$ = $c^{2-2/n}/G^{-1/n}$      & $[K]_\mathrm{pu}$ = $K_\mathrm{gu}$
\\
Length             & $[L]_\mathrm{cpu}$ = $\alpha$~~ (Eq.~(\ref{alpha}))      & $[L]_\mathrm{gu}$ = $1$     & $[L]_\mathrm{pu}$ = ${K_\mathrm{gu}}^{n/2}$   \\
Density            & $[D]_\mathrm{cpu}$ = $\varrho_\mathrm{c}$ & $[D]_\mathrm{gu}$ = $c^2/G$ & $[D]_\mathrm{pu}$ = ${K_\mathrm{gu}}^{-n}$ \\
Pressure           & $[P]_\mathrm{cpu}$ = $K_\mathrm{cgs} \, \varrho_\mathrm{c}^\Gamma$  & $[P]_\mathrm{gu}$ = $c^4/G$  & $[P]_\mathrm{pu}$ = ${K_\mathrm{gu}}^{-n}$  \\
Mass               & $[M]_\mathrm{cpu}$ = $4 \, \pi \, \alpha^3 \, \varrho_\mathrm{c}$ & $[M]_\mathrm{gu}$ = $c^2/G$  & $[M]_\mathrm{pu}$ = ${K_\mathrm{gu}}^{n/2}$ \\
Energy             & $[E]_\mathrm{cpu}$ =  $16 \, \pi^2 \, G \, \alpha^5 \, \varrho_\mathrm{c}^2$    & $[E]_\mathrm{gu}$ = $c^4/G$ & $[E]_\mathrm{pu}$ = ${K_\mathrm{gu}}^{n/2}$ \\
Angular velocity   & $[\Omega]_\mathrm{cpu}$ = $\sqrt{4 \, \pi \, G \, \varrho_\mathrm{c}}$  & $[\Omega]_\mathrm{gu}$ = $c$ & $[\Omega]_\mathrm{pu}$ = ${K_\mathrm{gu}}^{-n/2}$ \\
Angular momentum   & $[J]_\mathrm{cpu}$ = $8 \, \pi^{1.5} \, G^{0.5} \, \alpha^5 \, \varrho_\mathrm{c}^{1.5}$ & $[J]_\mathrm{gu}$ = $c^3/G$ & $[J]_\mathrm{pu}$ = ${K_\mathrm{gu}}^n$ \\
Moment of inertia  & $[I]_\mathrm{cpu}$ = $4 \, \pi \, \alpha^5 \, \varrho_\mathrm{c}$  & $[I]_\mathrm{gu}$ = $c^2/G$ & $[I]_\mathrm{pu}$ = ${K_\mathrm{gu}}^{3n/2}$ \\ 
[1.5ex]
\hline
\end{tabular}
\end{center}
\end{table}

When studying specific star models (e.g. neutron stars), we need to apply a `fine tuning' on the polytropic constant. In detail, we solve the relation for the polytropic unit of length,
\begin{equation}
\alpha = \left[\frac{(n+1)K_\mathrm{cgs}~\varrho_\mathrm{c}^{1/n-1}}{4~\pi~G}\right]^{1/2}, 
\label{alpha}
\end{equation}
in the variable $K_\mathrm{gu}$, 
\begin{equation}
    K_\mathrm{gu} = \left[\frac{4~ \pi~ \alpha^2~ \bar{\sigma}^{n-1}}{(n+1)}\right]^{1/n},
    \label{Kgu}
\end{equation}
and then we substitute $\alpha$ by the value
\begin{equation}
    \alpha = \frac{R}{(\xi_1)_n},
    \label{aRn}
\end{equation} 
where $R$ is the star radius and $(\xi_1)_n$ its radius in cpu for an appropriate polytropic index $n$. Taking for example a neutron star model with $R \simeq 12.5 \, \mathrm{km} = 1.25 \times 10^6 \, \mathrm{cm}$, $n=1$ and $\rho_\mathrm{c} = 1.28 \times 10^{-3}\, \mathrm{pu}$, hence $(\xi_1)_{n=1} \simeq 1.2489 \, \mathrm{pu} \simeq 3.13 \, \mathrm{cpu}$, we find
\begin{equation}
   \left(K_\mathrm{gu}\right)_{n=1} = \frac{2\,\pi R^2}{\xi_1^2} \simeq 100\times 10^{10}~\mathrm{cm^2} = 100~\mathrm{km^2}.    
    \label{KguR}
\end{equation}

All the computations related to our numerical method deal with cpu measures of physical quantities, since this method is inherently oriented to cpu. To compare our results with those of other investigations, we quote the computed values in pu, since pu are the units mostly used in the bibliography.
In this study, we do not intend to quote any values in cgs; we therefore use `direct' coefficients for converting cpu values to respective pu values. In detail,
Eqs.~\eqref{cgs-cpu},\eqref{cgs-gu}, and \eqref{gu-pu} provide proper conversion coefficients, 
\begin{equation}
    q_\mathrm{pu}=\left( \frac{[Q]_\mathrm{cpu}}{[Q]_\mathrm{gu}[Q]_\mathrm{pu}} \right) \times q_\mathrm{cpu}.
\label{cpu-pu}
\end{equation}
 After some algebra, we verify that such conversion coefficients, shown in Table \ref{direct}, depend only on the relativistic perturbation parameter $\bar{\sigma}$ (Eq.~\eqref{sigma-par-cpu}) and the polytropic index $n$. This conclusion offers the freedom of choosing a ``virtual polytropic constant'' (one not corresponding to a fine-tuning value for a given model),  
 \begin{equation}
 K_\mathrm{gu}=1,   
 \label{Keq1}
 \end{equation}
 for the needs of computations and comparisons with  codes of other investigators, provided that the obtained results are (shall be) quoted in pu. 

\begin{table}[ht]
\begin{center}
\renewcommand*{\arraystretch}{1.3}
\caption{Conversion coefficients from cpu to pu. The polytropic index $n$ and the relativity parameter $\bar{\sigma}$ ( Eq.~\eqref{sigma-par-cpu}) are basic input parameters for our models.
\\}
\label{direct}
\begin{tabular}{l|l} 
\hline
\\ [-1.5ex]
Physical quantity  & Conversion coefficients: symbols and definitions          \\
\\ [-1.5ex]
\hline
\\ [-1.5ex]
Density            & $[D]_\mathrm{cpu\_pu}$ = ${\bar{\sigma}}^{n}$\\
Pressure           & $[P]_\mathrm{cpu\_pu}$ = ${\bar{\sigma}}^{n+1}$ \\
Angular velocity   & $[\Omega]_\mathrm{cpu\_pu}$ = $\left[4\pi{\bar{\sigma}}^{n}\right]^{1/2}$  \\
Length             & $[L]_\mathrm{cpu\_pu}$ = $\left[(n+1)/(4\pi{\bar{\sigma}}^{n-1})\right]^{1/2}$  \\
Mass               & $[M]_\mathrm{cpu\_pu}$ = $\left[(n+1)^{2}/(4\pi{\bar{\sigma}}^{n-2})\right]^{1/2}$ \\
Energy             & $[E]_\mathrm{cpu\_pu}$ =  $\left[(n+1)^{5}/(4\pi{\bar{\sigma}}^{n-5})\right]^{1/2}$ \\
Angular momentum   & $[J]_\mathrm{cpu\_pu}$ = $\left[(n+1)^5/(16\pi^2{\bar{\sigma}}^{2n-5})\right]^{1/2}$ \\
Moment of inertia  & $[I]_\mathrm{cpu\_pu}$ = $\left[(n+1)^5/(64\pi^3{\bar{\sigma}}^{3n-5})\right]^{1/2}$  \\
[1.5ex]
\hline
\end{tabular}
\end{center}
\end{table}
    
It is finally worth noting that in the cpu system of units, Eqs.\eqref{upsilon-par} and \eqref{sigma-par} become
\begin{align}
    \bar{\upsilon}=2~\Omega_{c}^2 \quad &\rightarrow \quad \Omega_{c}=\sqrt{\frac{\bar{\upsilon}}{2}}, \label{upsilon-par-cpu} \\ 
    \bar{\sigma}=\frac{1}{(n+1)c^2} \quad &\rightarrow \quad \frac{1}{c^2}=(n+1)~\bar{\sigma}.
    \label{sigma-par-cpu}
\end{align}

\section{The Numerical Method}
\label{sec3}
As said above, our numerical method, hereinafter referred to as ``Post-Newtonian Rotating Polytropes'' (PNRP), is a combination of CIT \citep{G1991}, HAS \citep{GK2014}, and the method described by Seguin \citep{S1973}. In our study, we employ the generalized Lane-Emden variables. All the expressions and quantities, hereafter, are modified appropriately in order to be expressed in the cpu system of units.

\subsection{Generalized Lane-Emden function}
The generalized Lane-Emden function $\Theta(\xi,\mu)$
is defined as
\begin{align}
    \varrho &= \varrho_{c}\Theta(\xi,\mu)^n, \;\; \mathrm{hence} \label{rho_pol} \\
    P &= K\varrho_{c}^{\Gamma}\Theta(\xi,\mu)^{n+1}, \;\; \mathrm{where} \label{P-pol} \\
    \xi &= r/\alpha, \; \; \mathrm{and} 
    \label{xi} \\
    \mu &= \mathrm{cos}(\theta) \label{mu};
\end{align}
$\varrho_{c}$ is the central rest-mass density and $P_{c}$ the central pressure. Apparently, $\xi$ is the dimensionless length and $\Theta^n$ the dimensionless rest mass density, with
\begin{equation}
\begin{split}
    \Theta = 1 \quad & \mathrm{at} \quad \xi=0,\\
    \Theta = 0 \quad & \mathrm{at} \quad \xi=\xi_{s},
    \label{theta-initial-conditions}
\end{split}    
\end{equation}
where $\xi_{s}$ denotes the star radius at a point of its surface. In view of these definitions, Eq.~\eqref{seg-64} becomes 
\begin{equation}
\begin{split}
    \Theta =\mathcal{U}+D = \left\{ H 
    +\frac{1}{c^2}\left[2\Phi+W+2s^2\Omega^2U-4s\Omega
    U_{\phi}+\frac{(H+\delta)^2}{2}\right]\right\} \, + D.
    \label{theta-seg}
\end{split}
\end{equation}

When considered at a point $(\mu_{i}$, $r_{j})$ of an appropriate grid, Eq.~\eqref{theta-seg}  can be written as
\begin{equation}
    \Theta_{ij} =\mathcal{U}_{ij}+D_{ij}.
\end{equation}
The evaluation of the integration constant, $D$, can be achieved at the center of the configuration. With $B_c=W_c=U_{\phi c}=0$ and $\Theta(0)=1$, we find
\begin{equation}
    D = 1-U_{c}-\frac{1}{c^2}\left[ 2\Phi_{c}+\frac{(H_c+\delta)^2}{2} \right].
    \label{D-seg}
\end{equation}
Given a grid of points $(\mu_{i}$, $r_{j})$ of size $\texttt{KAP} \times \texttt{KRP}$, an alternative manner to compute $D$ is to express it as the mean value \citep{G1991} 
\begin{equation}
    \Tilde{D}=\langle D_{ij} \rangle = \langle \Theta_{ij}-\mathcal{U}_{ij} \rangle,
    \label{const-U0}
\end{equation}
by taking into account only the points inside the configuration, i.e., $\Theta_{ij}>0$. Compared to Eq.~\eqref{D-seg}, Eq.~\eqref{const-U0} can be seen as an `improved' computation of the integration constant in the sense that it takes into account all the points of the grid inside the star; while the computation via Eq.~\eqref{D-seg} $D$ takes into account only one point, the center of the star.

\subsection{Rotation law}
In order to proceed to the numerical implementation of our method, we have to choose a rotation law. In this study, we adopt the ``generalized Clement's rotation law'' (see e.g. \cite{G1990ApJ...350..355G})
\begin{equation}
    \omega(s)=\frac{\Omega(s)}{\Omega_{c}}=\left[\sum_{k=1}^{6}a_k e^{-b_{k}F_{r}s^2}\right]^{1/2},
    \label{rot-law}
\end{equation}
where $s$ is the cylindrical coordinate, $\Omega_{c}$ the angular velocity at the center, $a_{i}$ and $b_{i}$ the so-called ``non-uniformity parameters'', improved values of which are shown in Table~\ref{tab:a-b}, and $F_{r}$ a parameter denoting the `strength' of the differential rotation; the value $F_{r}=0$ denotes rigid rotation, while a value $0 < F_{r} \leq 1$ denotes differential rotation of respective strength.
\newcolumntype{C}[1]{>{\centering\arraybackslash}m{#1}}
\begin{table*}[ht]
\begin{center}
\renewcommand*{\arraystretch}{1.4}
\caption{Non-uniformity parameters $a_k$ and $b_k$ for several polytropic indices.}
\label{tab:a-b}

\begin{tabular}{C{1.7cm}C{1.7cm}C{1.7cm}C{1.7cm}C{1.7cm}C{1.7cm}C{1.7cm}} 
\hline \\ [-3.25ex]
& $k=1$ & $k=2$ & $k=3$ & $k=4$ & $k=5$ & $k=6$
\\ [0.25ex]
\hline 
& \multicolumn{6}{c}{$n=1.00$}
\\
\hline
$a_k$ & $0.177936$ & $0.177496$ & $0.167989$ & $0.138172$ & $0.160903$ & $0.177504$ \\
$b_k$ & $0.130844$ & $0.130844$ & $0.130844$ & $0.130844$ & $0.130844$ & $0.130844$
\\
\hline
& \multicolumn{6}{c}{$n=2.50$}
\\
\hline
$a_k$ & $0.247941$ & $0.152209$ & $0.021559$ & $0.247470$ & $0.241111$ & $0.089710$ \\
$b_k$ & $0.169188$ & $0.410591$ & $1.031256$ & $0.169180$ & $0.410598$ & $0.051024$
\\
\hline
& \multicolumn{6}{c}{$n=2.90$}
\\
\hline
$a_k$ & $0.434648$ & $0.251696$ & $0.008049$ & $0.223664$ & $0.068581$ & $0.013362$ \\
$b_k$ & $0.294168$ & $0.136841$ & $1.585230$ & $0.610060$ & $0.064196$ & $0.020817$
\\
\hline
\end{tabular}
\end{center}
\end{table*}
We consider two alternative cases for the rotation law. Namely, we first request the ratio $\Omega_{c}/\Omega_{e}$ to remain constant and equal to an `input value'. So, during the iterative process, the configuration in the current iteration will have equatorial radius $\xi_{e}$ different to the equatorial radius of the previous iteration. As the equatorial radius changes, To keep same $\Omega_{c}/\Omega_{e}$, we  `correct' $F_r$ accordingly. Denoting by $F_{r\mathrm{(old)}}$ the $F_r$ of the previous model and $F_{r\mathrm{(new)}}$ the $F_r$ of the current model, we take
\begin{equation}     F_{r\mathrm{(new)}}=F_{r\mathrm{(old)}}\left(\frac{\xi_{1}}{\xi_{e}}\right)^{2}.
\end{equation}
So, the rotation law \eqref{rot-law} is modified as
\begin{equation}
    \omega(s)=\frac{\Omega(s)}{\Omega_{c}}=\left[\sum_{k=1}^{6}a_k e^{-b_{k}F_{r}\left(\frac{\xi_{1}}{\xi_{e}}\right)^{2}s^2}\right]^{1/2}.
    \label{rot-law-2}
\end{equation}

The first case of differential rotation is referred to as ``constant ratio model''. In the second case, referred to as ``variable ratio model'', the ratio $\Omega_{e}/ \Omega_{c}$ of the previous model does not remain equal to the ratio $\Omega_{e}/ \Omega_{c}$ of the current model. This happens because, as $\Omega_{c}$ increases in each iteration, the equatorial radius $\xi_{e}$ of the new model is different to that of the previous one. The sequence of models, produced in this case, is one with same central density and higher $\Omega_{c}$ in each iteration until the maximum value of $\Omega_{c}$ is reached. 

It is worth clarifying here that a critically rotating configuration obtained by applying the variable ratio model coincides with the respective configuration obtained by applying the constant ratio model, provided that the computed ratios $\Omega_{e}/ \Omega_{c}$ of these two configurations become equal to each other.

\subsection{Computing the guess density function and the potentials}
CIT is an iterative method that computes significant quantities on a predefined grid of ($\mu,\xi$) points, where $\xi$ and $\mu$ are defined by Eqs.~\eqref{xi} and \eqref{mu}. The grid is then constracted as \cite[Eqs.~(14-16)]{PG2002}
\begin{equation}
\begin{split}
    \mu_{i}=&(i-1)/(\texttt{KAP}-1),  \qquad\qquad  i=1,...,\texttt{KAP}, \\
    \xi_{j}=&(j-1)/(\texttt{KRP}-1) \times \xi_{\mathrm{end}}, \quad j=1,...,\texttt{KRP},
    \label{mesh-points}
\end{split}
\end{equation}
where $\xi_{\mathrm{end}}$ is an appropriate upper limit for $\xi$. Since the cylindrical coordinate $\Tilde{\omega}$ appears in several relations, it is necessary to express it on the grid points. The dimensionless cylindrical coordinate $s$, $\Tilde{\omega}=\alpha s$, is related to the dimensionless spherical coordinates $\mu$  and $\xi$ as
\begin{equation}
    s_{ij}=\xi_{j}\sqrt{1-\mu_{i}^{2}}.
    \label{s-ij}
\end{equation} 
Any quantity then, that depends on $\mu$ and $\xi$, will be assigned a value at any grid point. 

At a first step, we have to supply CIT with an ``initial guess'' for the function $\Theta(\xi,\mu)$, which is necessary for starting the iterative procedure. To accomplish that, we use HAS \citep[\S2, Eqs.~(4-10); \S4]{GK2015} to calculate $\Theta(\xi,\mu)$ of the undistorted configuration on the grid.

To be able to continue, we have to express the quantities \eqref{pot-B-int}, \eqref{pot-W-int}, \eqref{pot-U-int}, \eqref{pot-Phi-int}, \eqref{pot-Uphi-int} and \eqref{pot-U*-int} on the grid.  

The dimensionless angular velocity $\Omega(s)$ at each grid point can be computed by using Eqs.~\eqref{rot-law} and \eqref{upsilon-par-cpu}, 
\begin{equation}
    \Omega_{ij}\equiv\Omega_{ij}(s)=\Omega_{c} \, \omega_{ij}(s)=\sqrt{\frac{\bar{\upsilon}}{2}}\left[\sum_{k=1}^{6}a_k e^{-b_{k}F_{r}s_{ij}^{2}}\right]^{1/2}.
    \label{omega-cpu}
\end{equation}
Knowing the values of $\Omega_{ij}$ at each grid point, the computation of the dimensionless potentials $B$ and $W$ on the grid can be achieved as follows. First, we calculate the potentials in cylindrical coordinates for each $j$-point of the grid by direct integration, 
\begin{equation}
    B_{j} = \int_0^{s_{1j}} s \, \Omega^2ds; \quad   W_{j} = 
            \int_0^{s_{1j}} s^3 \, \Omega^4ds.
            \label{BW}
\end{equation}
Since the quantities $B$ and $W$ are independent of the cylindrical coordinate $z$, they have same values for any $z$. Consequently, to determine the values of the potentials in spherical coordinates, $B_{ij}$ and $W_{ij}$, at each grid point, we simply interpolate the functions $B_{j}$ and $W_{j}$ accordingly.
        
As we have assumed axial symmetry, all the quantities of interest depend only on two variables, and the respective integrals end up to be double integrals. To integrate such integrals, we adopt a method based on Simpson's formula, proposed and applied by \cite{H1986}. For instance, the gravitational potential $U$ can be calculated as \cite[Eq.~(2,3)]{H1986}
\begin{equation}
\begin{split}
    U(\mathbf{r'}) =& G\int \frac{\varrho(\mathbf{r'})}{|\mathbf{r}-\mathbf{r'}|} d^3\mathbf{r'} \\
    =&-4\pi G \int_{0}^{\infty}dr'\int_{0}^{1}d\mu' \times \sum_{n=0}^{\infty}f_{2n}(r',r)P_{2n}(\mu)P_{2n}(\mu')\varrho(\mu',r'),
\end{split}
\end{equation}
where
\begin{equation}
    F_{2n}(r',r)=
    \begin{cases}
    r'^{2n+2}/r^{2n+1},\quad &r' < r,\\
    r^{2n}/r'{2n-1},  &r' > r,\\
    r,  &r' = r.
    \end{cases}
    \label{F2n}
\end{equation}
As a first step to calculate this, we construct a grid  ($\mu_{i},r_{j}$) of size $\texttt{KAP} \times \texttt{KRP}$ in accordance with the relations \eqref{mesh-points}. Then, the gravitational potential is calculated from the following three steps \cite[Eq.~(54-56)]{H1986}:
\begin{equation}
\begin{split}
    D_{k,n}^{(1)}=\sum_{i=1(2)}^{\texttt{KAP}-2}\left(\frac{1}{6}\right)\left(\mu_{i+2}-\mu_{i}\right)  \left[P_{2n}(\mu_{i})\varrho_{i,k} 
    +4P_{2n}(\mu_{i+1})\varrho_{i+1,k}+P_{2n}(\mu_{i+2})\varrho_{i+2,k}\right],
    \label{Hac-D1}
\end{split}
\end{equation}
\begin{equation}
\begin{split}
    D_{n,j}^{(2)}=\sum_{i=1(2)}^{\texttt{KRP}-2}\left(\frac{1}{6}\right)\left(r_{k+2}-r_{k}\right)\left[f_{2n}(r_{k},r_{j})D_{k,n}^{(1)}
    +4f_{2n}(r_{k+1},r_{j})D_{k+1,n}^{(1)}+f_{2n}(r_{k+2},r_{j})D_{k+2,n}^{(1)}\right],
    \label{Hac-D2}
\end{split}
\end{equation}
\begin{equation}
    U_{i,j}=-\sum_{n=0}^{\texttt{LMAX}}D_{n,j}^{(2)}P_{2n}(\mu_{i}),
    \label{Hac-Uij}
\end{equation}
where the notation $i=1(2)$ means that $i$ increases by 2 starting from 1, and $\texttt{LMAX}$ denotes the cutoff number of Legendre polynomials (usually $\texttt{LMAX} = 8 \: \mathrm{or} \: 16$). In \eqref{Hac-Uij}, the term $4\pi$ has been incorporated into the units. Similar quantities can be computed in the same way by replacing the respective terms to be integrated.  

Furthermore, quantities such as the rest mass, $M$, can be integrated in a similar way \cite[Eqs.~(57,58)]{H1986},
\begin{equation}    
    Q_{j}^{(1)}=\sum_{i=1(2)}^{\texttt{KAP}-2}\left(\frac{1}{6}\right)\left(\mu_{i+2}-\mu_{i}\right)  \left[\varrho_{i,j}+4\varrho_{i+1,j}+\varrho_{i+2,j}\right],
    \label{Hac-Q1}
\end{equation}
\begin{equation}
\begin{split}
    M=\sum_{i=1(2)}^{\texttt{KRP}-2}\left(\frac{1}{6}\right)&\left(r_{j+2}-r_{j}\right) \left[r_{j}^{2}Q_{j}^{(1)}+4r_{j+1}^{2}Q_{j+1}^{(1)}+r_{j+2}^{2}Q_{j+2}^{(1)}\right].
\end{split}
\end{equation}

\section{Outline of the Numerical Method}
\label{outline}
All the relations, appearing in this section, are considered to involve quantities referred to the cpu system of units.
\begin{enumerate}   
    \item \label{firstinput} 
    We choose input values for the polytropic index, $n$, the relativity parameter, $\bar{\sigma}$, and the grid size, $\texttt{KAP} \times \texttt{KRP}$, which directly affects the accuracy of the computations (in order for the results to be reliable, the grid size must be appropriately large; typically, but not exclusively, we employ a size of $\texttt{KAP} \times \texttt{KRP} = 131 \times 261$). 
    
    Since we are not going to quote, or to compare, any results in the cgs system of units, we use exclusively the virtual polytropic constant $K_\mathrm{gu}=1$ (Eq.~\ref{Keq1})) when implementing  other available public domain codes for computing results and comparing with respective ones of our code.

    \item \label{secondinput}
    We also choose input values for the following parameters:
    \begin{enumerate}        
        \item \label{FrRatio}
        The strength $F_r$ of the differential rotation, or, equivalently, the ratio $\Omega_{c}/ \Omega_{e}$.
        \vspace{1.5mm}        
        \item \label{upsilons}
        An `initial step' $\bar{\upsilon}_\mathrm{start}$, a `working step' $\bar{\upsilon}_\mathrm{step}$, and a `minimum step' $\bar{\upsilon}_\mathrm{min}$ permitted for the working step. The latter plays the role of a `global termination criterion' for the iterative procedure when, after successive reductions of the working step, it becomes less than the minimum step, 
        $\bar{\upsilon}_\mathrm{step}<\bar{\upsilon}_\mathrm{min}$. We also initialize a `working rotation parameter', $\bar{\upsilon}_\mathrm{now}$, 
        \begin{equation}            \bar{\upsilon}_\mathrm{now}=\bar{\upsilon}_\mathrm{start}.
        \end{equation}  
        \item \label{zetadescription}
        A requested percentage limit, $\delta\zeta$, which plays the role of a `criterion on reducing $\bar{\upsilon}_\mathrm{step}$' (see Eq.~\eqref{zeta}) for the iterative procedure to keep or reduce the value of the working step in the following iterative procedure.
        \item \label{deltaU}
        A requested accuracy, $\delta U$, which plays the role of a `partial termination criterion' (see Eqs.~\eqref{100U}) for the current iteration; so, provided that this criterion is satisfied, the current iteration is terminated and a new iteration starts.
    \end{enumerate} 
    
    \item 
    We calculate and save in appropriate arrays the coordinates $\mu_{i}$, $\xi_{j}$, and $s_{ij}$ on the grid by using Eqs.~(\ref{mesh-points}) and (\ref{s-ij}). 
    The upper limit $\xi_\mathrm{end}$  of $\xi$ is taken to be $\xi_\mathrm{end} \simeq 4\xi_{1}$, $\xi_{1}$ being the radius of the undistorted configuration. Likewise, 
    we construct appropriate arrays of values of the Legendre polynomials $P_{2q}(\mu)$ and of the function $F_{2q}$ (Eq.~(\ref{F2n}); $r'$ and $r$ are substituted by $\xi_i$ and $\xi_j$, respectively),
    \begin{equation}
        F_{2q}(\xi_i,\xi_j)=
        \begin{cases}
        \frac{\xi_i^{2q+2}}{\xi_j^{2q+1}},\quad &\mathrm{if} \quad \xi_i < \xi_j, \\
        \frac{\xi_j^{2q}}{\xi_i^{2q-1}},  &\mathrm{if} \quad \xi_i > \xi_j, \\
        \xi_i,   &\mathrm{if} \quad\xi_i = \xi_j.
       \end{cases}
   \end{equation}   
   \item \label{itm:theta}
   By applying HAS \citep[\S2, Eqs.~(4-10); \S4]{GK2015} with
   \begin{equation}
      \bar{\upsilon}=\bar{\upsilon}_\mathrm{start}, 
   \end{equation}
   we construct the array $\Theta_{ij}$ of the initial guess function $\Theta(\xi_j,\mu_i)$ on the grid. 
   \item \label{itm:theta2}
   It can be verified that, for each $i$, there exists a value $j=\nu(i)$ for which 
 $\Theta$ becomes for the first time negative, $\Theta_{i\nu(i)}<0$; in addition, there exists a value $j=\ell(i)>\nu(i)$ for which $\Theta_{i\ell(i)}$ becomes a minimum. Then, we `correct' the values of $\Theta$ beyond $\ell(i)$ by setting
    \begin{equation}
        \Theta_{ij} = \Theta_{i\ell(i)}, \quad \forall \, i \And \left( \forall \, j \; \mathrm{such~that:} \; \; \nu(i)<\ell(i)<j\leq\texttt{KRP} \right).
    \end{equation}
    Moreover, to be certain that the central value $\Theta_{c}$ is the same for all angles, i.e. for all the $\Theta_{i1}$ elements, we replace these elements by their mean value,
    \begin{equation}
        \left(\Theta_c = \langle\Theta_{i1}\rangle; \; \mathrm{then} \;\; \Theta_{i1} = \Theta_c \right), \;\; \forall\:i.
    \end{equation}
    In accordance with the initial conditions \eqref{theta-initial-conditions}, however, the central value $\Theta_c$ must remain equal to unity, $\Theta_{c}=1$. Thus, we `normalize' all the elements $\Theta_{ij}$ by setting
    \begin{equation}
        \Theta_{ij}=\frac{\Theta_{ij}}{\Theta_{c}}, \quad \forall\:i \And \forall \, j.
    \end{equation}

    \item \label{itm:first-iter}
    The iterative procedure goes as follows.
    \begin{enumerate}
        \item \label{setupsilon}
        We set
        \begin{equation}
            \bar{\upsilon}=\bar{\upsilon}_\mathrm{now} + \bar{\upsilon}_\mathrm{step}.
        \end{equation}
        \item \label{itm:omega}
        We compute the array $\Omega_{ij}$ on the grid via Eq.~\eqref{omega-cpu}.
        If the ratio $\Omega_{c}/ \Omega_{e}$ has been given instead of $F_r$, we solve Eq.~\eqref{rot-law} with $s=\xi_e$ for the `variable' $F_r$. Then, we calculate $\Omega_{ij}$ on the grid via Eq.~\eqref{omega-cpu}. At the same time, we compute the derivative $d\Omega/ds$, which is involved in the computation of $\Omega^{*}$. By differentiating Eq.~\eqref{omega-cpu}, we find 
        \begin{equation}
        \begin{split}
            \Omega^{'}_{ij} \equiv \left(d\Omega/ds\right)_{ij} = \, -\Omega_{ij}(s_{ij}) \, F_{r} \, s_{ij} \, \left[ \sum_{k=1}^{6}a_k b_k e^{\left({-b_{k}F_{r}s_{ij}^{2}}\right)}\right],
        \end{split}    
        \end{equation}
        
        \item \label{itm:second_iter}
        We then proceed with the second iteration.
        
        (Remark: As the procedure goes on, the first iteration is generally referred to as `previous iteration', while the second iteration is generally referred to as `new or primed iteration'.)
        
        \begin{enumerate}
            \item\label{itm:rho}
             We compute the dimensionless rest-mass density, $\varrho_{ij}$, as follows
        \begin{equation}
        \begin{split}
             \varrho_{ij}&=\Theta_{ij}^{n}\quad \text{for} \quad \Theta_{ij}>0,\\
             \varrho_{ij}&=0 \qquad \text{for} \quad \Theta_{ij}<0.
        \end{split}
        \end{equation}
            
            Using Eqs.~\eqref{Hac-D1}-\eqref{Hac-Uij}, we compute the potentials $U_{ij}$, $U_{\phi\,ij}$ and $\Phi_{ij}$ on the grid, taking as integrant(s) the following quantities, respectively, 
            \begin{gather}
                \Theta^n_{ij}, \nonumber \\
                \Theta^n_{ij}\upsilon_{ij}=\Theta^n_{ij}s_{ij}\Omega_{ij}, \\
                \Theta^n_{ij}\phi_{ij}=\Theta^n_{ij} \left( s_{ij}^2 \Omega_{ij}^{2} + U_{ij} + \frac{(n+3)}{2(n+1)}\Theta_{ij} \right), \nonumber
            \end{gather}
            where, for writing this way the last expression, we use Eqs.~\eqref{phi}, \eqref{np} and \eqref{Omega-star}.
            Note that $U_{\phi}=0$ at the center of the star; thus, we reformulate accordingly Eqs.~\eqref{Hac-D1}-\eqref{Hac-Uij}.          
            
            The computation of the potentials $B_{ij}$ and $W_{ij}$ is achieved by properly interpolating the functions (see remark following Eq.~(\ref{BW}))
            \begin{gather}
                B_{j}=\int_0^{s_{1j}}s_{1j}\Omega_{1j}^{2}ds, \\
                W_{j}=\int_0^{s_{1j}}s_{1j}^3\Omega_{1j}^4ds.
            \end{gather}            
            
            The efficient potential $\mathcal{U}_{ij}$ is computed via Eq.~\eqref{Ueff},
            \begin{equation}
                \mathcal{U}_{ij} = H_{ij} + \frac{1}{c^2}\left[2\Phi_{ij} + W_{ij} + 2s_{ij}^2\Omega_{ij}^2 U_{ij} - 4s_{ij}\Omega_{ij} U_{\phi\,ij}\right],
            \end{equation}
            where
            \begin{equation}
                H_{ij} \equiv U_{ij} + B_{ij}.
            \end{equation}
            
            The integration constant, hereinafter $U_{0} \equiv \Tilde{D}$, is determined by Eq.~\eqref{const-U0},
            \begin{equation}
                U_{0} = \langle \Theta_{ij}-\mathcal{U}_{ij} \rangle.
            \end{equation}
            
            \item
           The new array of values $\Theta^{'}_{ij}$ on the grid is computed via the relation
            \begin{equation}
                \Theta^{'}_{ij}=\mathcal{U}_{ij}-U_{0}.
            \end{equation}
            (Remark: The primed $\Theta^{'}_{ij}$ is simply a symbol for the new iteration).            
            At this point, we again `correct' and `normalize' the values of $\Theta^{'}_{ij}$ as in step~[\ref{itm:theta2}].

            \item \label{cAB}
            We check the reliability of the new array of values $\Theta^{'}_{ij}$ by examining if:
            \begin{itemize}
                \item \label{labelA} 
                A.~The equation  $\Theta^{'}_i(\xi) = 0$ has a root $\Xi_i$ for any $i$. By $\Theta^{'}_i(\xi)$ we denote a function in the `independent variable' $\xi$, which properly interpolates  the values $\Theta^{'}_{ij}$ over all points $\xi_j$. The meaning of this criterion is that the density of the configuration becomes zero at any $\Xi_i$, and thus a boundary can be determined for the configuration. Note that the particular root $\Xi_i(i=\texttt{KAP})$ coincides with the equatorial radius, $\xi_e$, of the configuration; while the particular root $\Xi_i(i=1)$ coincides with the polar radius, $\xi_p$.              
                \item \label{labelB}
                B.~The number of the positive elements of the array ($\texttt{NoE}$) $\Theta^{'}_{ij}$ does not exceed a prescribed percentage, $\delta\zeta$, 
                \begin{equation}
                 100 \times \left[   \frac{\texttt{NoE}\, \Theta^{'}_{ij}>0}{\texttt{KAP} \times \texttt{KRP}} \right] < \delta\zeta.
                 \label{zeta}
                \end{equation}
                We usually set the limit at $\delta\zeta=90\%$. Extended numerical experiments have shown that this condition assigns further stability to the iterative procedure. 
                \end{itemize}
                
                If the criteria A and/or B are not satisfied, we properly reduce the value of $\bar{\upsilon}_\mathrm{step}$, by taking, for instance,
                \begin{equation}
                   \bar{\upsilon'}_\mathrm{step} = \frac{1}{2}\,\bar{\upsilon}_\mathrm{step}; \; \mathrm{then} \;\; \bar{\upsilon}_\mathrm{step} = \bar{\upsilon'}_\mathrm{step}, 
                \end{equation}
                and we return to the step~[\ref{setupsilon}].
                
                If the criteria A and B are both satisfied, we go to the following step.  
            
            \item \label{findxie}
            We check if the new iteration satisfies the condition 
            \begin{equation}
                100 \times \left| \frac{U^{'}_{0}-U_{0}}{U^{'}_{0}} \right| < \delta U,
                \label{100U}
            \end{equation}
            where $U_{0}$ is the integration constant found in the previous iteration, $U^{'}_{0}$ the integration constant found in the new iteration, and $\delta U$ is a prescribed accuracy (see step [\ref{deltaU}]). Typically, but not exclusively, we select for $\delta U$ a value in the interval $[0.1,1.0]$.
            
            If the new iteration fails to satisfy the condition \eqref{100U}, we make the update 
            \begin{equation}
            \Theta_{ij} = \Theta^{'}_{ij},
            \label{peqn}
            \end{equation}
            and return to the step [\ref{itm:second_iter}].
            
            If the new iteration satisfies the condition \eqref{100U}, we set 
            \begin{equation}
               \xi_e = \Xi_i(i=\texttt{KAP}) 
            \end{equation}
            (see step [\ref{cAB}]) and, provided that 
        \begin{equation}
            \bar{\upsilon}_\mathrm{step} > \bar{\upsilon}_\mathrm{min},
            \label{gtv}
        \end{equation}
        we update the value of $\bar{\upsilon}_\mathrm{now}$,
        \begin{equation}
            \bar{\upsilon}_\mathrm{now} = \bar{\upsilon},
            \label{update}
        \end{equation}
        and return to the step~[\ref{setupsilon}]. 
        
        If the condition \eqref{gtv} is not satisfied, we go to the next step.
        \end{enumerate}   
        \end{enumerate}
        \item \label{procedureterminate}
        The iteration procedure (step~[\ref{itm:first-iter}]) is terminated, because the working step has become less than the minimum step, 
        \begin{equation}
           \bar{\upsilon}_\mathrm{step} < \bar{\upsilon}_\mathrm{min}.
           \label{minupsilon}
        \end{equation}
       We take the critical rotation parameter, $\bar{\upsilon}_\mathrm{c}$, to be equal to the last working value of $\bar{\upsilon}$,
       \begin{equation}
         \bar{\upsilon}_\mathrm{c} = \bar{\upsilon}.
         \label{critical}
       \end{equation}       
    We make the update
            \begin{equation}
            \Theta_{ij} = \Theta^{'}_{ij},
            \label{peqn2}
            \end{equation}
    and we compute the array $\Omega_{ij}$ (step~\ref{itm:omega}), we repeat the computations involved in the step~[\ref{itm:rho}], and, finally, we  compute the array $\Omega^{*}_{ij}$ (Eq.~\eqref{Omega-star}),
    \begin{equation}
        \Omega^{*}_{ij}=\left[\Omega_{ij}^{2}+\frac{1}{c^2}4\Omega^{'}_{ij}\left(s_{ij}\Omega_{ij} U_{ij} - U_{\phi ij}\right)\right]^{1/2}.
        \label{Omega-star_ij}
    \end{equation}
\end{enumerate}

After the whole procedure has been terminated, we can repeat the procedure for a different $\bar{\sigma}$ and/or a different $F_\mathrm{r}$. Apparently, to repeat the procedure for a different $\bar{\sigma}$, we first need to recompute the array $\Theta_{ij}$ (step~[\ref{itm:theta}]) of the initial guess function.

\section{Physical Characteristics}
\label{phys-char}
All the relations, appearing in this section, are considered to involve quantities referred to the cpu system of units.

The boundary of the configuration is determined as described in the step[\ref{labelA}/part~A], and, accordingly, the equatorial radius, $\xi_e$, and the polar radius, $\xi_p$.

The rest mass of the star can be computed by Eq.~\eqref{rest_mass},
\begin{equation}
    M_{0} = \int \Theta^{n} \left( 1+\frac{3U}{c^2} \right) d^3x.
\end{equation}
The binding energy of the star, given by \eqref{bind_nrg}, is written via Eq.~\eqref{sigma-par} as
\begin{gather}
\begin{split}
    E_{b}=-\int \frac{1}{2} s^{2}\Omega^{2}\Theta^{n} +& \frac{n\Theta^{n+1}}{n+1} - \frac{1}{2}U^{*}\Theta^{n} 
    \\ 
    +& \frac{1}{c^2} \Theta^{n} \left[ \frac{5}{8}s^{4}\Omega^{4}
    + \frac{11}{4}s^{2}\Omega^{2}U 
    + s^{2}\Omega^{2}\Theta 
    + \frac{2nU\Theta}{n+1} - U^2 - 2s\Omega U_{\phi} \right] d^3x.
    \label{goth_E_cpu}
\end{split}
\end{gather}
The gravitational mass of the star is then computed as
\begin{equation}
    M=M_{0}+\frac{E_{b}}{c^2}.
\end{equation}
The proper mass, $M_p$, can be evaluated by integrating the quantity $\epsilon=\varrho (1+\Pi/c^2)$ over the proper volume
\begin{equation}
    M_p=\int \epsilon dV = \int \varrho \left[ 1+\frac{1}{c^2} \left( \frac{n\Theta}{n+1}+3U \right) \right] d^3x.
\end{equation}

Given that the density $\sigma$ (Eq.~\eqref{sigma}) is expressed as
\begin{equation}
    \sigma = \Theta^n \left[ 1 + \frac{1}{c^2} \left( s^2\Omega^2+2U+\Theta \right) \right],
\end{equation}
the angular momentum (Eq.~\eqref{J-dV}) becomes
\begin{equation}
\begin{split}
    J = \int s^2\Omega^{*}\Theta^{n}  
    + \frac{\Theta^{n}}{c^2} \left[ s^{4}\Omega^{3} + s^{2}\Omega\left( 6U + \Theta \right) - 4sU_{\phi} \right] \, dV.
\end{split}
\end{equation}
Consequently, the rotational kinetic energy (Eq.~\eqref{T-dV})  is given by
\begin{equation}
    T = \frac{1}{2} \int s^2{\Omega^{*}}^{2}\Theta^{n}  
    + \frac{\Theta^{n}}{c^2} \left[ s^{4}\Omega^{4} + s^{2}\Omega^{2}\left( 6U + \Theta \right) - 4s\Omega U_{\phi} \right] \, dV.
\end{equation}
Finally, the gravitational mass density (Eq.~\eqref{rho-grav}) is expressed as
\begin{equation}
    \varrho_{g}=\Theta^{n} \left[ 1+\frac{1}{c^2} \left( 3U -\frac{1}{2} s^{2}\Omega^{2} - \frac{n\Theta}{n+1} + \frac{1}{2}U^{*} \right) \right];
\end{equation}
accordingly, the gravitational potential energy can be computed as 
\begin{equation}
    W = - \, \frac{1}{2} \int \varrho_{g}U_{g} \, dV.
\end{equation}


\begin{table*}[!ht]
\begin{center}
\renewcommand*{\arraystretch}{1.43}
\caption{Comparative table for different strengths of differential rotation and polytropic indices $n=1.00, 2.50,2.90$. All quantities are expressed in the pu system of units.}
\label{tab:n10}
\begin{tabular}{C{1.7cm}C{1.7cm}C{1.7cm}C{1.7cm}C{1.7cm}C{1.7cm}C{1.7cm}} 
\multicolumn{7}{c}{$n=1.00, \qquad \bar{\sigma}=1.2800 \times 10^{-3}, \qquad \rho_\mathrm{c}=1.2800 \times 10^{-3}$}, 
\\
\hline \hline \\ [-3.25ex]
& $\Omega_e (10^{-2})$ & $M (10^{-3})$ & $M_0 (10^{-3})$ & $J (10^{-4})$ & $|T/W|$ & $R_p/R_e$
\\ [0.25ex]
\hline \hline
\multicolumn{7}{c}{$A=3.00, \quad Fr=0.065, \quad \Omega_c/\Omega_e=1.112$}
\\
\hline
PNRP & $2.501$ & $4.522$ & $4.529$ & $0.902$ & $0.119$ & $0.525$ \\
RNSID & $2.503$ & $4.542$ & $4.549$ & $0.909$ & $0.134$ & $0.523$
\\
\hline
\multicolumn{7}{c}{$A=1.90, \quad Fr=0.132, \quad \Omega_c/\Omega_e=1.280$}
\\
\hline
PNRP & $2.362$ & $4.880$ & $4.887$ & $1.131$ & $0.144$ & $0.479$ \\
RNSID & $2.377$ & $4.914$ & $4.921$ & $1.145$ & $0.115$ & $0.476$
\\
\hline
\multicolumn{7}{c}{$A=1.30, \quad Fr=0.049, \quad \Omega_c/\Omega_e=1.538$}
\\
\hline
PNRP & $2.091$ & $5.622$ & $5.632$ & $1.662$ & $0.188$ & $0.399$ \\
RNSID & $2.171$ & $5.839$ & $5.849$ & $1.799$ & $0.170$ & $0.397$
\\
\hline
\\
\end{tabular}

\begin{tabular}{C{1.7cm}C{1.7cm}C{1.7cm}C{1.7cm}C{1.7cm}C{1.7cm}C{1.7cm}} 
\multicolumn{7}{c}{$n=2.50, \qquad \bar{\sigma} = 1.3403 \times 10^{-2}, \qquad \rho_\mathrm{c}=2.0798 \times 10^{-5}$} 
\\ 
\hline \hline \\ [-3.25ex]
& $\Omega_e (10^{-4})$ & $M$ & $M_0$ & $J$ & $|T/W| (10^{-2})$ & $R_p/R_e$
\\ [0.25ex]
\hline \hline
\multicolumn{7}{c}{$A=3.50, \quad Fr=0.0107, \quad \Omega_c/\Omega_e=1.086$}
\\
\hline
PNRP & $1.070$ & $1.267$ & $1.271$ & $0.903$ & $1.868$ & $0.639$ \\
RNSID & $1.059$ & $1.255$ & $1.258$ & $0.895$ & $1.837$ & $0.568$
\\
\hline
\multicolumn{7}{c}{$A=1.40, \quad Fr=0.0491, \quad \Omega_c/\Omega_e=1.538$}
\\
\hline
PNRP & $0.893$ & $1.291$ & $1.295$ & $1.099$ & $2.613$ & $0.554$ \\
RNSID & $0.890$ & $1.279$ & $1.283$ & $1.096$ & $2.497$ & $0.551$
\\
\hline
\multicolumn{7}{c}{$A=1.00, \quad Fr=0.0690, \quad \Omega_c/\Omega_e=2.049$}
\\
\hline
PNRP & $0.734$ & $1.308$ & $1.312$ & $1.235$ & $3.150$ & $0.478$ \\
RNSID & $0.747$ & $1.301$ & $1.305$ & $1.261$ & $3.047$ & $0.479$
\\
\hline
\\
\end{tabular}

\begin{tabular}{C{1.7cm}C{1.7cm}C{1.7cm}C{1.7cm}C{1.7cm}C{1.7cm}C{1.7cm}} 
\multicolumn{7}{c}{$n=2.90, \qquad \bar{\sigma} = 4.4159 \times 10^{-3}, \qquad \rho_\mathrm{c}=1.4810 \times 10^{-7}$} 
\\ 
\hline \hline \\ [-3.25ex]
& $\Omega_e (10^{-5})$ & $M$ & $M_0$ & $J$ & $|T/W| (10^{-2})$ & $R_p/R_e$
\\ [0.25ex]
\hline \hline
\multicolumn{7}{c}{$A=3.50, \quad Fr=0.0056, \quad \Omega_c/\Omega_e=1.083$}
\\
\hline
PNRP & $6.396$ & $3.328$ & $3.329$ & $8.024$ & $1.095$ & $0.640$ \\
RNSID & $6.400$ & $3.332$ & $3.333$ & $8.074$ & $1.091$ & $0.643$
\\
\hline
\multicolumn{7}{c}{$A=1.40, \quad Fr=0.0263, \quad \Omega_c/\Omega_e=1.518$}
\\
\hline
PNRP & $5.371$ & $3.365$ & $3.366$ & $9.531$ & $1.499$ & $0.564$ \\
RNSID & $5.368$ & $3.371$ & $3.372$ & $9.637$ & $1.462$ & $0.562$
\\
\hline
\multicolumn{7}{c}{$A=1.00, \quad Fr=0.0389, \quad \Omega_c/\Omega_e=2.014$}
\\
\hline
PNRP & $4.431$ & $3.395$ & $3.396$ & $1.065$ & $1.820$ & $0.489$ \\
RNSID & $4.485$ & $3.403$ & $3.404$ & $1.086$ & $1.766$ & $0.492$
\\
\hline
\end{tabular}
\end{center}
\end{table*}


\section{Results and Discussion}
\label{res}
All the physical quantities, to be discussed in this section, are considered to be expressed in the pu system of units.

The PNRP code, implementing our numerical method, is written in Fortran and, for its compilation, the GNU Fortran compiler ``gfortran'' is used; it belongs to the GNU Compiler Collection (http://gcc.gnu.org/) and is licensed under the GNU General Public License (http://www.gnu.org/licenses/gpl.html). This environment 
has been installed by the TDM-GCC ‘‘Compiler Suite for Windows’’ (http://tdm-gcc.tdragon.net/), which is free software distributed under the terms of the GPL. 
PNRP cooperates with the Fortran package dcrkf54.f95 \citep{GV2012}, a Runge–Kutta-Fehlberg code of fourth and fifth order appropriately modified for the solution of complex initial value problems with highly complex expressions for their ordinary differential equations along contours, not necessarily simple or closed, prescribed as continuous chains of straight-line segments. In addition, the presented plots of data are derived by using the GNU Plot (http://www.gnuplot.info).
Subroutines required for standard numerical procedures (e.g. interpolation of functions, rootfinding of algebraic equations, localizing extrema of functions, etc.) are taken from ``SLATEC Common Mathematical Library'' (SLATEC). This is an extensive public domain Fortran Source Code Library, incorporating several public domain packages. The full SLATEC release is available at the site https://netlib.org/slatec/.

To resolve a model, we need input values for several basic parameters, enumerated in the steps~[\ref{firstinput}] and~[\ref{secondinput}] of Section~\ref{outline}. Our code is automated so that to read a ``run-stream file'' with the requested values. Alternatively, concerning $\bar{\upsilon}_\mathrm{start}$, $\bar{\upsilon}_\mathrm{step}$,  $\bar{\upsilon}_\mathrm{min}$, and 
$\bar{\upsilon}_\mathrm{now}$,
described in the step~[\ref{upsilons}] of Section~\ref{outline}, the code can automatically assign respective values to these quantities.  

To examine the accuracy and reliability of PNRP, we have to confirm that the derived results are valid in the Newtonian limit (low values of $\bar{\sigma}$), but also to examine the behaviour of the results for cases with high relativistic effects (high values of $\bar{\sigma}$). To accomplish this task, we  compare PNRP results for rigidly and critically rotating polytropic configurations with respective results of the code ``Rotating Neutron Star'' (RNS) \citep[https://github.com/cgca/rns]{Sterg-1995,sterg-2003}, written by N.~Stergioulas. This code, efficiently computing equilibrium models of rigidly and critically rotating neutron stars, has been widely used by many authors.  

Furthermore, to compare PNRP results for diferentially and critically rotating polytropic configurations, we use  the code ``Rotating Neutron Star Initial Data'' (RNSID), written by N.~Stergioulas. RNSID is part of the thorn Hydro\_RNSID \citep{Sterg-1995,2015PhRvD..91f4057L,1996PhDT.........4S,1998LRR.....1....8S,2000MNRAS.313..678F}, which effectively employs RNSID and interpolates its output on a Cartesian grid, thus generating initial data for rotating stars obeying  either a zero-emperature tabulated EOS, or a polytropic EOS. The thorn Hydro\_RNSID is part of the well-known ``Einstein Toolkit'' \citep{EinsteinToolkit:2022_11}. 

\begin{figure}
     \centering
     \begin{subfigure}[t]{0.49\textwidth}
         \centering
         \includegraphics[width=\textwidth]{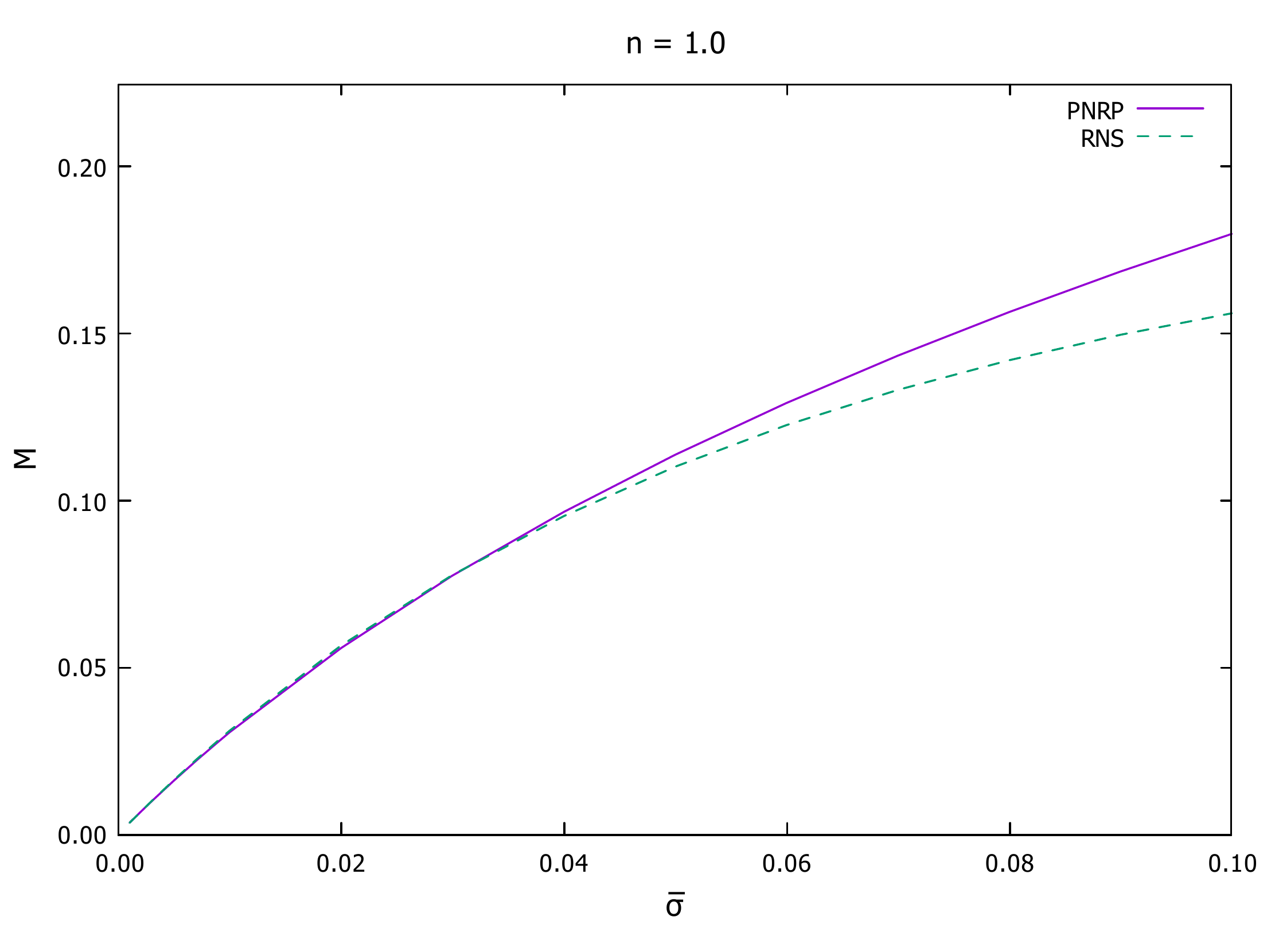}
         \includegraphics[width=\textwidth]{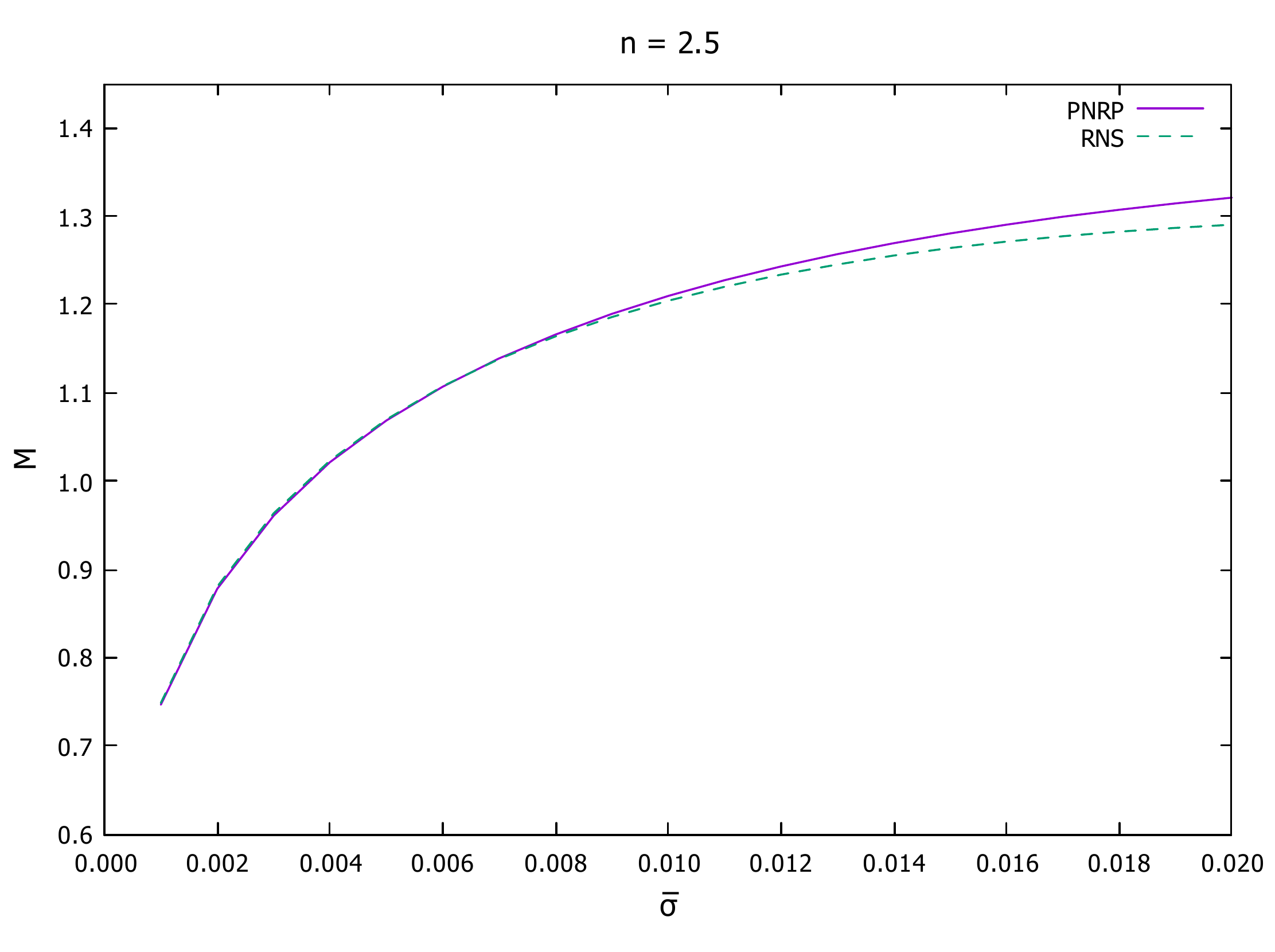}
         \includegraphics[width=\textwidth]{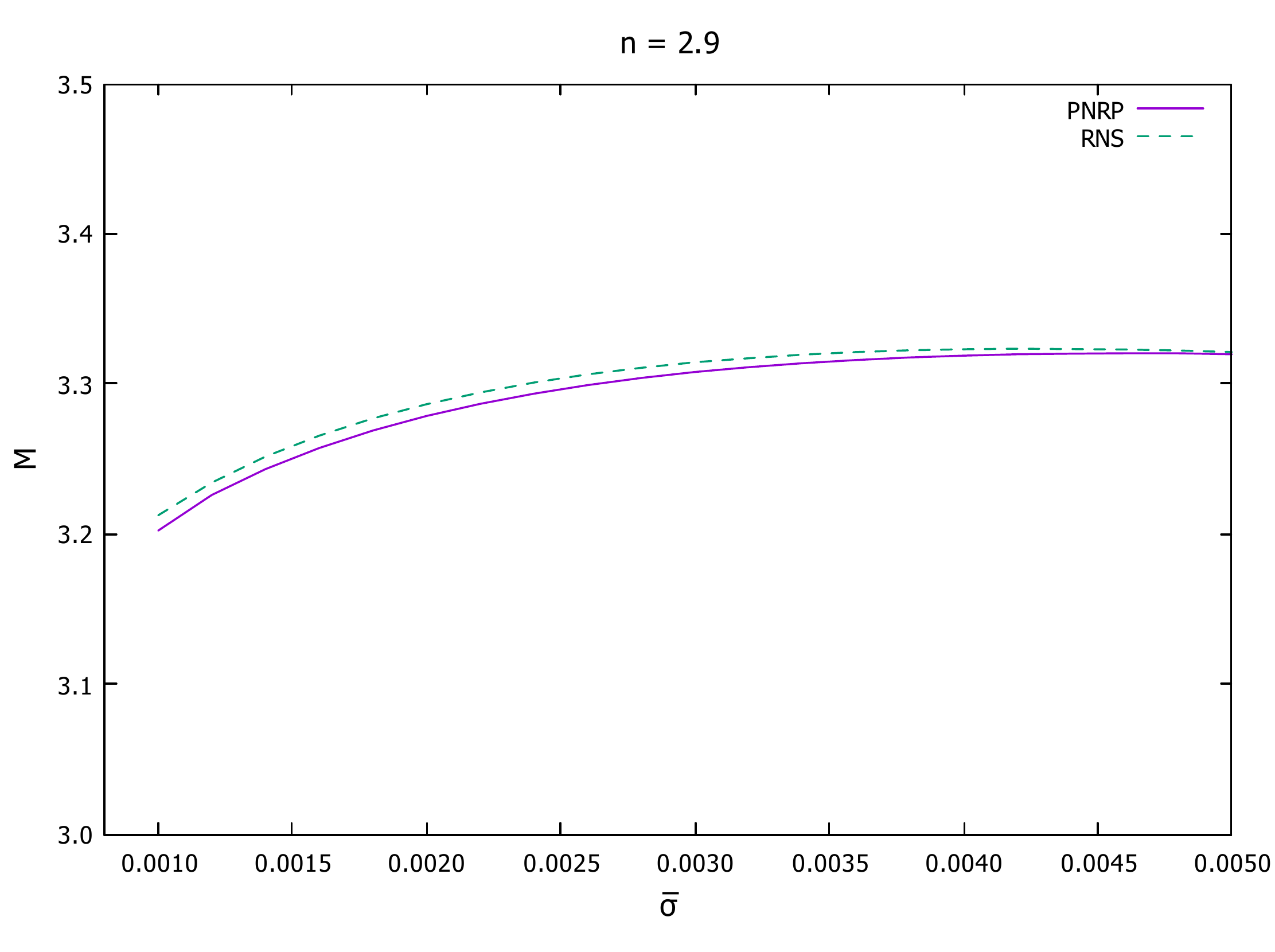}
         \captionsetup{width=0.8\textwidth}
         \caption{Gravitational mass $M$ vs. relativity parameter $\bar{\sigma}$. The polytropic index $n$ is assigned the values $n=1.0$ (upper diagram), $n=2.5$ (middle diagram), and $n=2.9$ (lower diagram). Comparisons are made between results of the PNRP code and the RNS code.}
         \label{fig_grav_mass_rigid}
    \end{subfigure}
    \begin{subfigure}[t]{0.49\textwidth}
         \centering
         \includegraphics[width=\linewidth]{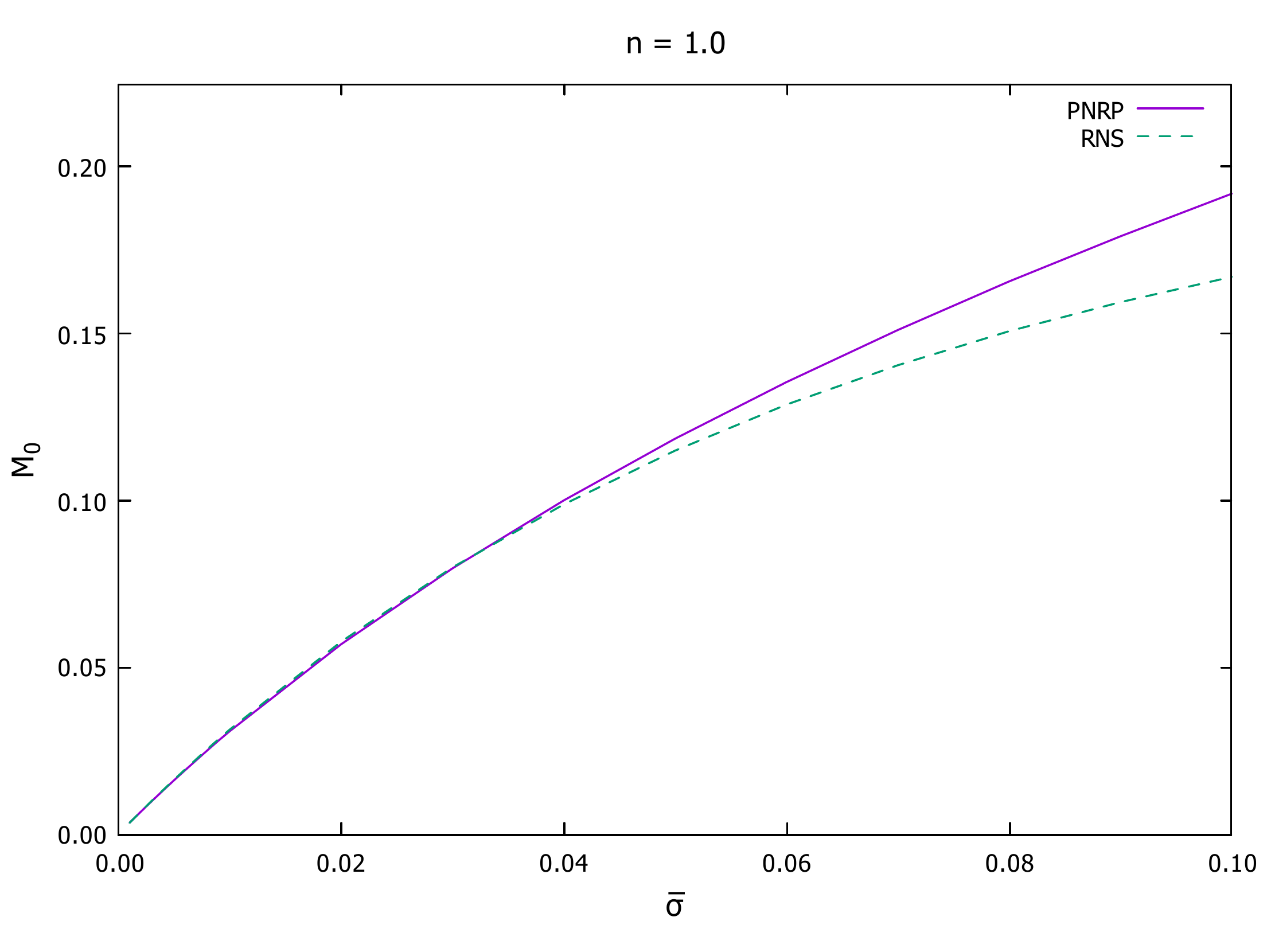}
         \includegraphics[width=\textwidth]{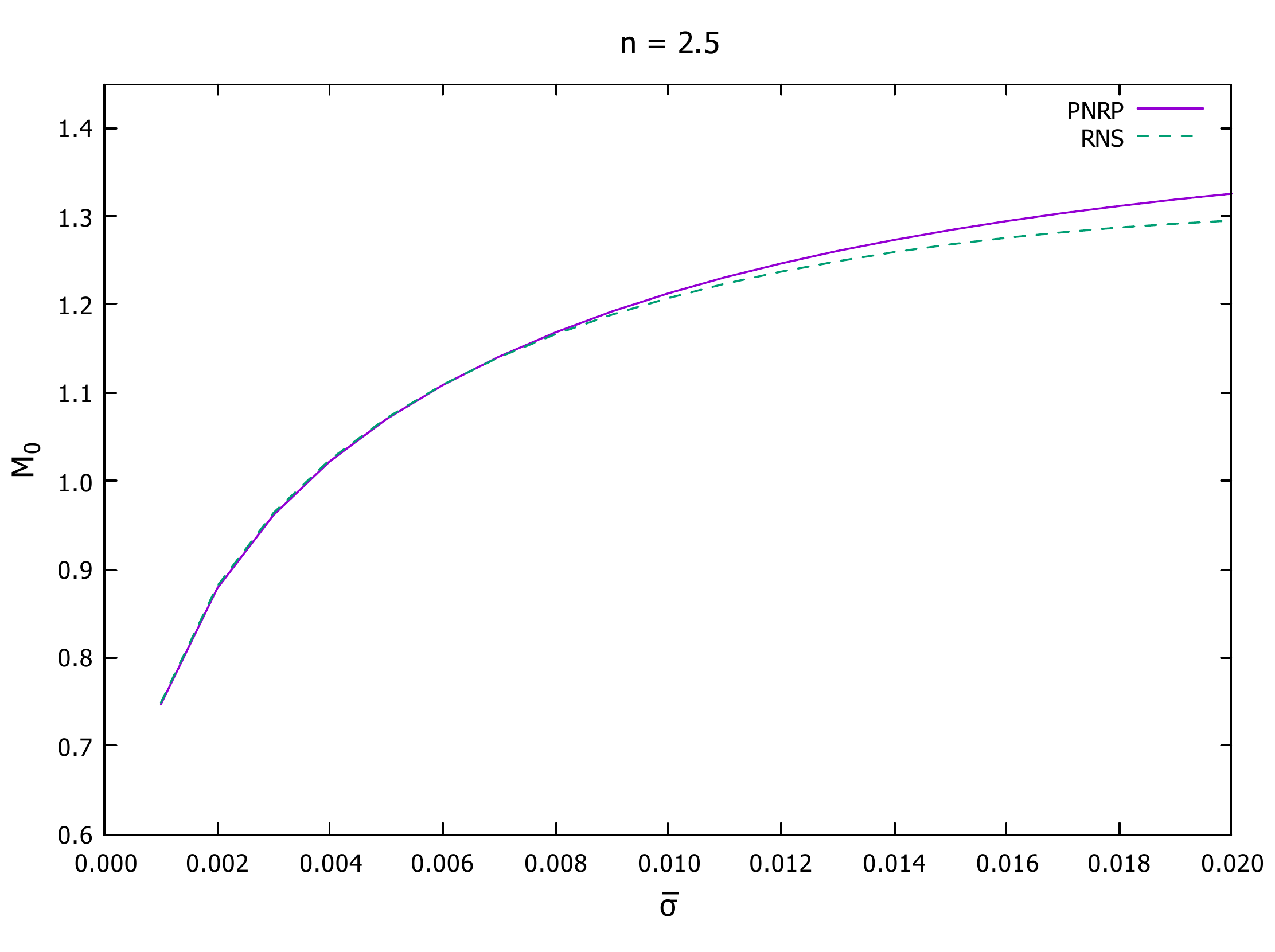}
         \includegraphics[width=\textwidth]{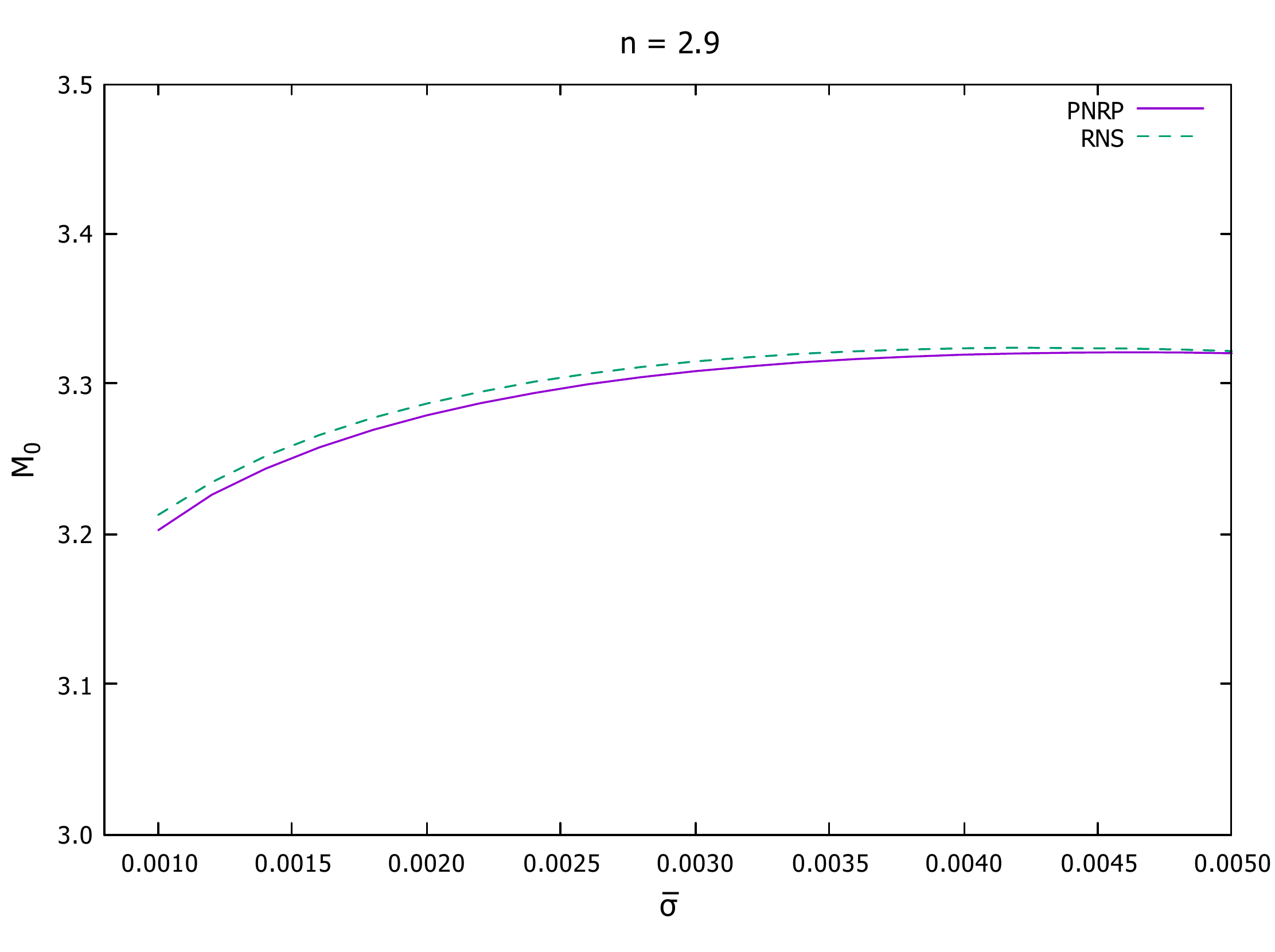}
         \captionsetup{width=0.8\textwidth}
         \caption{Rest mass $M_0$ vs. relativity parameter $\bar{\sigma}$. Details as in Fig.~\ref{fig_grav_mass_rigid}.}
         \label{fig_rest_mass_rigid}
    \end{subfigure}
    \caption{}
    \label{fig-rigid-first}
\end{figure}

\begin{figure}
     \centering
     \begin{subfigure}[t]{0.49\textwidth}
         \centering
         \includegraphics[width=\textwidth]{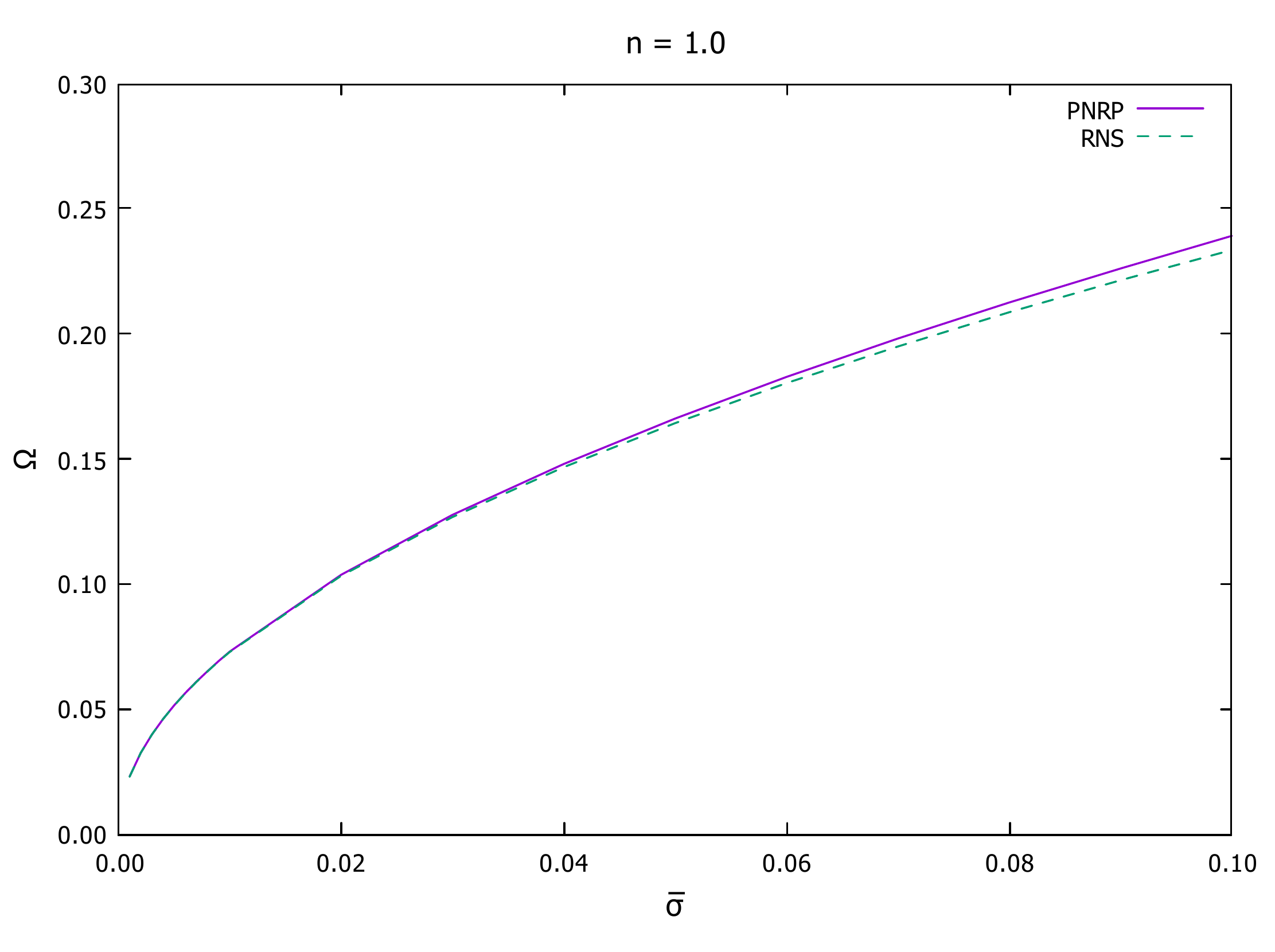}
         \includegraphics[width=\textwidth]{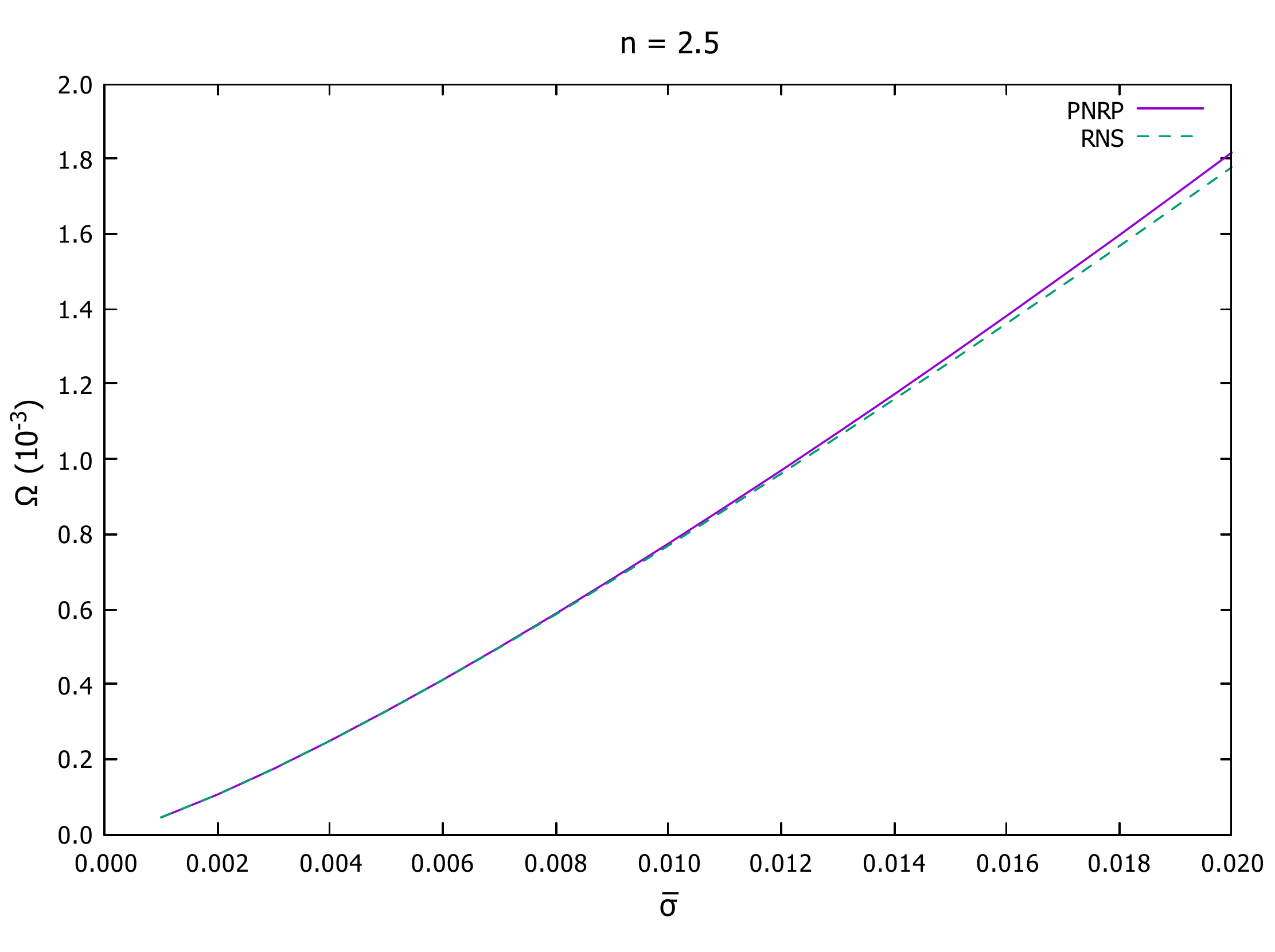}
         \includegraphics[width=\textwidth]{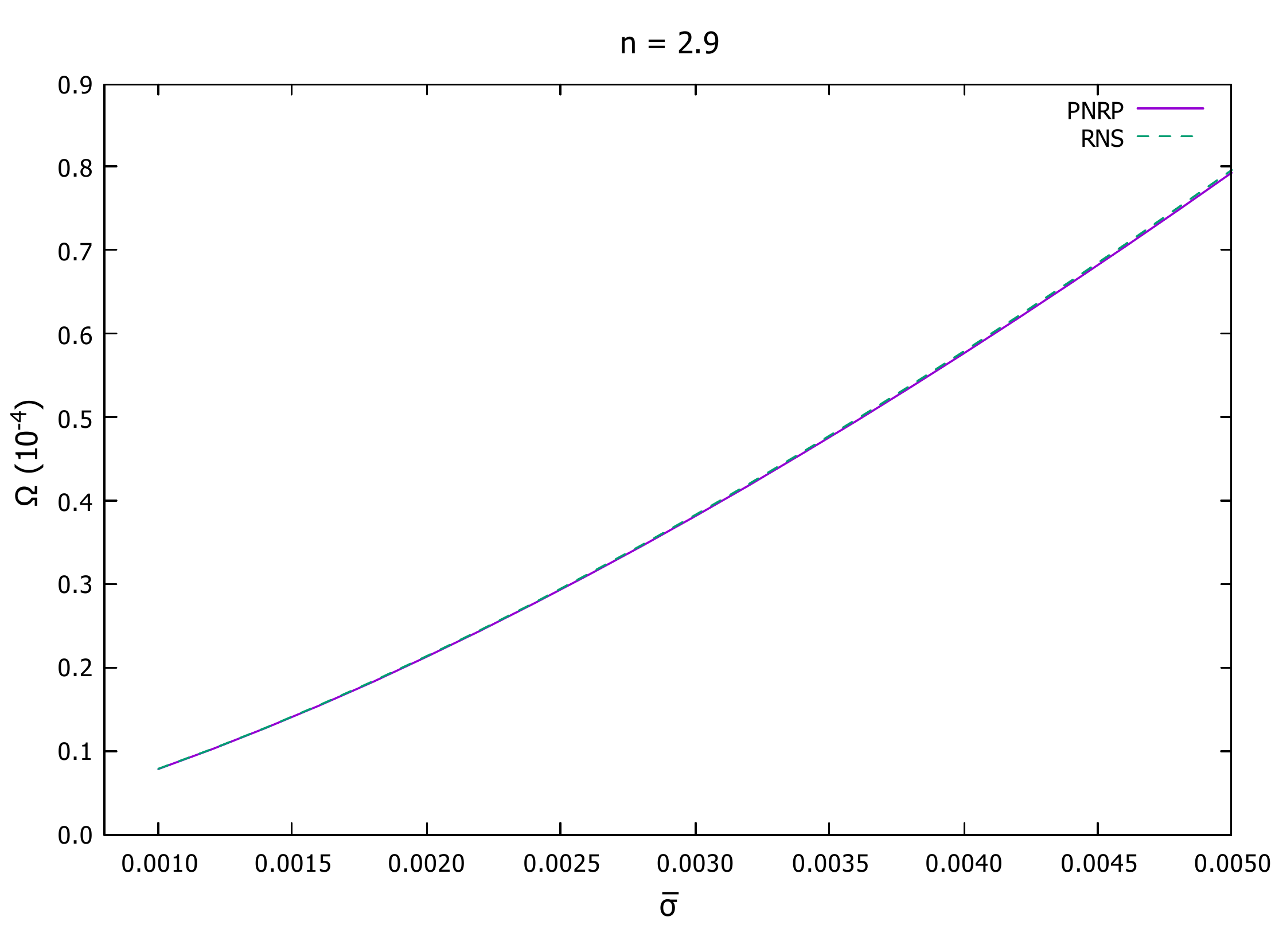}
         \captionsetup{width=0.8\textwidth}
         \caption{Keplerian angular velocity $\Omega$ vs. relativity parameter $\bar{\sigma}$. Details as in Fig.~\ref{fig_grav_mass_rigid}. }
         \label{fig_omega_rigid}
    \end{subfigure}
    \begin{subfigure}[t]{0.49\textwidth}
         \centering
         \includegraphics[width=\linewidth]{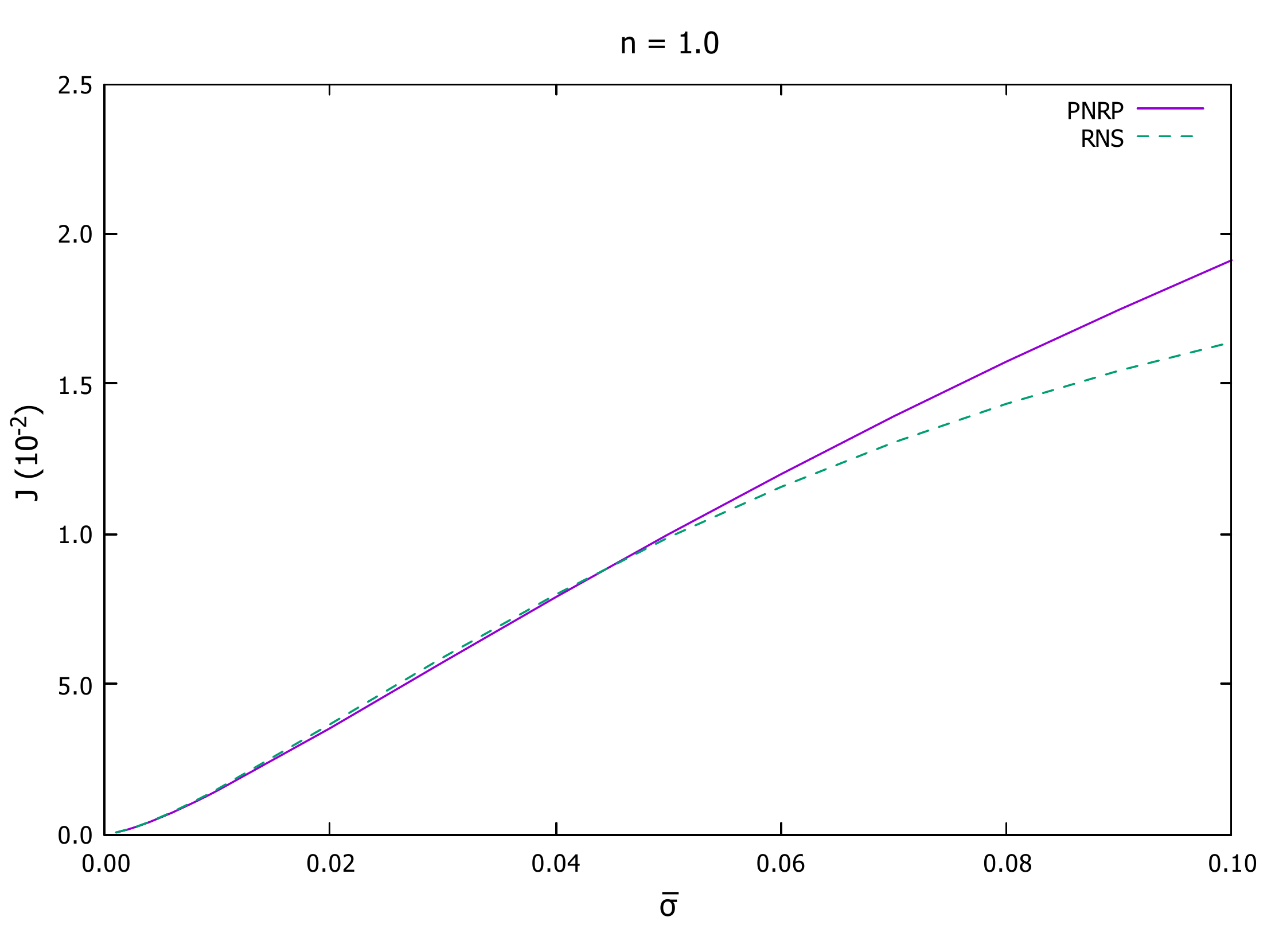}
         \includegraphics[width=\textwidth]{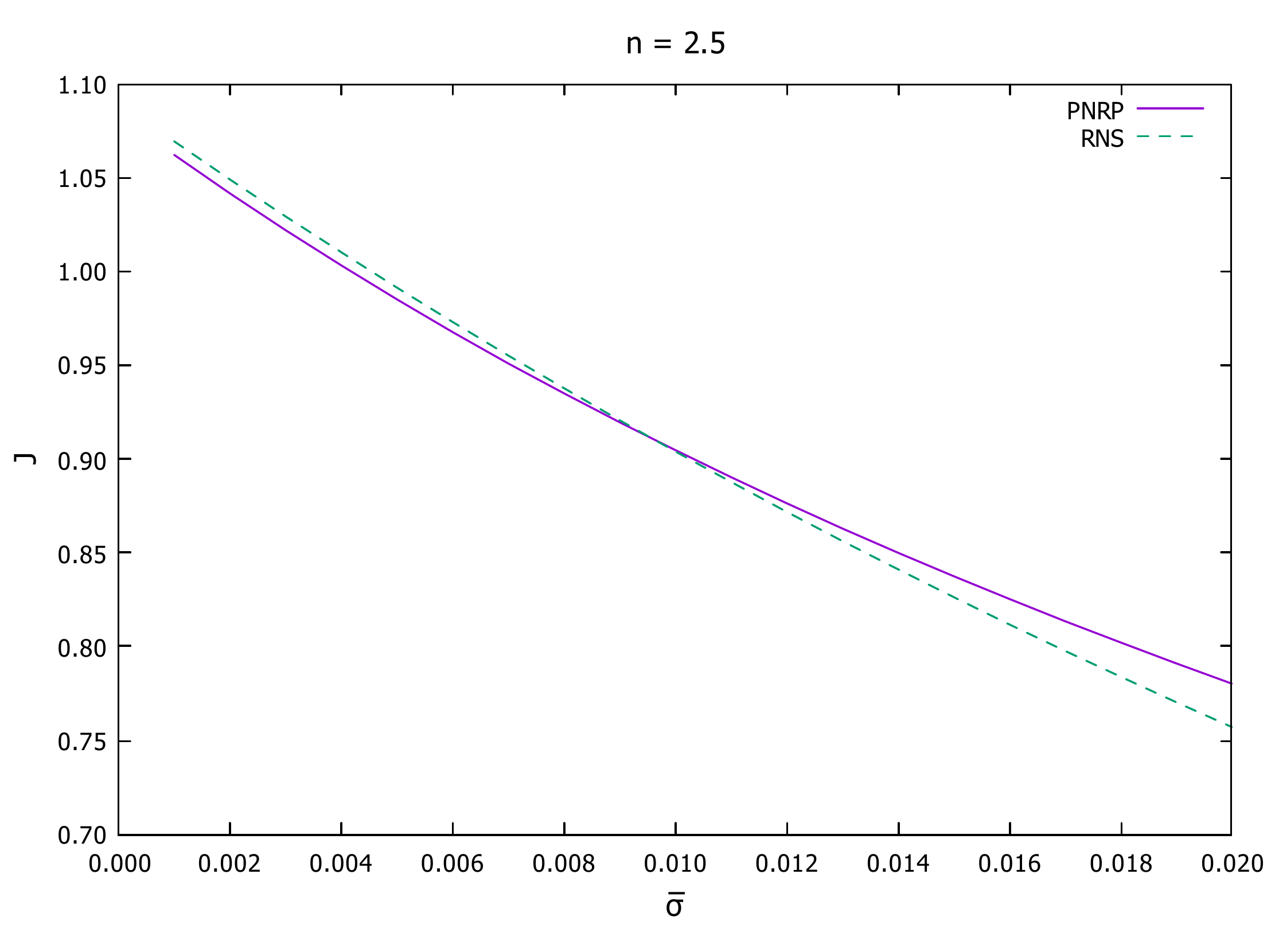}
         \includegraphics[width=\textwidth]{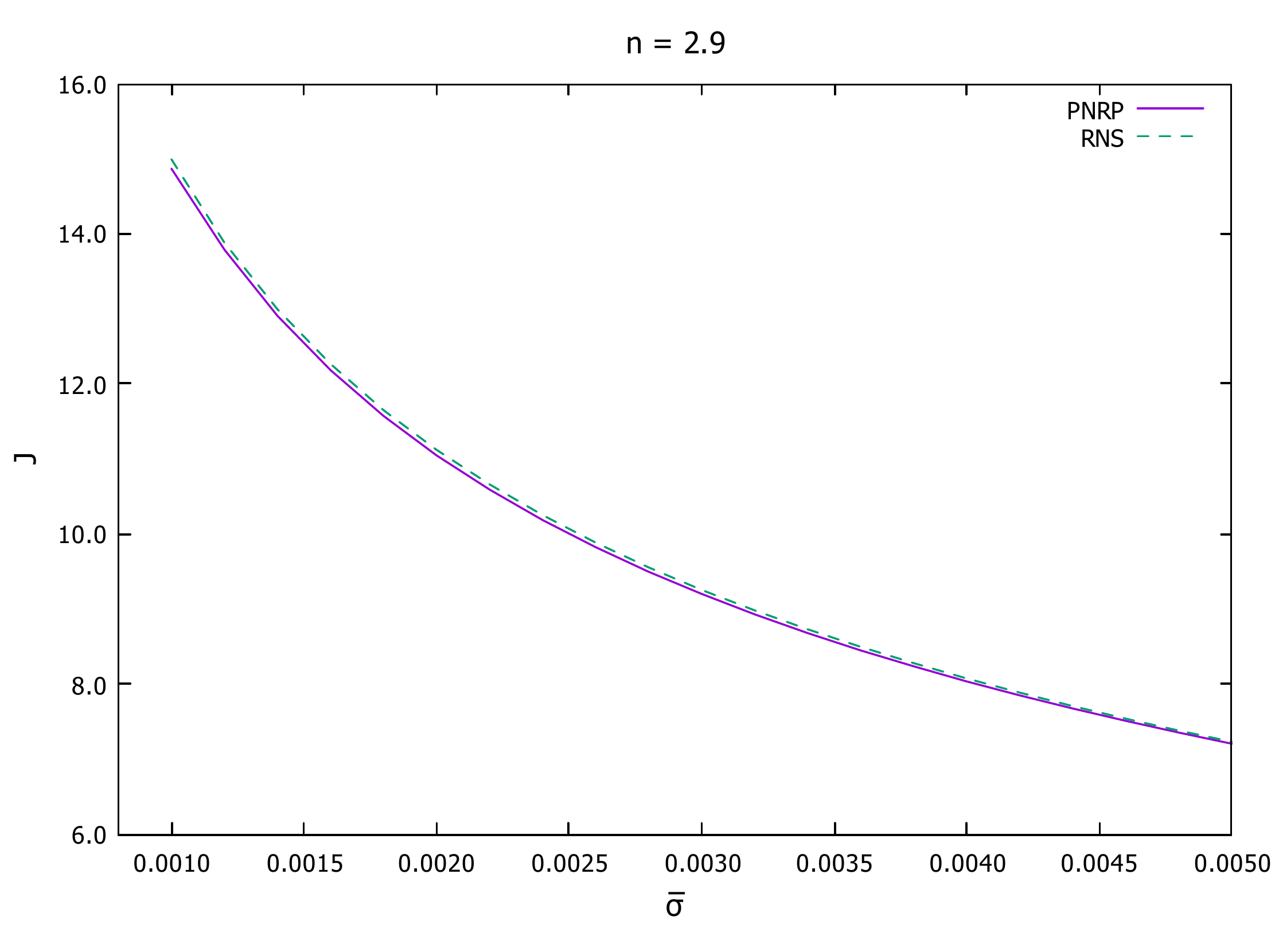}
         \captionsetup{width=0.8\textwidth}
         \caption{Angular momentum $J$ vs. relativity parameter $\bar{\sigma}$. Details as in Fig.~\ref{fig_grav_mass_rigid}.}
         \label{fig_J_rigid}
    \end{subfigure}
    \caption{}
\end{figure}

\begin{figure}
     \centering
     \begin{subfigure}[t]{0.49\textwidth}
         \centering
         \includegraphics[width=\linewidth]{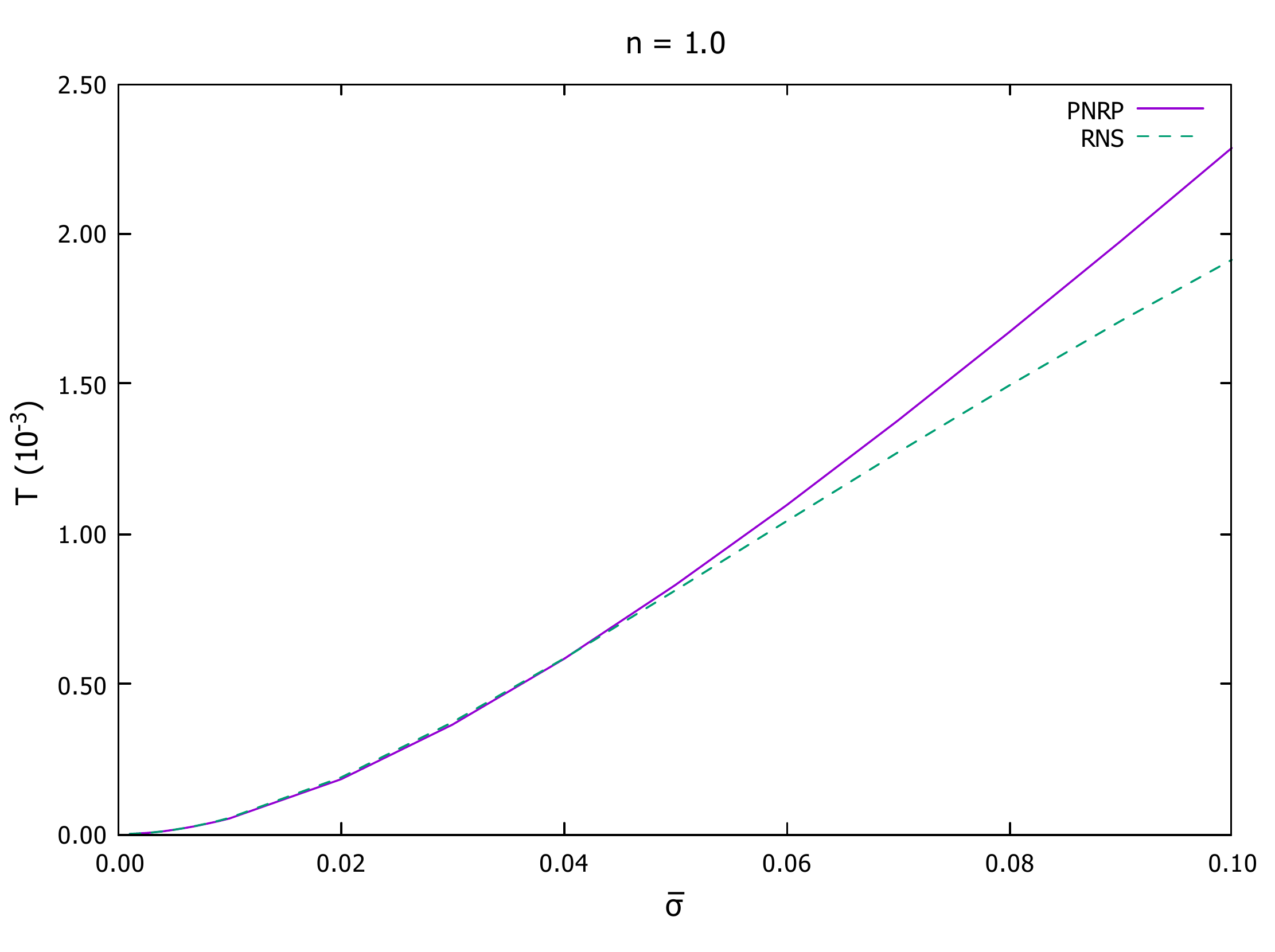}
         \includegraphics[width=\textwidth]{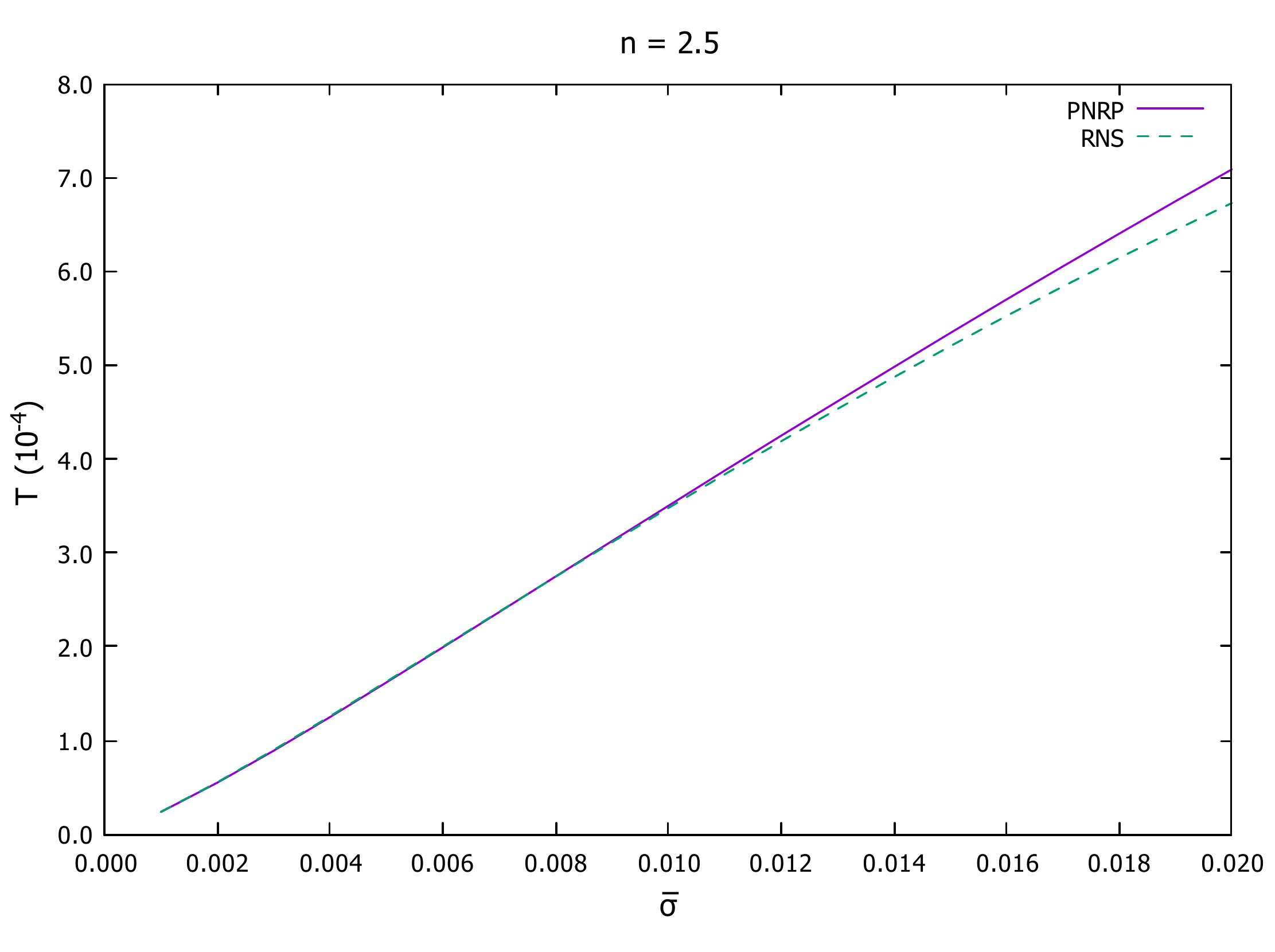}
         \includegraphics[width=\textwidth]{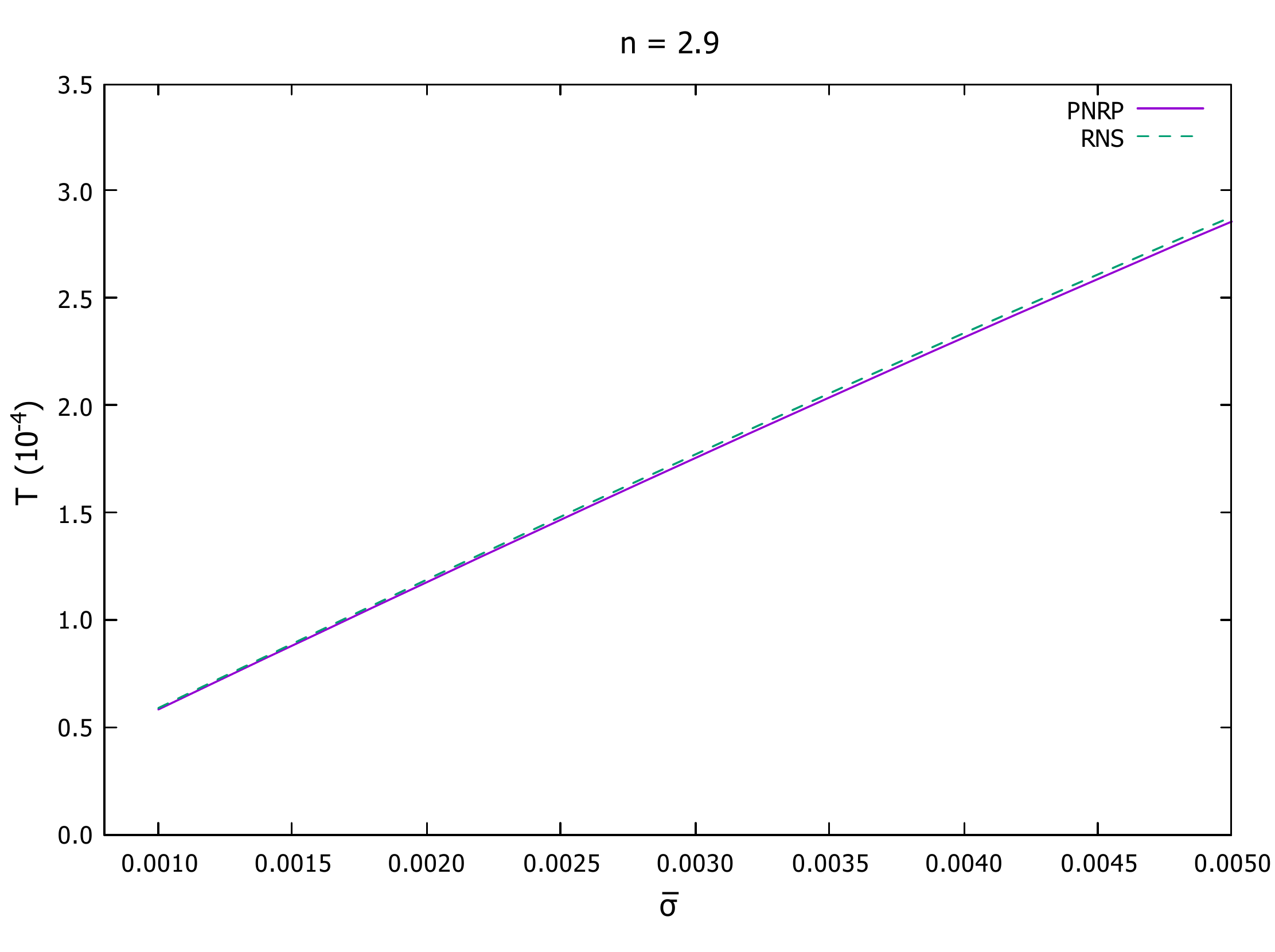}
         \captionsetup{width=0.8\textwidth}
         \caption{Rotational kinetic energy $T$ vs. relativity parameter $\bar{\sigma}$. Details as in Fig.~\ref{fig_grav_mass_rigid}.}
         \label{fig_t_rigid}
    \end{subfigure}
    \begin{subfigure}[t]{0.49\textwidth}
         \centering
         \includegraphics[width=\linewidth]{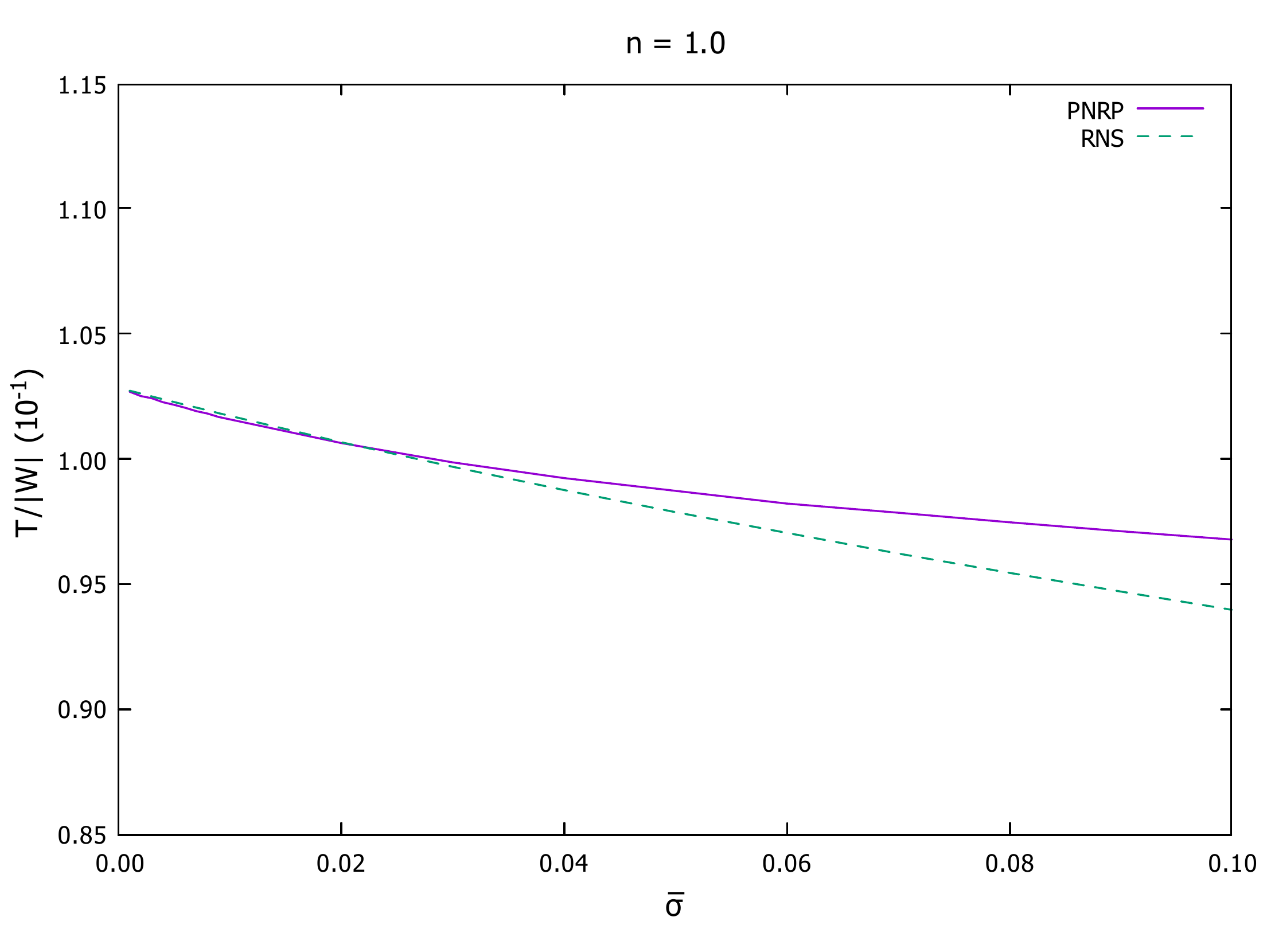}
         \includegraphics[width=\textwidth]{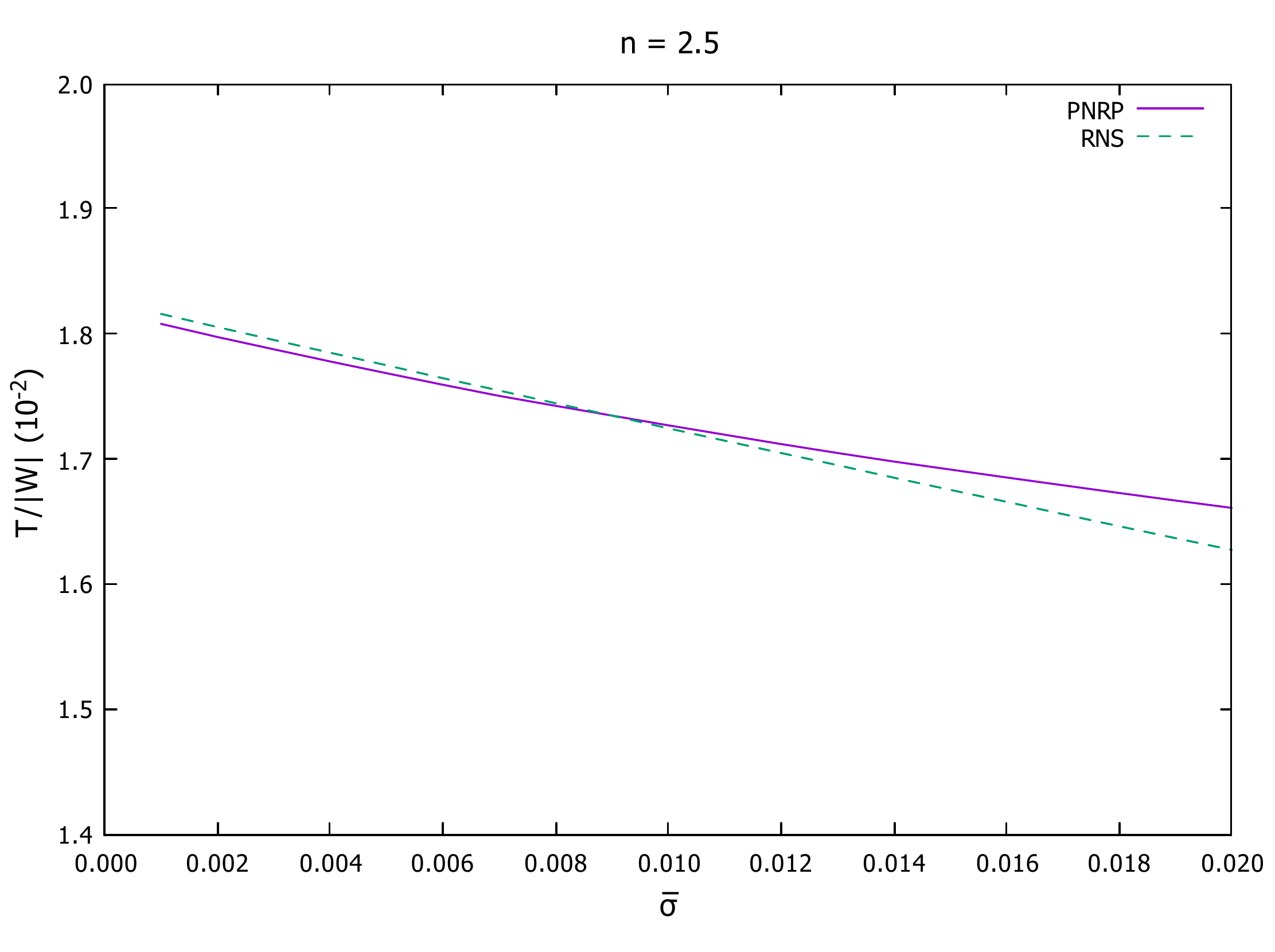}
         \includegraphics[width=\textwidth]{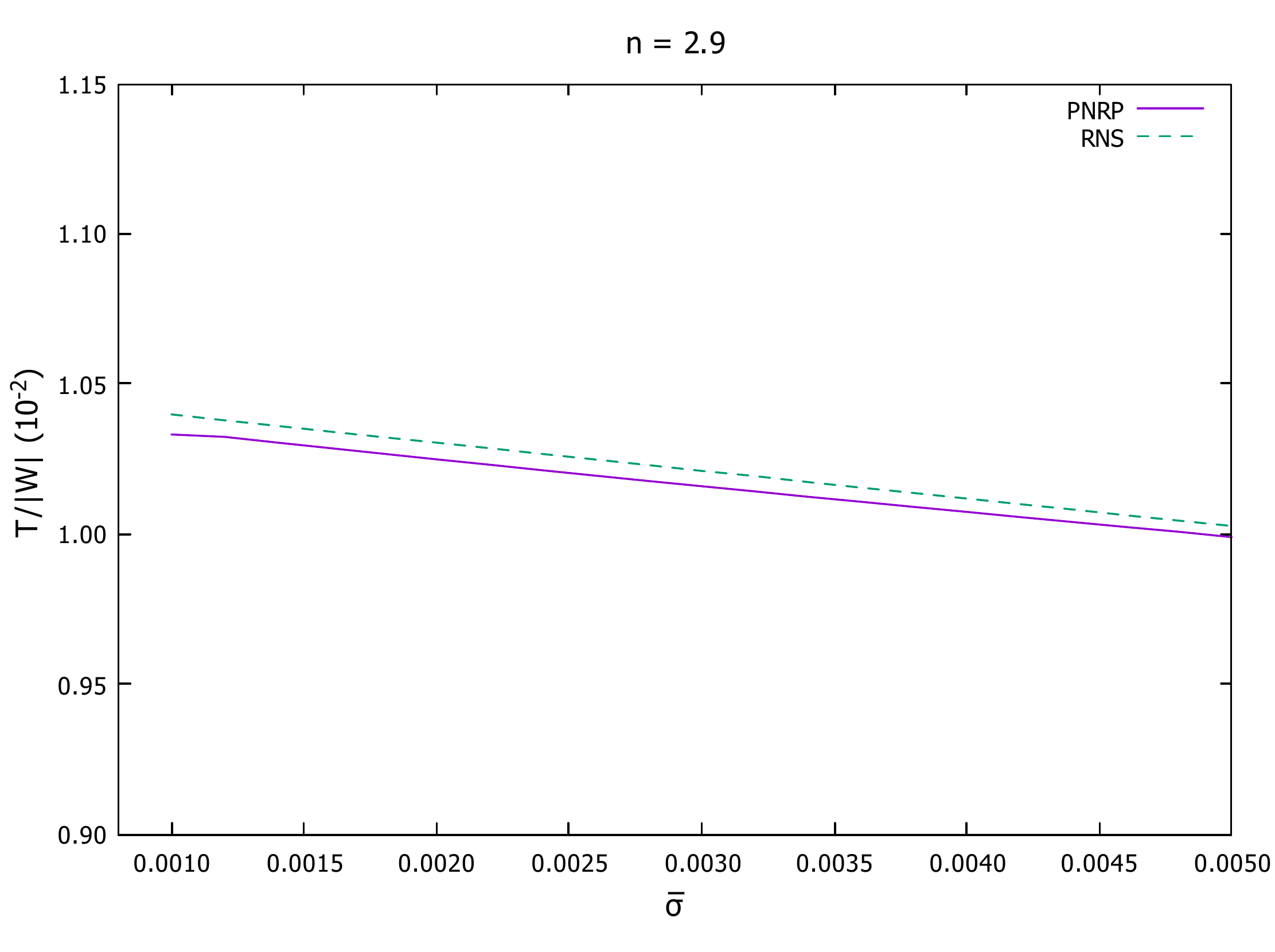}
         \captionsetup{width=0.8\textwidth}
         \caption{Ratio $T/W$ vs. relativity parameter $\bar{\sigma}$.Details as in Fig.~\ref{fig_grav_mass_rigid}.}
         \label{fig_t_w_rigid}
    \end{subfigure}
    \caption{}
    \label{fig-rigid-last}
\end{figure}

In case of rigidly and critically rotating polytropic configurations, we present diagrams showing the variation of several physical quantities with the relativity parameter $\bar{\sigma}$ for the polytropic indices $n=1.00,2.50,2.90$.
In Figures \ref{fig-rigid-first} - \ref{fig-rigid-last}, we give three plots for each examined quantity; the upper one refers to the polytropic index $n=1.00$, the middle to $n=2.50$ and the lower to $n=2.90$. In each diagram, there are two curves; the solid one denotes PNRP results, and the dotted one RNS results for the corresponding central mass-energy density, $\epsilon_c$. The two curves deviate slightly each other for relatively low values of $\bar{\sigma}$, a fact that confirms the validity of our method not only in the Newtonian limit, but also in cases with small or moderate relativistic effects. Specifically, for the ``soft'' polytropic index $n=2.90$, our results are in satisfactory agreement with those of RNS, as they differ each other less than $1\%$ even beyond $\bar{\sigma}_\mathrm{max}$. 
For the ``nearly soft'' polytropic index $n=2.50$, the two curves fit well for small values of $\bar{\sigma}$, and then they start deviating for progressively increasing values of $\bar{\sigma}$. We observe that our results are in good agreement (e.g. mass difference within $1\%$) up to a $\bar{\sigma}$ of about $\bar{\sigma}_\mathrm{max}/2$. Since the deviation between the two curves gets high enough around $\bar{\sigma}_\mathrm{max}$, we show results up to  $\bar{\sigma}=0.02$. For the ``stiff'' polytropic index $n=1.00$, the values of $\bar{\sigma}$ for which the differences remain within $1\%$ are less than $0.1$, i.e., less than $\bar{\sigma}_\mathrm{max}/3$. 
Eventually, the deviation between the two curves increases as $n$ decreases (equivalently, as $n$ becomes more stiff). It is worth remarking here that the value of $\bar{\sigma}$ for which the two curves start deviating, is not the same for all the  examined quantities.

\begin{figure}
     \centering
     \begin{subfigure}[t]{0.49\textwidth}
         \centering
         \includegraphics[width=\textwidth]{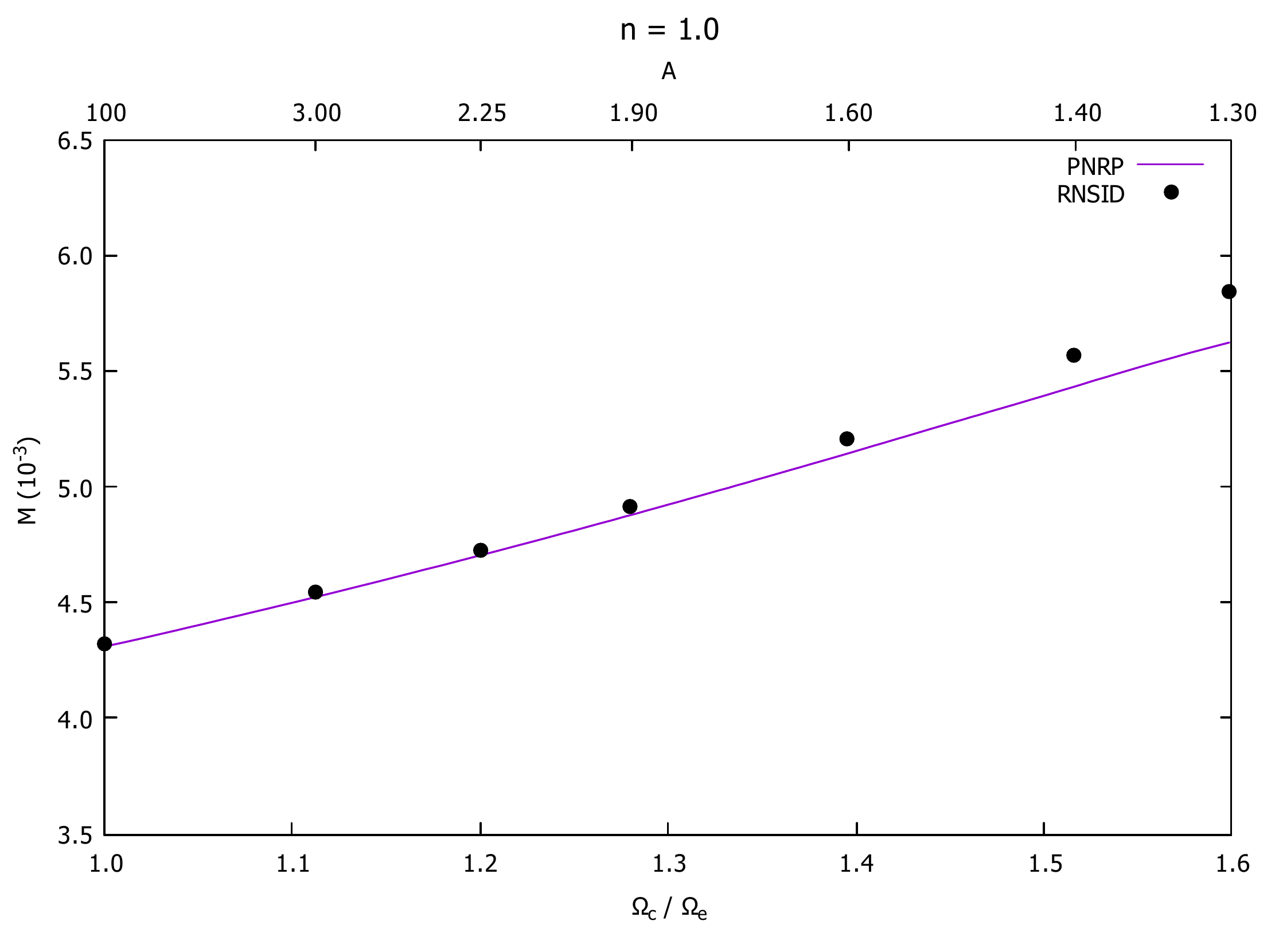}
         \includegraphics[width=\textwidth]{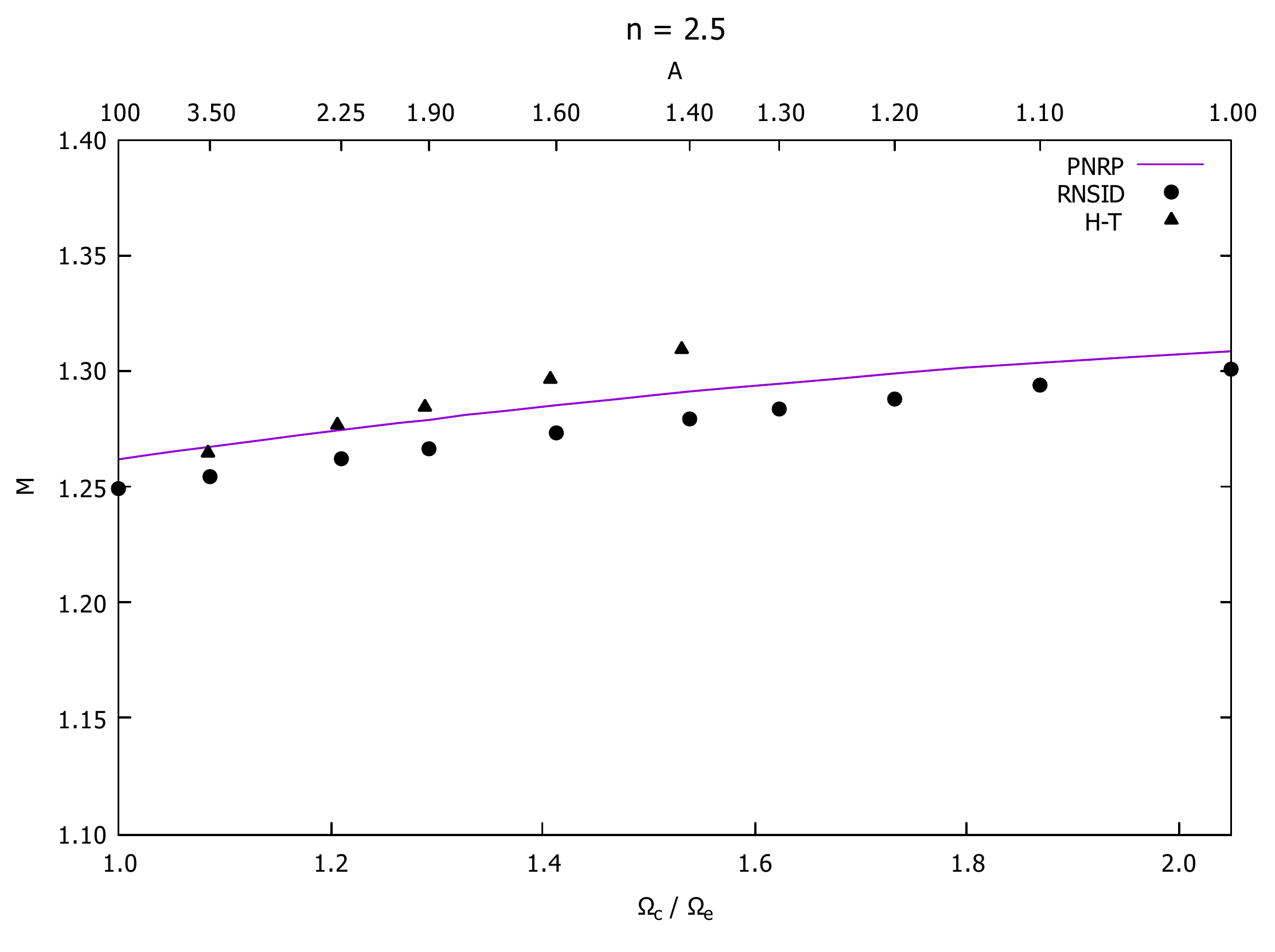}
         \includegraphics[width=\textwidth]{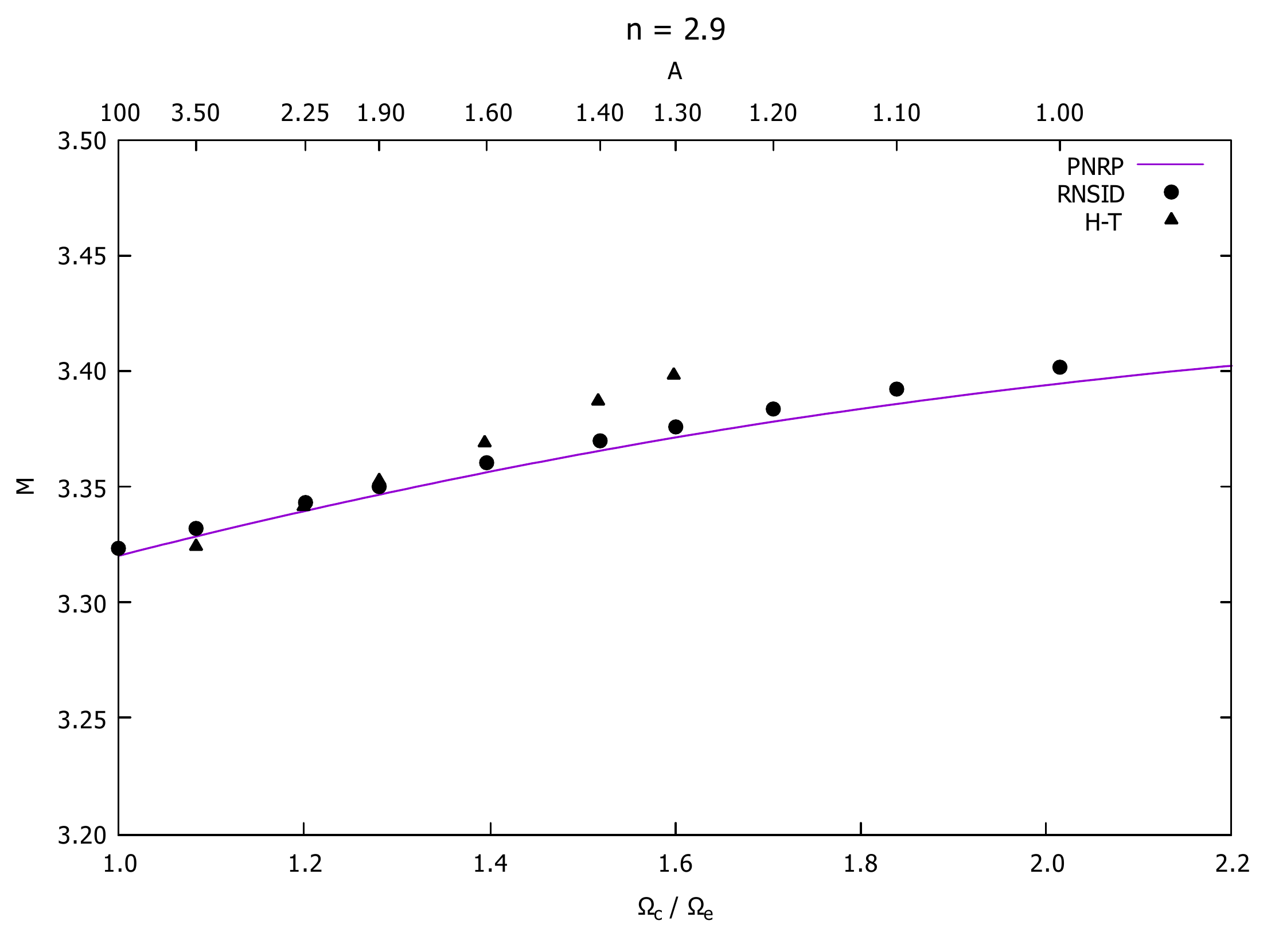}
         \captionsetup{width=0.8\textwidth}
         \caption{Gravitational mass $M$ vs. $\Omega_c/\Omega_e$ (lower horizontal axis; the upper horizontal axis shows respective values of the parameter $A$, involved in the RNSID's rotation law. Comparisons are made between results of PNRP and RNSID; in some cases, results computed by implementing the Hartle-Thorne method (H-T) are also quoted. Further details as in Fig.~\ref{fig_grav_mass_rigid}.}
         \label{fig_gm_diff}
    \end{subfigure}
    \begin{subfigure}[t]{0.49\textwidth}
         \centering
         \includegraphics[width=\linewidth]{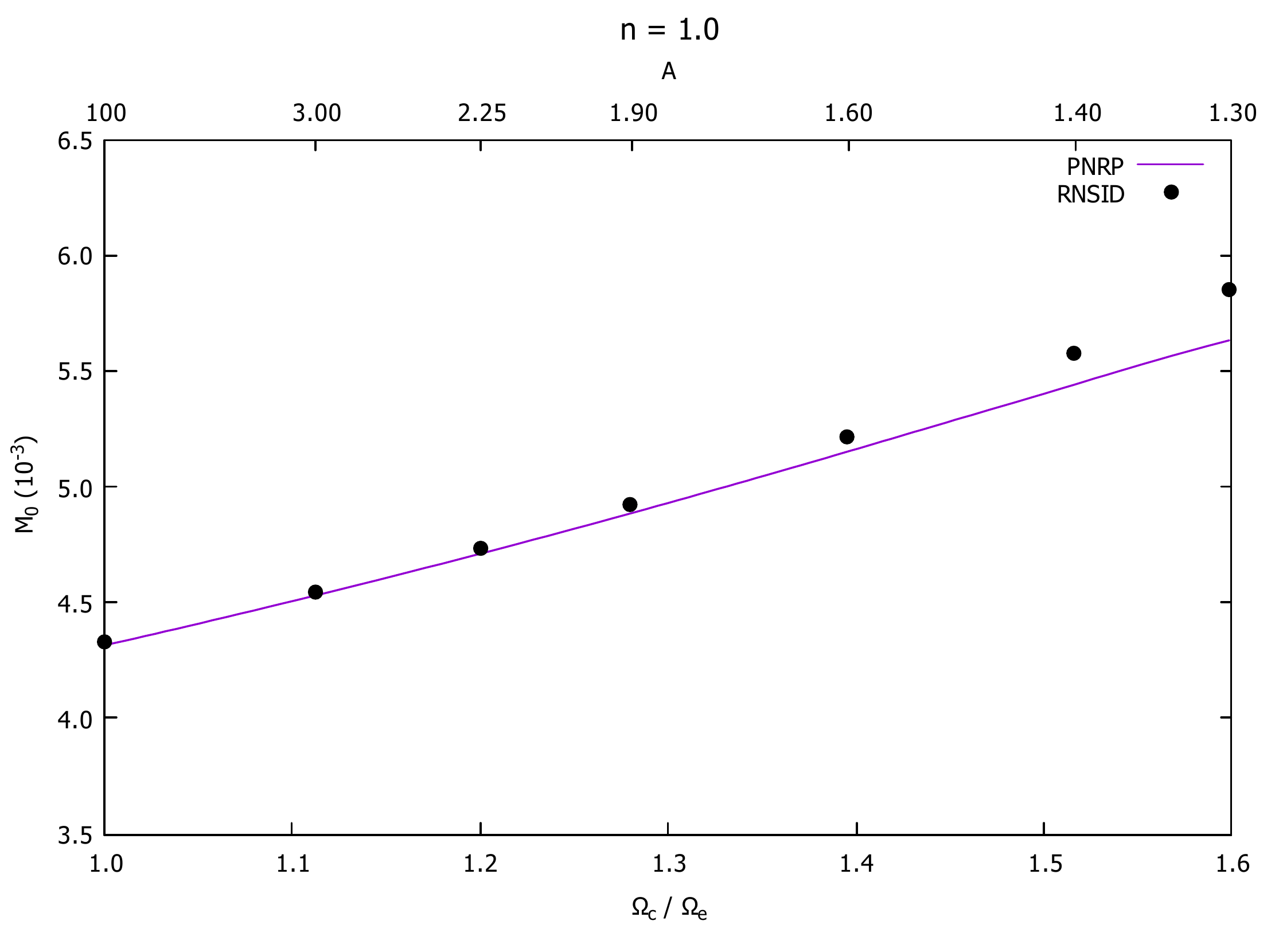}
         \includegraphics[width=\textwidth]{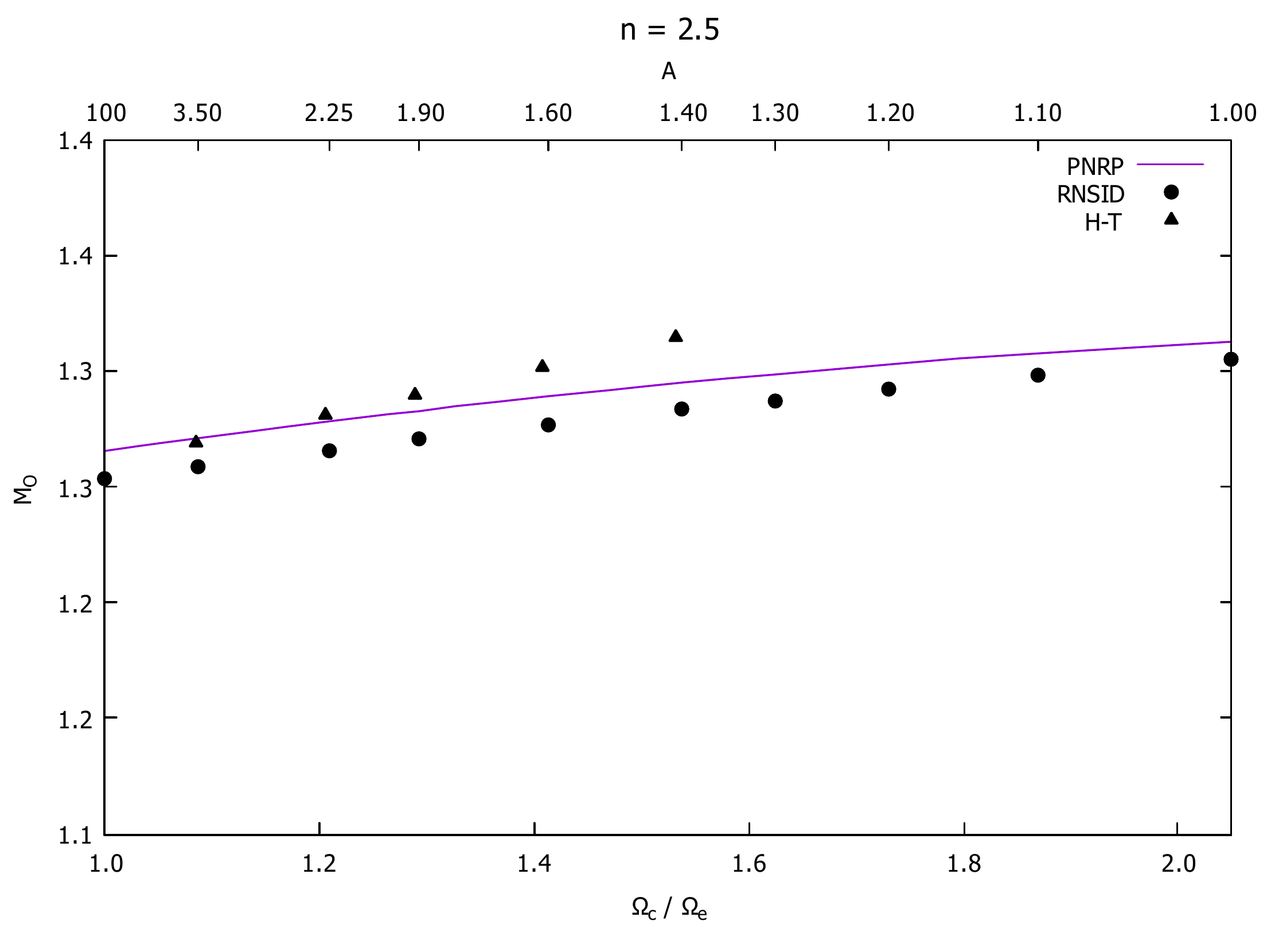}
         \includegraphics[width=\textwidth]{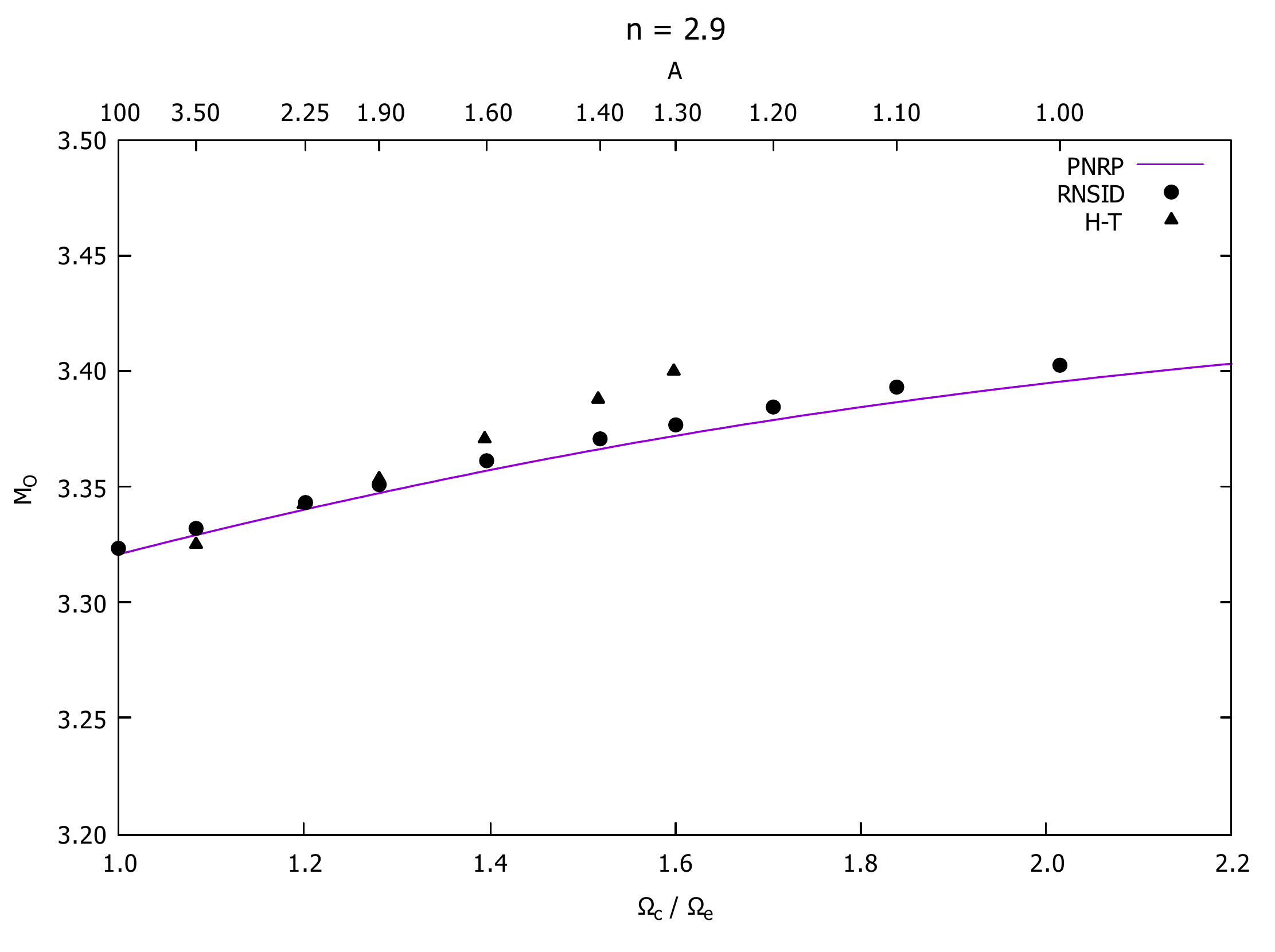}
         \captionsetup{width=0.8\textwidth}
         \caption{Rest mass $M_0$ vs. $\Omega_c/\Omega_e$. Details as in Fig.~\ref{fig_gm_diff}. }
         \label{fig_rm_diff}
    \end{subfigure}
    \caption{}
    \label{fig-diff-first}
\end{figure}

\begin{figure}
     \centering
     \begin{subfigure}[t]{0.49\textwidth}
         \centering
         \includegraphics[width=\textwidth]{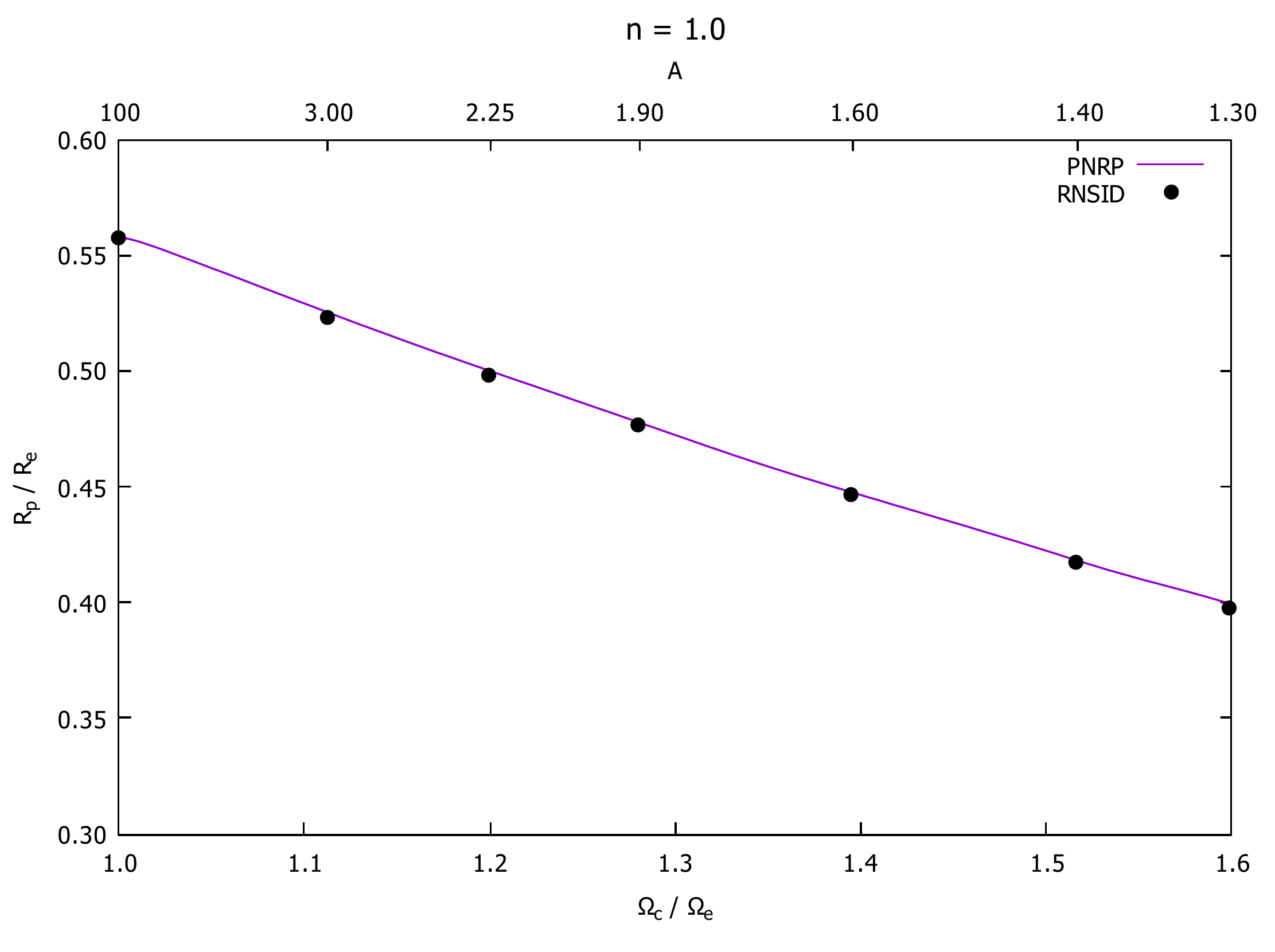}
         \includegraphics[width=\textwidth]{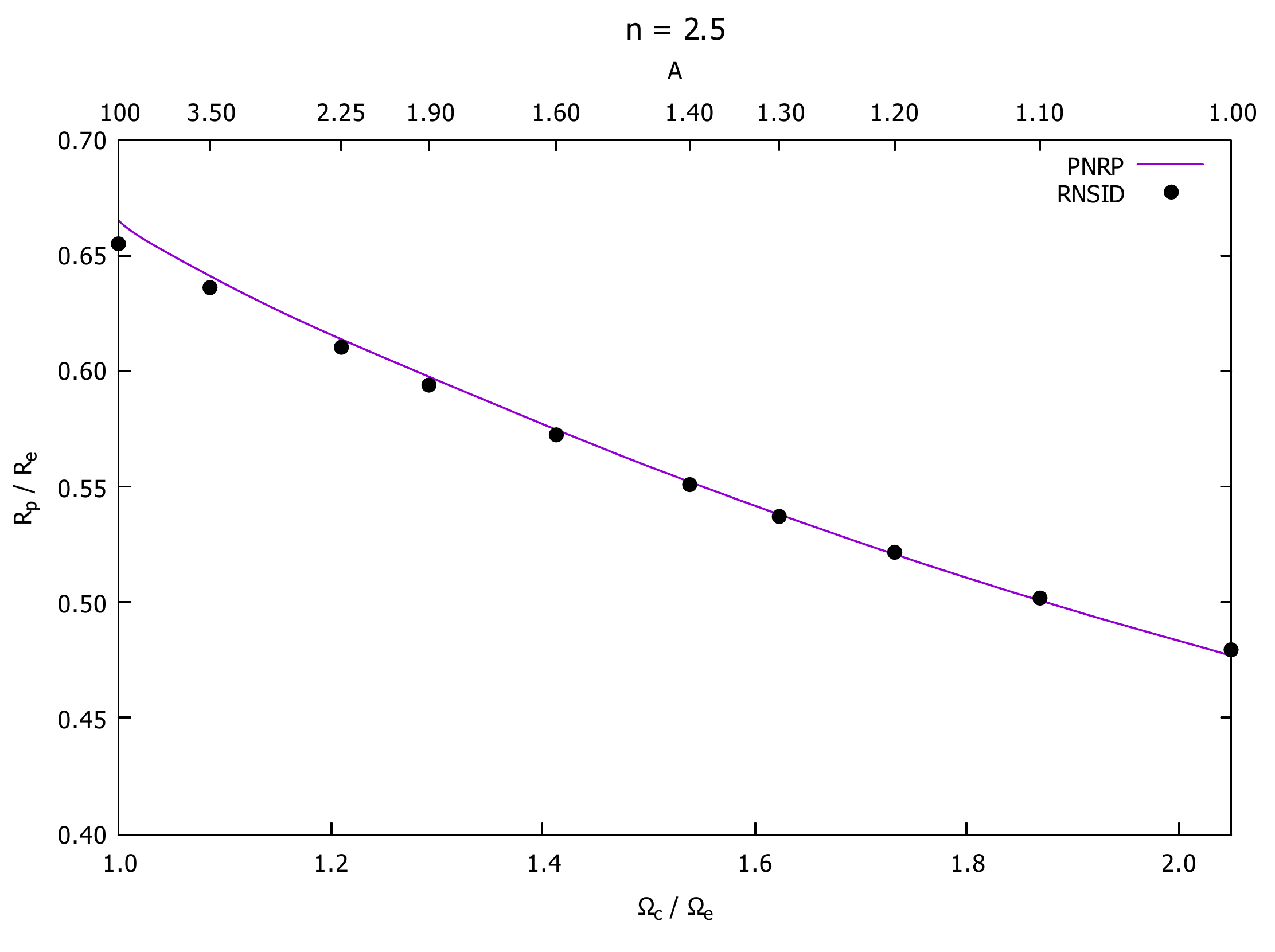}
         \includegraphics[width=\textwidth]{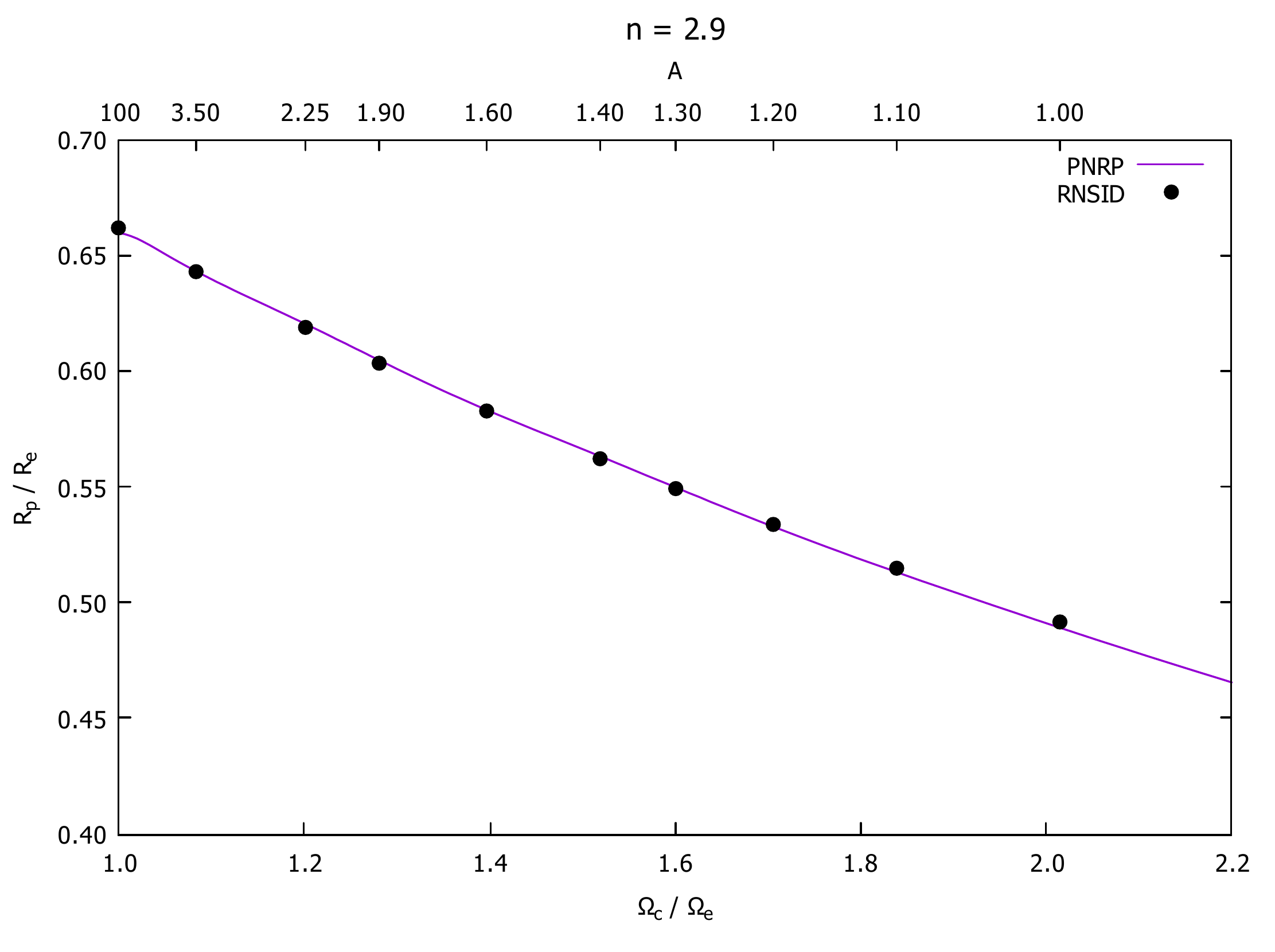}
         \captionsetup{width=0.8\textwidth}
         \caption{Ratio $R_p/R_e$ vs. $\Omega_c/\Omega_e$. Details as in Fig.~\ref{fig_gm_diff}.}
         \label{fig_Omg_c_diff}
    \end{subfigure}
    \begin{subfigure}[t]{0.49\textwidth}
         \centering
         \includegraphics[width=\linewidth]{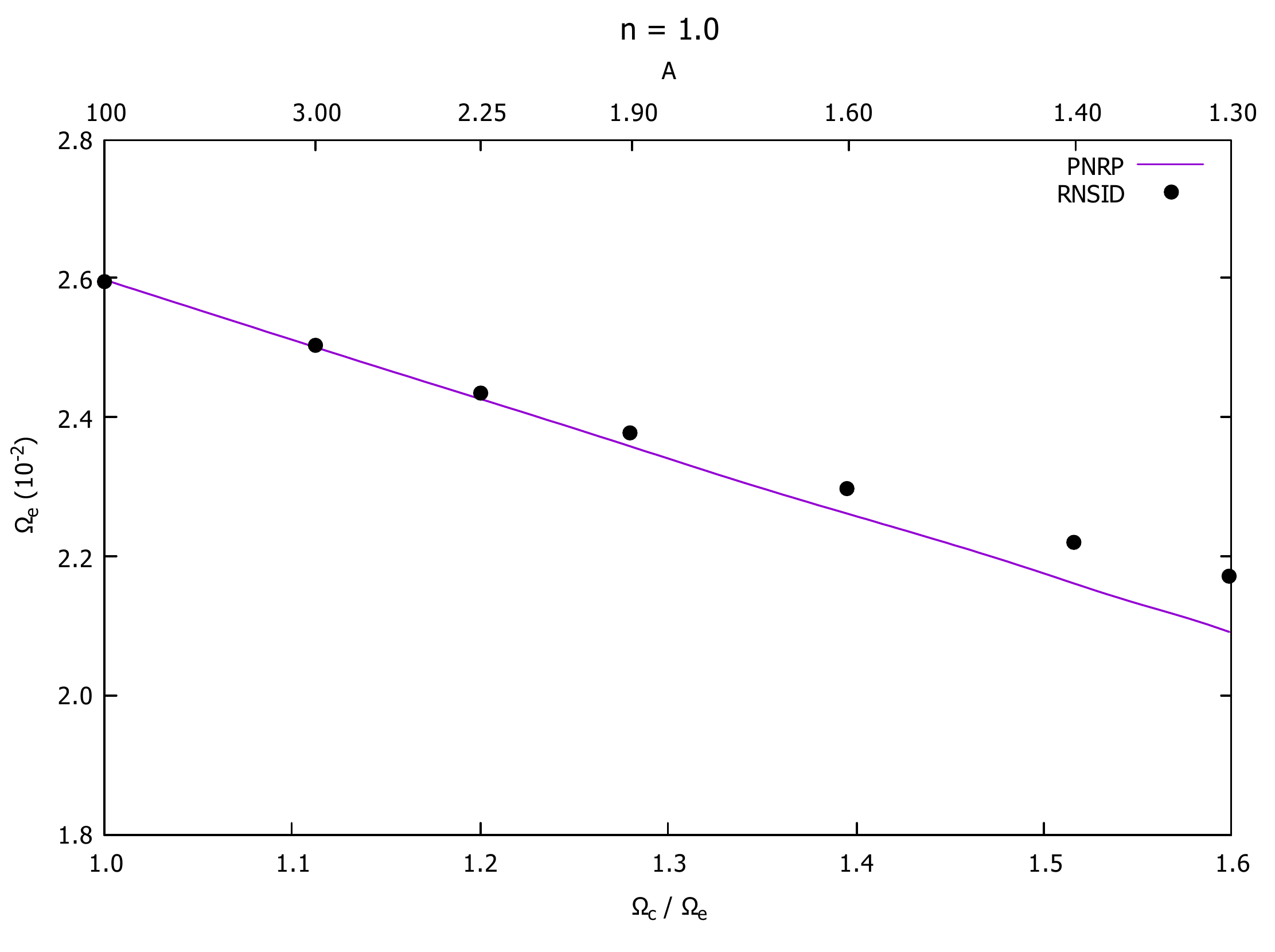}
         \includegraphics[width=\textwidth]{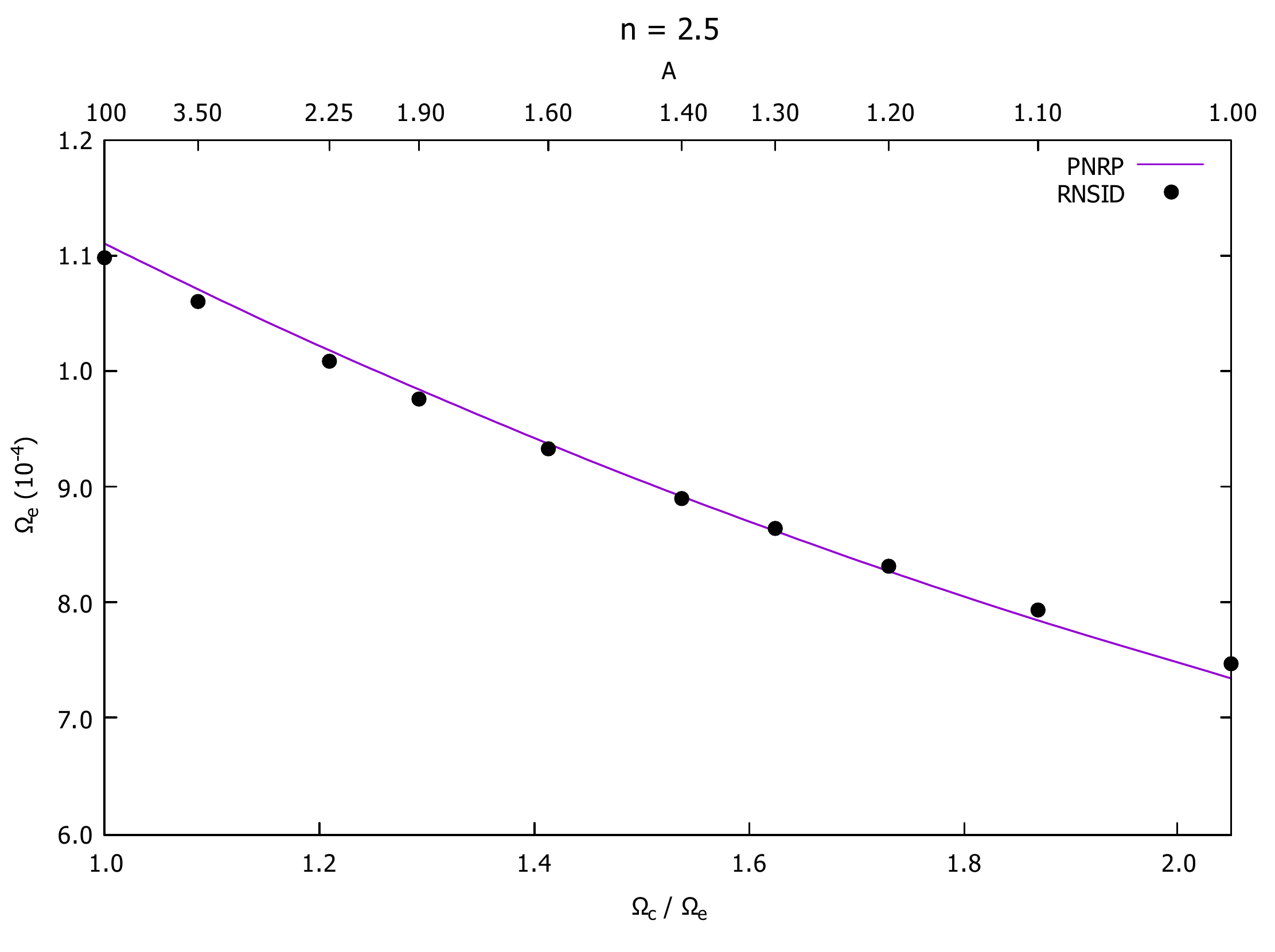}
         \includegraphics[width=\textwidth]{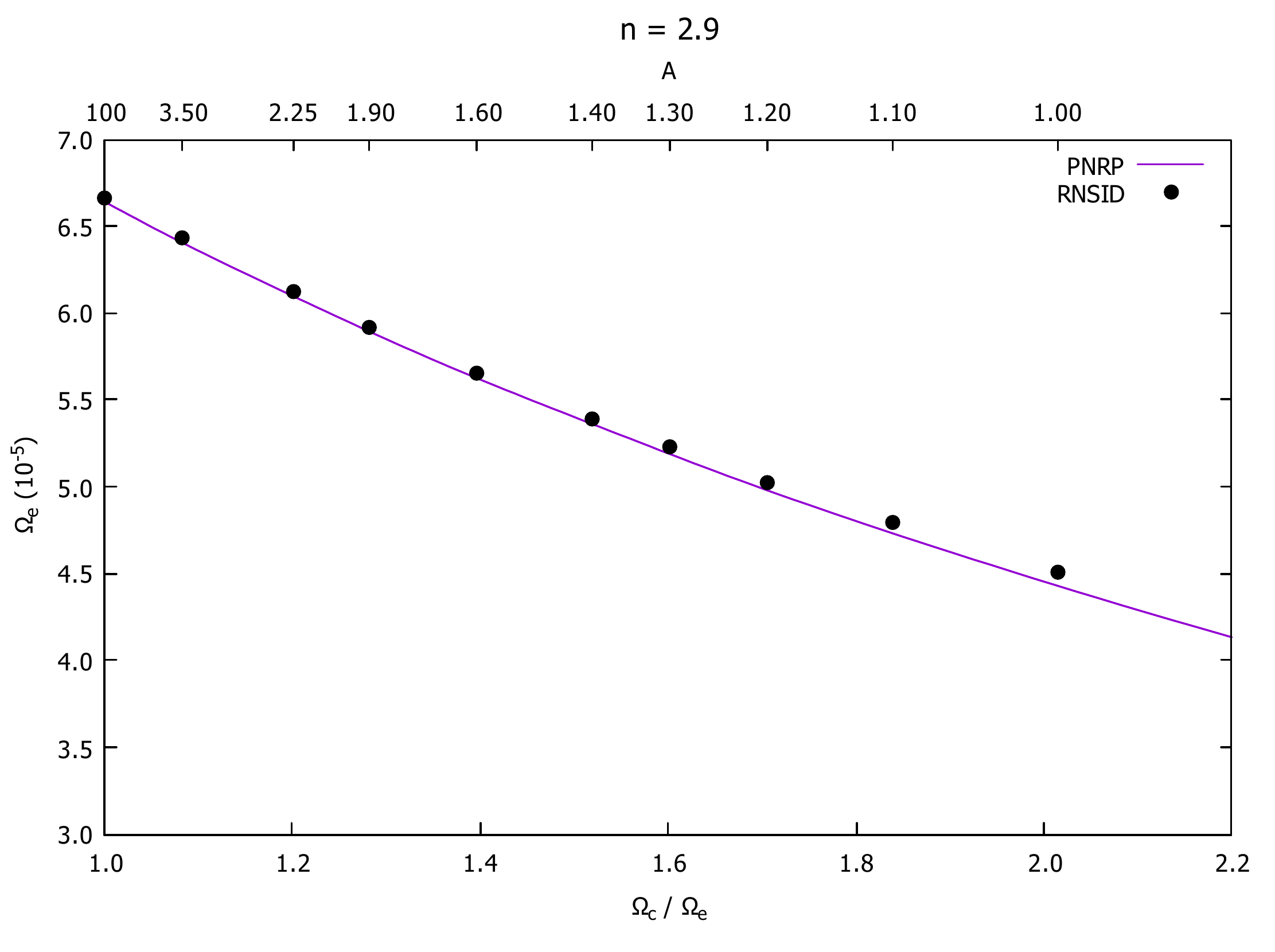}
         \captionsetup{width=0.8\textwidth}
         \caption{Equatorial angular velocity $\Omega_e$ vs. $\Omega_c/\Omega_e$. Details as in Fig.~\ref{fig_gm_diff}.}
         \label{fig_Omg_e_diff}
    \end{subfigure}
    \caption{}
\end{figure}

\begin{figure}
     \centering
     \begin{subfigure}[t]{0.49\textwidth}
         \centering
         \includegraphics[width=\textwidth]{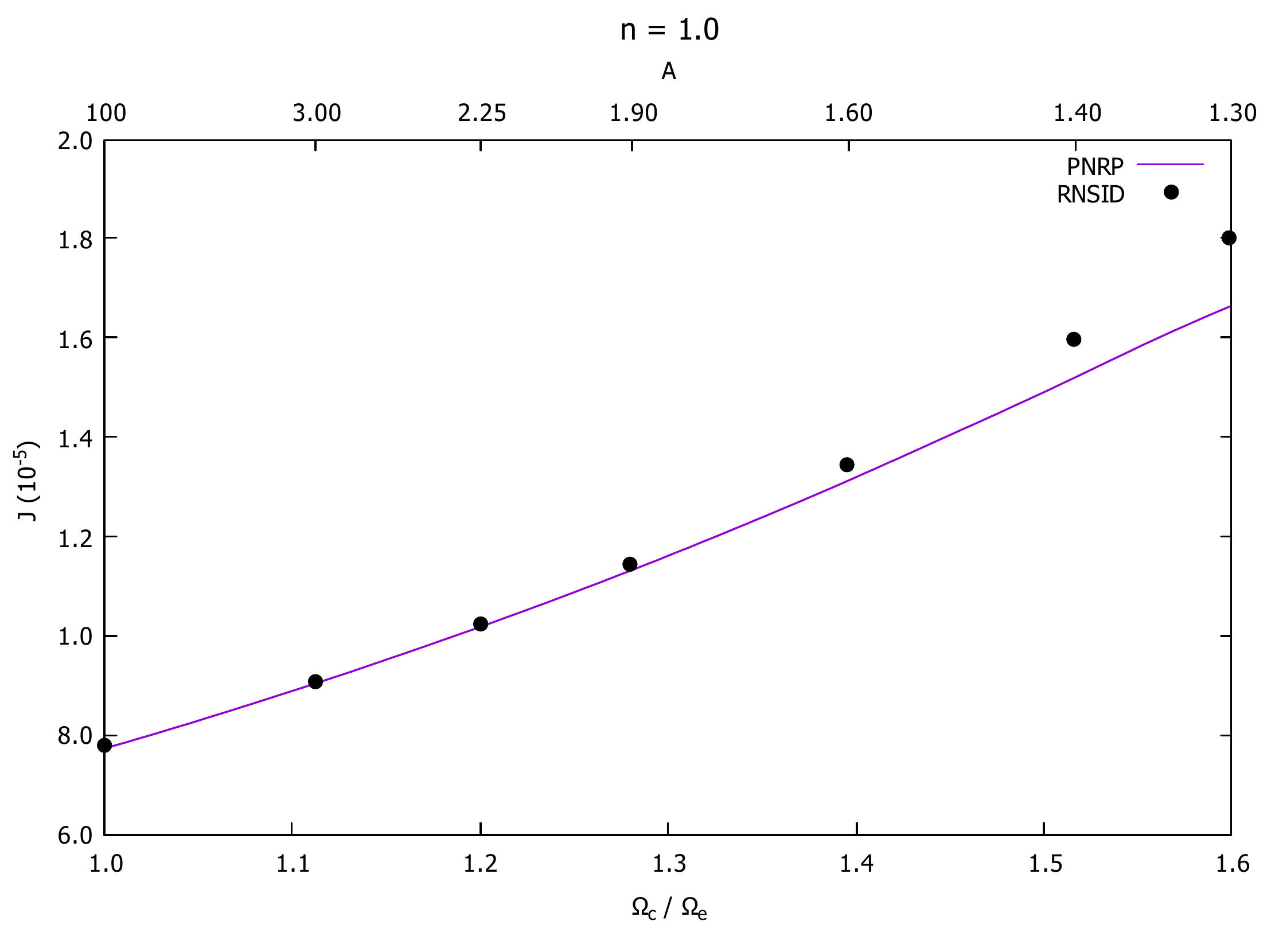}
         \includegraphics[width=\textwidth]{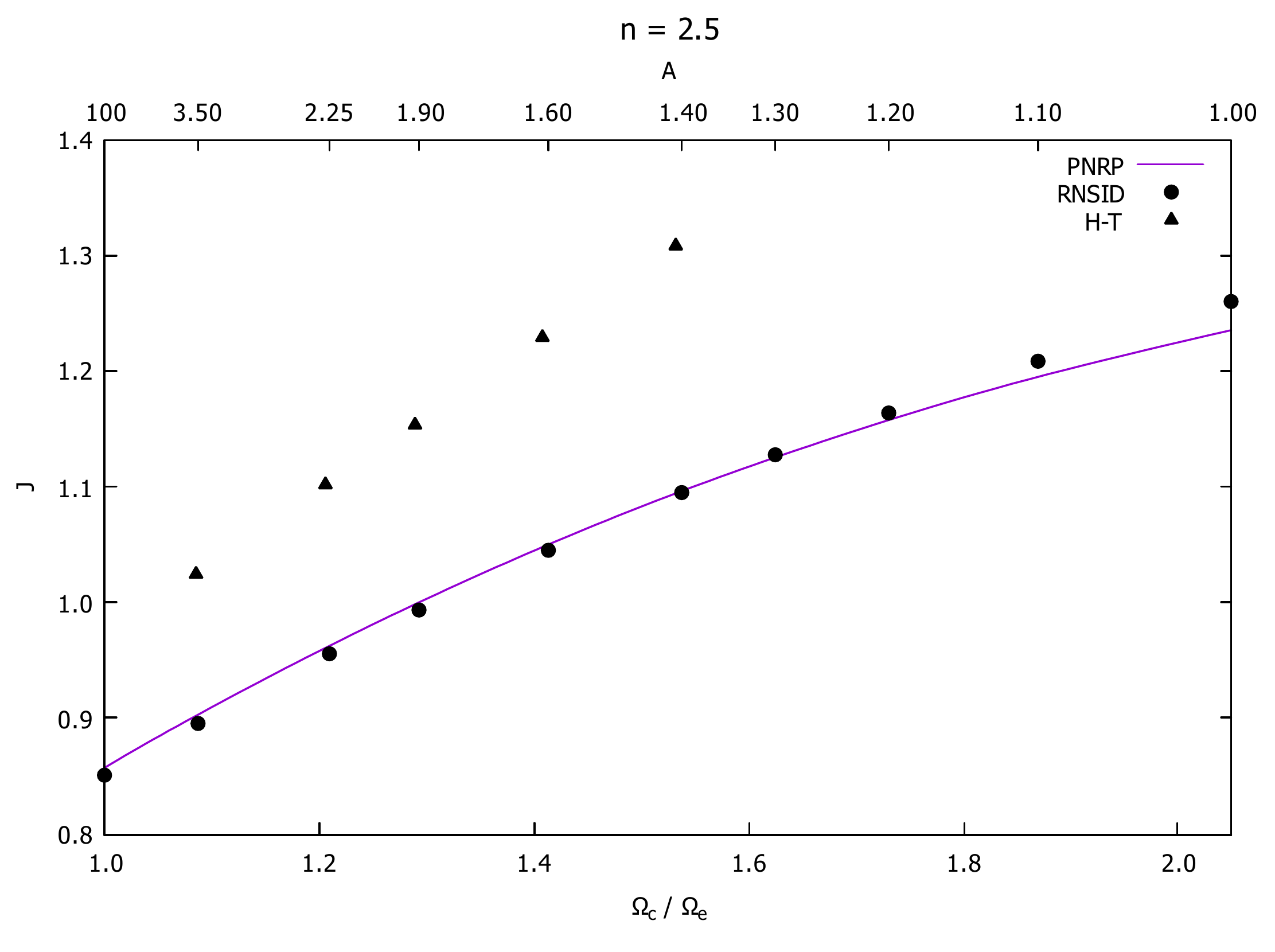}
         \includegraphics[width=\textwidth]{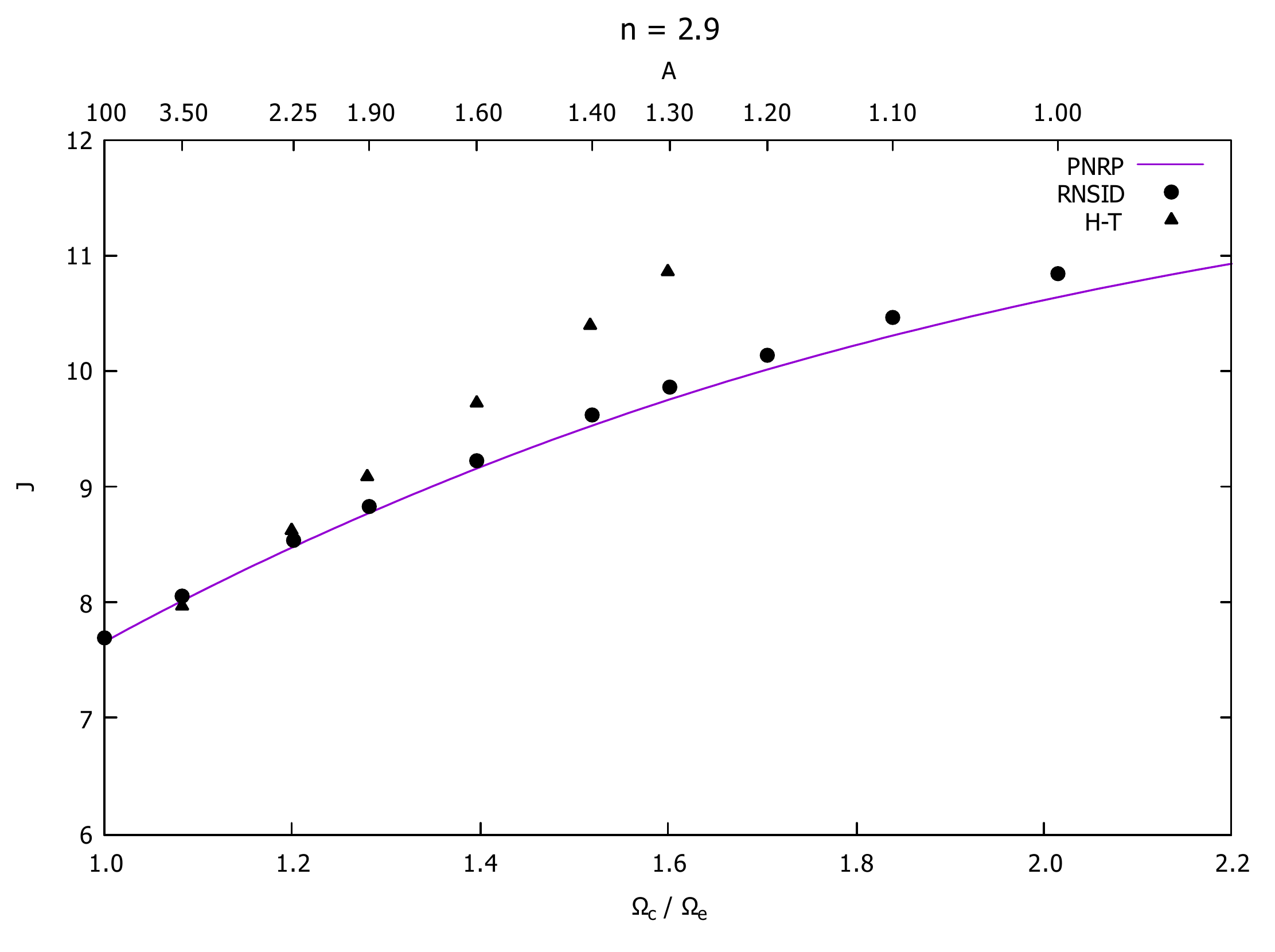}
         \captionsetup{width=0.8\textwidth}
         \caption{Angular momentum $J$ vs. $\Omega_c/\Omega_e$. Details as in Fig.~\ref{fig_gm_diff}.}
         \label{fig_J_diff}
    \end{subfigure}
    \begin{subfigure}[t]{0.49\textwidth}
         \centering
         \includegraphics[width=\linewidth]{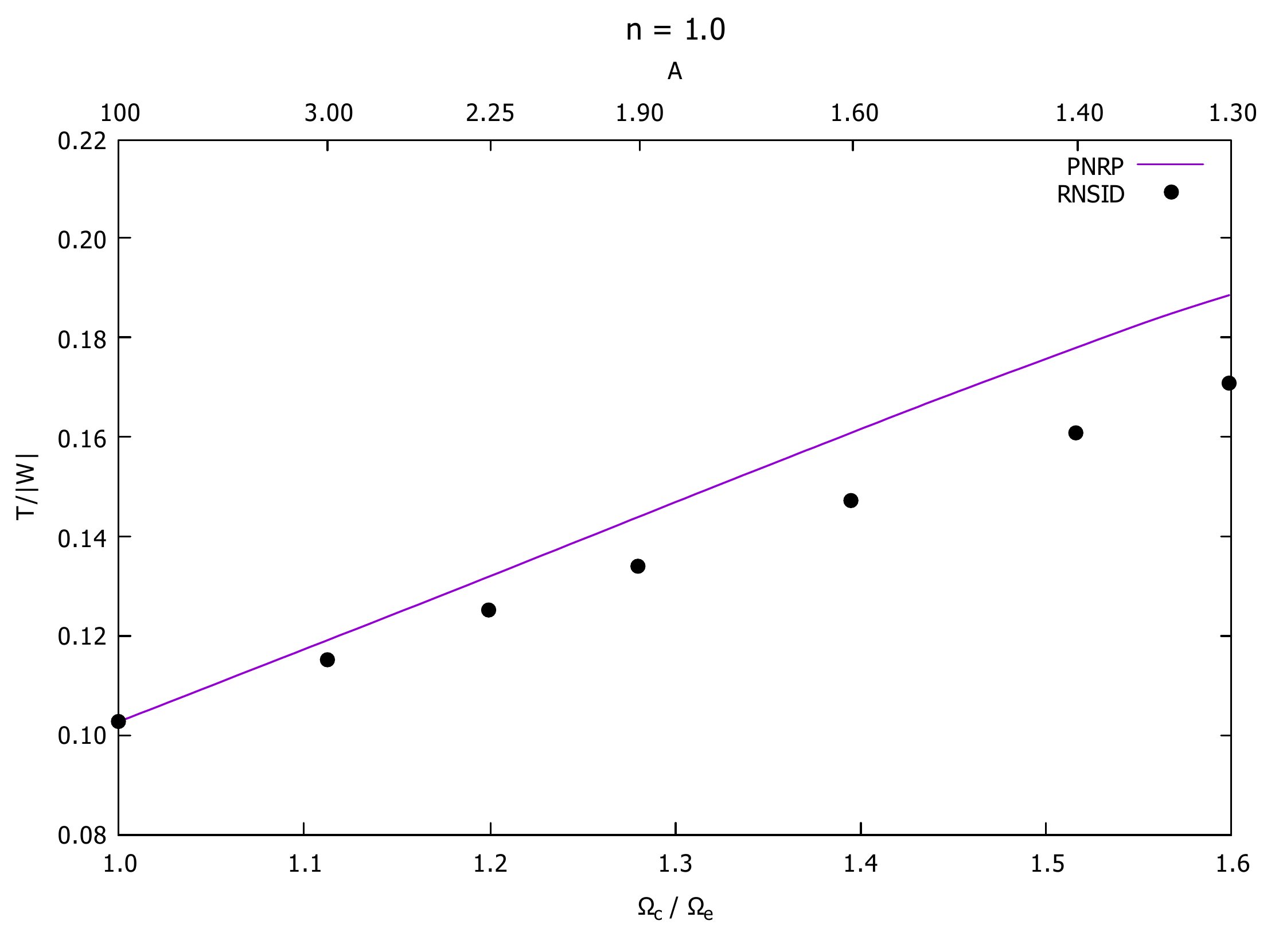}
         \includegraphics[width=\textwidth]{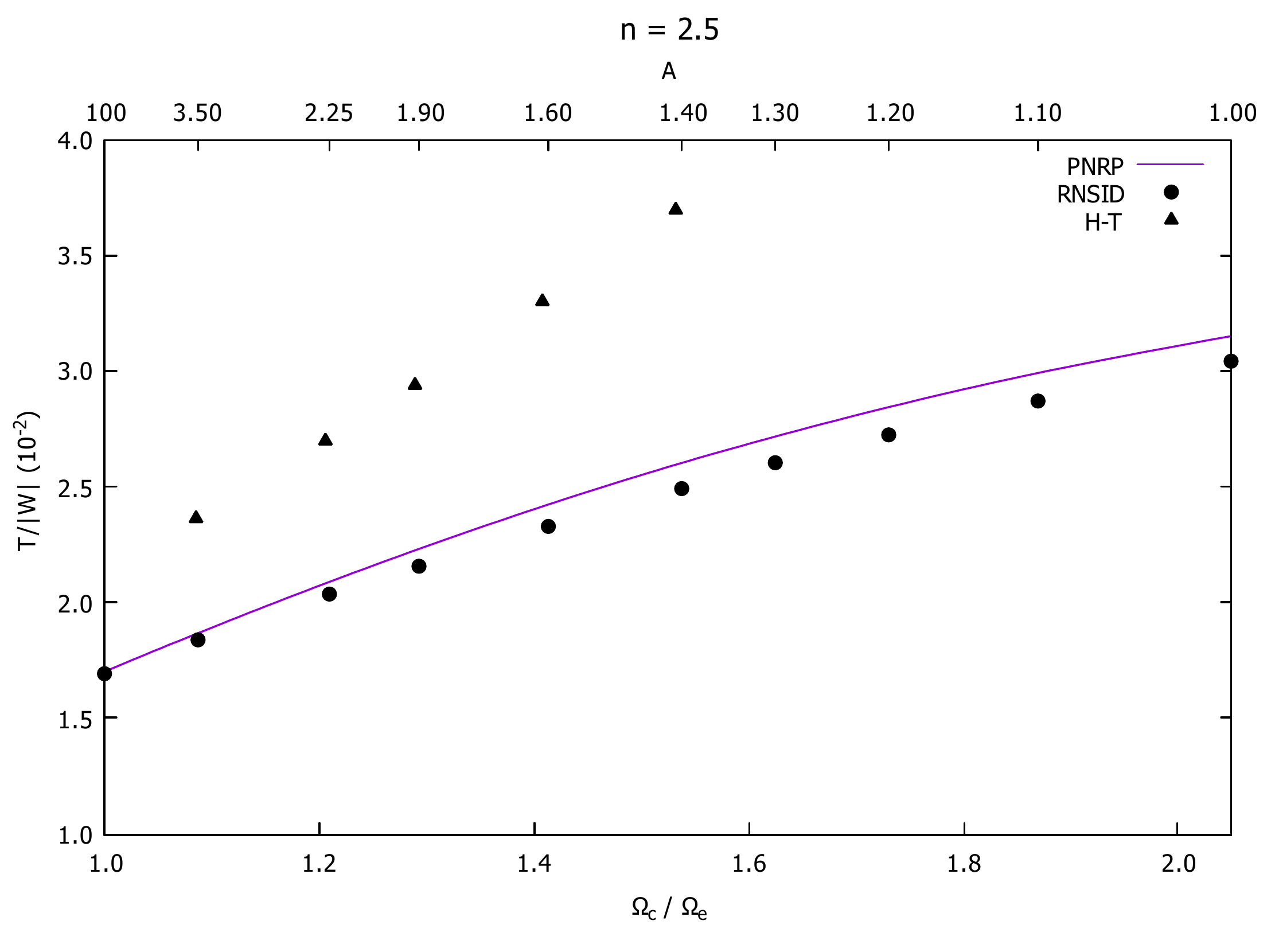}
         \includegraphics[width=\textwidth]{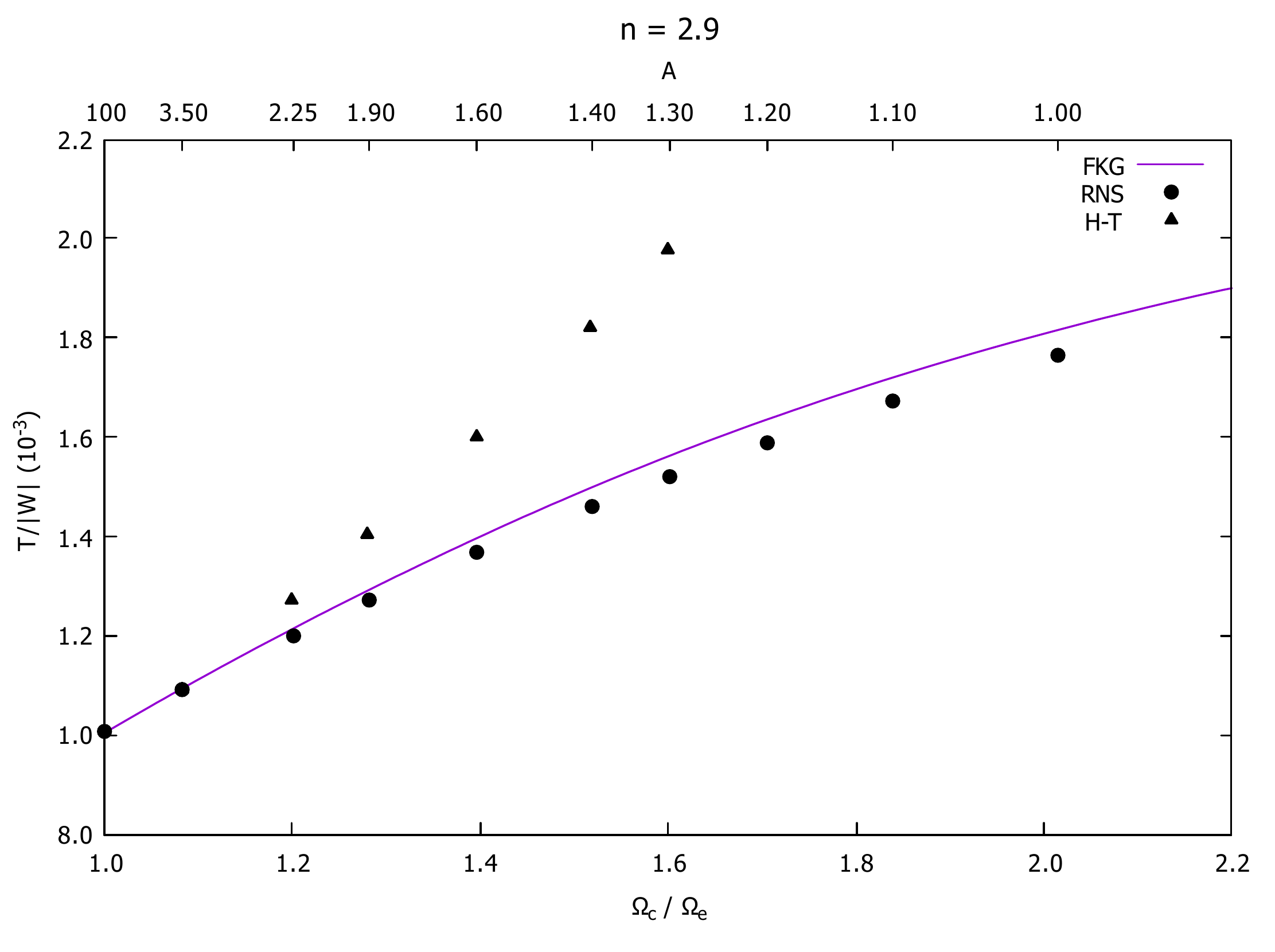}
         \captionsetup{width=0.8\textwidth}
         \caption{Ratio $T/|W|$ vs. $\Omega_c/\Omega_e$. Details as in Fig.~\ref{fig_gm_diff}.}
         \label{fig_T_W_diff}
    \end{subfigure}
    \caption{}
    \label{fig-diff-last}
\end{figure}

In case of differentially and critically rotating polytropic configurations, Figures \ref{fig-diff-first} - \ref{fig-diff-last} show the variation of several physical quantities with the ratio $\Omega_c/\Omega_e$; the lower horizontal axis shows values of $\Omega_c/\Omega_e$, while the upper one shows counterpart values of the parameter $A$, which is involved in the RNSID's rotation law \citep[Eq.~(9)]{2004STERG_AP-FONT}. In these figures, we show PNRP results as solid curves, and respective RNSID results as filled circles. For $n=2.50,2.90$, we also include in the diagrams some results obtained by a code \citep{2008IJMPC..19.1863G}, hereinafter ``H-T code'', implementing Hartle's perturbation method \citep{Hartle_1_1967ApJ...150.1005H} developed further by J. Hartle and K. Thorne \citep{Hartle_2_1968ApJ...153..807H} and for that reason also called ``Hartle-Thorne perturbation method'' by many authors. H-T results are denoted by filled rhombuses.
Note that we present PNRP results corresponding to the values of $\bar{\sigma}$ for which the respective rigid-rotation results have been found to be in satisfactory agreement with the RNS results. This particular $\bar{\sigma}$ varies with $n$. In detail, for $n=1.00,2.50,2.90$, we take $\bar{\sigma}=1.28 \times 10^{-3}$, $\bar{\sigma} = \bar{\sigma}_\mathrm{max}/2$, and $\bar{\sigma} = \bar{\sigma}_\mathrm{max}$, respectively. In addition, we quote in Table~\ref{tab:n10} PNRP results and respective RNSID ones for  $n=1.00,2.50,2.90$, so that comparisons can be directly and clearly made.

Compared to the differential rotation law used in RNSID \citep{2004STERG_AP-FONT}, our law~\eqref{rot-law}, established via newtonian arguments, turns out to have limits when applied to relativistic configurations (high $\bar{\sigma}$).
In particular, this law resembles the Newtonian limit expressed by Eq.~(20) of \cite{LBS-2003ApJ...583..410L} (see also the discussion following this equation) but for relativistic configurations, this relation holds only approximately. 
In Figures~\ref{fig-diff-first}-\ref{fig-diff-last}, we plot several physical quantities up to a maximum strength of differential rotation, for which the rotation law keeps holding when describing a spheroidal configuration. 
 
At first sight, the law~\eqref{rot-law} seems to give satisfactory results, especially for relatively low values of $\Omega_c/\Omega_e$. As we can observe in Figures~\ref{fig-diff-first}-\ref{fig-diff-last}, the slope of the two curves is not the same. The RNSID curve has a higher slope than the PNRP curve; and this occurs for all three polytropic indices and for all the examined quantities.

Aiming at a clear view about the question `if this behaviour is an outcome of the different expressions of the rotation law only, or if it is also due to relativistic effects', we have computed same results not only in the Newtonian limit (i.e. $\bar{\sigma} \sim 10^{-3}-10^{-2}$) but also in highly relativistic cases (i.e. $\bar{\sigma} \sim \bar{\sigma}_\mathrm{max}$). 
As it turned out, the general behaviour between the two curves remains the same, but the deviation from each other increases for higher values of $\bar{\sigma}$.
Consequently, we could say that this behaviour is due to the different expressions of the rotation law.

The increased value of $\bar{\sigma}$ contributes in a different way: The maximum strength of differential rotation for a configuration depends not only on the polytropic index but also on the relativity parameter. A lower polytropic index describes more compact objects, as well as higher values of $\bar{\sigma}$ signify higher values of central density and thus more compact configurations. Consequently, for given $n$ and $F_r$ (alternatively, $A$), higher values of $\bar{\sigma}$ lead to  configurations with lower $\Omega_c/\Omega_e$ ratios.

As it is apparent from the presented diagrams, PNRP can accurately resolve fully relativistic configurations at critical rotation for the soft polytropic index $n=2.90$. Next, as we move to lower (i.e. stiffer) polytropic indices, we see that our method computes accurate results for less relativistic configurations. As PNRP seems to work adequately on soft and moderately stiff cases, we understand that, due to the order of the post-Newtonian approximation used, it is difficult for this method to efficiently resolve highly relativistic configurations. 
To be specific, in the post-Newtonian approximation an involved quantity, say $Q$, is expressed as
\begin{equation}
    Q=Q_0+\frac{1}{c^2}Q_1+\frac{1}{c^4}Q_2+ \dots \quad,
    \label{PNA-exp}
\end{equation}
or, by using Eq.~\eqref{sigma-par-cpu} and the cpu system of units,
\begin{equation}
    Q=Q_0 + (n+1) \,\bar{\sigma}\,Q_1+\left[(n+1)\,\bar{\sigma}\right]^2 \, Q_2+... \quad .
    \label{PNA-exp-sigma}
\end{equation}
A relevant case for a quantity, say $S$, is to be expressed as 
\begin{equation}
    S=S_0+\frac{1}{c^1}S_1+\frac{1}{c^3}S_2+ \dots \quad,
    \label{PNA-expS}
\end{equation}
but, without loss of generality, we discuss here the relations~\eqref{PNA-exp}-\eqref{PNA-exp-sigma}. $Q_0$ is the leading term, $(1/c^2) Q_1$ the ``approximative term'' (equivalently, the ``perturbation term'') of first order, with $1/c^2 =  (n+1) \, \bar{\sigma}$, $(1/c^4) Q_2$ the perturbation term of second order, with $1/c^4 = [(n+1) \,\bar{\sigma}]^2$, etc. 
In the approximative methods, however, the higher-order perturbation term taken into account has to be small when compared to the leading term in order for the results to be accurate and reliable. Accordingly, we expect PNRP to give accurate and reliable results as long as the involved first-order perturbation term is kept small in comparison with the leading term. Assuming, without loss of generality, that the respective ``kernels'' $Q_0$, $Q_1$, \dots, are more or less of same order of magnitude, this can happen for values of  $\bar{\sigma}$ of order, say, $10^{-3} - 10^{-2}$. For higher values of $\bar{\sigma}$ and stiff cases, an approximative term $(1/c^2) Q_1$ can even approach $60\%$ to $70\%$ of the value of the leading term $Q_0$. Apparently, in such cases, approximative terms of higher order need to be involved in the computations. To give some relevant numbers, we consider the stiff case $n=1.00$ with $\bar{\sigma}=\bar{\sigma}_\mathrm{max}/2 \simeq 1.5 \times 10^{-1}$. Then, $1/c^2 \simeq 0.3$ and $1/c^4 \simeq 0.1$. So, keeping the assumption made above on the magnitudes of the involved kernels, we conclude that terms of higher order should be taken into account for the accuracy of the computations  to be improved for stiff cases.

To discuss the possibility of having a second-order post-Newtonian approximation in the form of an iterative numerical method, we first mention that \cite{S1973} uses as theoretical framework for describing his method the formalism developed by \cite{C1965a}. In his investigation, Chandrasechar includes higher-order terms in the  relativistic equations, but the formalism itself aims to go deeper into the theory of general relativistic hydrodynamics than to develop a numerical method. Furthermore, detailed analysis on the second-order post-Newtonian approximation is given by \cite{Chandra-Nutku-1969}. To the extend of our knowledge, this work has not be modified properly so that to obtain the form of a numerical method (as done by Seguin for the first-order approximation). Thus, a direct way could be to develop a second-order post-Newtonian approximation in the form of an iterative method. Nevertheless, a less complicated way seems to be feasible: to introduce to our numerical scheme some ``artificial second-order terms'', such terms being (probably)  optimum combinations of terms already involved  in the first-order approximation. This issue seems to be interesting as an alternative treatment of the problem; so, we will proceed to its study in a subsequent investigation.          

\appendix
\section{Equations of motion}
\label{app1}
Rearranging the terms, Eq.~\eqref{EoM_static} is written as
\begin{multline}
    \frac{\partial}{\partial x_{\alpha}} \left[ \left( 1 + \frac{2U}{c^2} \right) P \right] + \frac{\partial}{\partial x_{\mu}}(\sigma \upsilon_{\alpha} \upsilon_{\mu}) - \varrho \frac{\partial U}{\partial x_{\alpha}} + \frac{4}{c^2} \varrho \left( \upsilon_{\mu}\frac{\partial U_{\mu}}{\partial x_{\alpha}} - \upsilon_{\mu}\frac{\partial U_{\alpha}}{\partial x_{\mu}} \right) 
     \\ 
     - \frac{2}{c^2} \varrho \left( \phi \frac{\partial U}{\partial x_{\alpha}} + \frac{\partial \Phi}{\partial x_{\alpha}} \right) + \frac{4}{c^2}\varrho \upsilon_{\mu} \frac{\partial}{\partial x_{\mu}} \left( \upsilon_{\alpha}U\right) = 0.
\end{multline}
Using Eqs.~\eqref{subt1}, \eqref{subt2} and \eqref{subt3}, we have
\begin{multline}
    \frac{\partial}{\partial x_{\alpha}} \left[ \left( 1 + \frac{2U}{c^2} \right) P \right] - \sigma \Tilde{\omega} \Omega^{*2} \hat{\Tilde{\omega}} - \varrho \frac{\partial U}{\partial x_{\alpha}} + \frac{4}{c^2}\varrho\left( \nabla \left( \Tilde{\omega} \Omega U_{\phi} \right) - \Tilde{\omega} U_{\phi} \frac{d\Omega}{d\Tilde{\omega}} \hat{\Tilde{\omega}} \right) 
    \\ 
    - \frac{2}{c^2} \varrho \left( \phi \frac{\partial U}{\partial x_{\alpha}} + \frac{\partial \Phi}{\partial x_{\alpha}} \right) - \frac{4}{c^2}\varrho\Tilde{\omega} \Omega^{2} U \hat{\Tilde{\omega}} = 0,
\end{multline}
Now, using Eqs.~\eqref{sigma}, \eqref{phi} and dividing by $\varrho$, 
the latter transforms to
\begin{multline}
    \frac{1}{\varrho}\nabla{\left[\left( 1 + \frac{2U}{c^2} \right) P \right]} 
    - \frac{\sigma}{\varrho} \Tilde{\omega} \Omega^{*2} \hat{\Tilde{\omega}} - \nabla{U} + \frac{4}{c^2}\nabla{\left(\Tilde{\omega}\Omega U_{\phi} \right)} 
    - \frac{4}{c^2}\Tilde{\omega}U_{\phi}\frac{d\Omega}{d\Tilde{\omega}}\hat{\Tilde{\omega}}  
    \\
    - \frac{2}{c^2}\left( \Tilde{\omega}^{2}\Omega^{2}+U+\left(\frac{\Pi}{2}+\frac{3P}{2\varrho}\right)\right)\nabla{U}
    - \frac{2}{c^2}\nabla{\Phi} - \frac{4}{c^2}\varrho\Tilde{\omega} \Omega^{2} U \hat{\Tilde{\omega}} = 0, 
\end{multline}
\begin{multline}
    \frac{1}{\varrho}\nabla{\left[\left( 1 + \frac{2U}{c^2} \right) P \right]} 
    - \frac{\sigma}{\varrho} \Tilde{\omega} \Omega^{*2} \hat{\Tilde{\omega}} - \nabla{U} - \frac{1}{c^2}\bigg[ -\nabla{\left( 4\Tilde{\omega}\Omega U_{\phi}\right)} 
    + 4\Tilde{\omega}U_{\phi}\frac{d\Omega}{d\Tilde{\omega}}\hat{\Tilde{\omega}} 
    \\
    + 2\Tilde{\omega}^{2}\Omega^{2}\nabla{U} + 2U\nabla{U} +2\left(\frac{\Pi}{2}+\frac{3P}{2\varrho}\right)\nabla{U} 
    + 2\nabla{\Phi} + 4\Tilde{\omega}\Omega^{2}U\hat{\Tilde{\omega}} \bigg] = 0, 
\end{multline}
\begin{multline}
    \frac{1}{\varrho}\nabla{\left[\left( 1 + \frac{2U}{c^2} \right) P \right]} 
    - \left[ 1+\frac{1}{c^2}\left( \Tilde{\omega}^{2}\Omega^{2} + 2U + \left(\Pi+\frac{P}{\varrho}\right) \right) \right] \Tilde{\omega} \Omega^{*2} \hat{\Tilde{\omega}} - \nabla{U} 
    - \frac{1}{c^2}\bigg[ -\nabla{\left(4\Tilde{\omega}\Omega U_{\phi}\right)} 
    \\ 
    + 4\Tilde{\omega}U_{\phi}\frac{d\Omega}{d\Tilde{\omega}}\hat{\Tilde{\omega}} + 2\Tilde{\omega}^{2}\Omega^{2}\nabla{U} 
    + 2U\nabla{U} +2\left(\frac{\Pi}{2}+\frac{3P}{2\varrho}\right)\nabla{U} + \nabla{\left(2\Phi\right)} + 4\Tilde{\omega}\Omega^{2}U\hat{\Tilde{\omega}} \bigg] = 0.
\end{multline}
Next, using the relation
\begin{align}
    \nabla{(\Tilde{\omega}^{2}\Omega^{2}U)}&=\Tilde{\omega}^{2}\Omega^{2}\nabla{U} + U\nabla{(\Tilde{\omega}^{2}\Omega^{2})} \nonumber \\ 
    &=\Tilde{\omega}^{2}\Omega^{2}\nabla{U} + U\left( 2\Tilde{\omega}\Omega^{2}\hat{\Tilde{\omega}} +2\Tilde{\omega}^{2}\Omega\frac{d\Omega}{d\Tilde{\omega}}\hat{\Tilde{\omega}} \right) \\
    &=\Tilde{\omega}^{2}\Omega^{2}\nabla{U} + 2\Tilde{\omega}\Omega^{2}U\hat{\Tilde{\omega}} + 2U\Tilde{\omega}^{2}\Omega\frac{d\Omega}{d\Tilde{\omega}}\hat{\Tilde{\omega}}, \nonumber
\end{align}
on the above equation, we take
\begin{multline}
    \frac{1}{\varrho}\nabla{\left[\left( 1 + \frac{2U}{c^2} \right) P \right]} 
    - \Tilde{\omega} \Omega^{*2} \hat{\Tilde{\omega}} - \nabla{U} 
    - \frac{1}{c^2}\bigg[ -\nabla{\left(4\Tilde{\omega}\Omega U_{\phi}\right)} 
    + 4\Tilde{\omega}U_{\phi}\frac{d\Omega}{d\Tilde{\omega}}\hat{\Tilde{\omega}} 
    + 2\nabla{(\Tilde{\omega}^{2}\Omega^{2}U)} 
    \\ 
    - 4\Tilde{\omega}\Omega^{2}U\hat{\Tilde{\omega}} 
    - 4U\Tilde{\omega}^{2}\Omega\frac{d\Omega}{d\Tilde{\omega}}\hat{\Tilde{\omega}} 
    + 2U\nabla{U} +\left(\Pi+\frac{3P}{\varrho}\right)\nabla{U} + \nabla{\left(2\Phi\right)} + 4\Tilde{\omega}\Omega^{2}U\hat{\Tilde{\omega}} 
     \\
    + \Tilde{\omega}^{3}\Omega^{4}\hat{\Tilde{\omega}} +2\Tilde{\omega}\Omega^{2}U\hat{\Tilde{\omega}} + \Tilde{\omega}\Omega^{2}\hat{\Tilde{\omega}}\left(\Pi+\frac{P}{\varrho}\right) \bigg] = 0, 
\end{multline}
\begin{multline}
    \frac{1}{\varrho}\nabla{\left[\left( 1 + \frac{2U}{c^2} \right) P \right]} 
    - \Tilde{\omega} \Omega^{*2} \hat{\Tilde{\omega}} - \nabla{U} - \frac{1}{c^2}\left[ -\nabla{\left(4\Tilde{\omega}\Omega U_{\phi}\right)} 
    + 2\nabla{(\Tilde{\omega}^{2}\Omega^{2}U)} + 4\Tilde{\omega}\left( U_{\phi} - 4U\Tilde{\omega}\Omega \right)\frac{d\Omega}{d\Tilde{\omega}}\hat{\Tilde{\omega}} 
    \right. \\ \left.
    + 2U\nabla{U} +\left(\Pi+\frac{3P}{\varrho}\right)\nabla{U} + \nabla{\left(2\Phi\right)} 
    + \Tilde{\omega}^{3}\Omega^{4}\hat{\Tilde{\omega}} +2\Tilde{\omega}\Omega^{2}U\hat{\Tilde{\omega}} + \Tilde{\omega}\Omega^{2}\hat{\Tilde{\omega}}\left(\Pi+\frac{P}{\varrho}\right) \right] = 0.
\end{multline}
By using Eqs.~\eqref{pot-B}, \eqref{pot-W}, this equation is written as
\begin{multline}
    \frac{1}{\varrho}\nabla{\left[\left( 1 + \frac{2U}{c^2} \right) P \right]} 
    - \Tilde{\omega} \Omega^{*2} \hat{\Tilde{\omega}} - \nabla{U} - \frac{1}{c^2}\left[ -\nabla{\left(4\Tilde{\omega}\Omega U_{\phi}\right)} 
    + 2\nabla{(\Tilde{\omega}^{2}\Omega^{2}U)} + 4\Tilde{\omega}\left( U_{\phi} - 4U\Tilde{\omega}\Omega \right)\frac{d\Omega}{d\Tilde{\omega}}\hat{\Tilde{\omega}} 
    \right. \\ \left.
    + 2U\nabla{U} +\left(\Pi+\frac{3P}{\varrho}\right)\nabla{U} + \nabla{\left(2\Phi\right)} 
    + \nabla{W} +2U\nabla{B} + \left(\Pi+\frac{P}{\varrho}\right)\nabla{B} \right] = 0.
\end{multline}
Rearranging the terms, we take
\begin{multline}
    \frac{1}{\varrho}\nabla{\left[\left( 1 + \frac{2U}{c^2} \right) P \right]} 
    - \Tilde{\omega} \Omega^{*2} \hat{\Tilde{\omega}} - \nabla{U} - \frac{1}{c^2}\bigg[  \nabla{(2\Tilde{\omega}^{2}\Omega^{2}U)} 
    -\nabla{\left(4\Tilde{\omega}\Omega U_{\phi}\right)} + \nabla{W} + \nabla{\left(2\Phi\right)} 
    \\ 
    + 4\Tilde{\omega}\left( U_{\phi} - 4U\Tilde{\omega}\Omega \right)\frac{d\Omega}{d\Tilde{\omega}}\hat{\Tilde{\omega}} 
    + 2U\left( \nabla{U}+\nabla{B} \right) +\left(\Pi+\frac{P}{\varrho}\right)\left( \nabla{U}+\nabla{B} \right) + 2\frac{P}{\varrho}\nabla{U} \bigg] = 0,
\end{multline}
\begin{multline}
    \frac{1}{\varrho}\nabla{\left[\left( 1 + \frac{2U}{c^2} \right) P \right]} 
    - \Tilde{\omega} \Omega^{*2} \hat{\Tilde{\omega}} - \nabla{U} 
    - \frac{1}{c^2}\bigg[  \nabla{(2\Tilde{\omega}^{2}\Omega^{2}U)} 
    -\nabla{\left(4\Tilde{\omega}\Omega U_{\phi}\right)} + \nabla{W} + \nabla{\left(2\Phi\right)} 
    \\ 
    + 4\Tilde{\omega}\left( U_{\phi} - 4U\Tilde{\omega}\Omega \right)\frac{d\Omega}{d\Tilde{\omega}}\hat{\Tilde{\omega}} 
    + 2U\nabla{\left(U+B\right)} + \left(\Pi+\frac{P}{\varrho}\right) \nabla{\left(U+B\right)} 
    + \frac{2}{\varrho}\nabla{\left(UP\right)} - \frac{2}{\varrho}U\nabla{P} \bigg] = 0,
\end{multline}
\begin{multline}
    \frac{1}{\varrho}\nabla{P} 
    - \Tilde{\omega} \Omega^{*2} \hat{\Tilde{\omega}} - \nabla{U} 
    - \frac{1}{c^2}\bigg[ \nabla{(2\Tilde{\omega}^{2}\Omega^{2}U)} -\nabla{\left(4\Tilde{\omega}\Omega U_{\phi}\right)} + \nabla{W} + \nabla{\left(2\Phi\right)} 
    + 4\Tilde{\omega}\left( U_{\phi} - 4U\Tilde{\omega}\Omega \right)\frac{d\Omega}{d\Tilde{\omega}}\hat{\Tilde{\omega}} 
    \\
    + \left(\Pi+\frac{P}{\varrho}\right) \nabla{\left(U+B\right)} 
    + 2U\nabla{\left(U+B\right)} - \frac{2}{\varrho}U\nabla{P} \bigg] = 0.
\end{multline}
Using Eqs.~\eqref{Pi_p_U_B} and \eqref{kr-15} on the post-Newtonian terms, the last two terms cancel each other and we are left with
\begin{multline}
    \frac{1}{\varrho}\nabla{P} 
    - \Tilde{\omega} \Omega^{*2} \hat{\Tilde{\omega}} - \nabla{U} - \frac{1}{c^2}\bigg[ \nabla{(2\Tilde{\omega}^{2}\Omega^{2}U)} -\nabla{\left(4\Tilde{\omega}\Omega U_{\phi}\right)} 
    + \nabla{W} + \nabla{\left(2\Phi\right)} 
    \\ 
    + 4\Tilde{\omega}\left( U_{\phi} - 4U\Tilde{\omega}\Omega \right)\frac{d\Omega}{d\Tilde{\omega}}\hat{\Tilde{\omega}} 
    + \left(\Pi+\frac{P}{\varrho}\right) \nabla{\left(\Pi+\frac{P}{\varrho}\right)} \bigg] = 0.
\end{multline}
Using the relation~\eqref{kr-15}, this equation becomes
\begin{multline}
    \nabla{\left(\Pi+\frac{P}{\varrho}\right)} 
    - \Tilde{\omega} \Omega^{*2} \hat{\Tilde{\omega}} - \nabla{U} - \frac{1}{c^2}\bigg[ \nabla{(2\Tilde{\omega}^{2}\Omega^{2}U)} 
    -\nabla{\left(4\Tilde{\omega}\Omega U_{\phi}\right)} + \nabla{W} + \nabla{\left(2\Phi\right)} 
    \\ 
    + 4\Tilde{\omega}\left( U_{\phi} - 4U\Tilde{\omega}\Omega \right)\frac{d\Omega}{d\Tilde{\omega}}\hat{\Tilde{\omega}} 
    + \left(\Pi+\frac{P}{\varrho}\right) \nabla{\left(\Pi+\frac{P}{\varrho}\right)} \bigg] = 0,
\end{multline}
The last term, by using Eq.~\eqref{Pi_p_U_B}, can be written as
\begin{equation}
\begin{split}
    \left(\Pi+\frac{P}{\varrho}\right) \nabla{\left(\Pi+\frac{P}{\varrho}\right)} =& \left( U+B+\delta\right) \nabla{\left(U+B+\delta\right)}
    \\ 
    =& \left( H+\delta\right) \nabla{\left(H+\delta\right)}
    \\ 
    =&\nabla{\left(\frac{1}{2}{(H+\delta)}^2\right)},
\end{split}
\end{equation}
and the equation becomes
\begin{multline}
    \nabla{\left(\Pi+\frac{P}{\varrho}\right)} 
    - \Tilde{\omega} \Omega^{*2} \hat{\Tilde{\omega}} - \nabla{U}
    - \frac{1}{c^2}\bigg[ + \nabla{(2\Tilde{\omega}^{2}\Omega^{2}U)}
    -\nabla{\left(4\Tilde{\omega}\Omega U_{\phi}\right)} + \nabla{W} + \nabla{\left(2\Phi\right)} 
    \\ 
    + \nabla{\left(\frac{1}{2}{(H+\delta)}^2\right)} 
    + 4\Tilde{\omega}\left( U_{\phi} - 4U\Tilde{\omega}\Omega \right)\frac{d\Omega}{d\Tilde{\omega}}\hat{\Tilde{\omega}} \bigg] = 0.
\end{multline}
Now, using the relation.~\eqref{Omega-star} for $\Omega^*$, the equation reads
\begin{multline}
    \nabla{\left(\Pi+\frac{P}{\varrho}\right)} 
    - \nabla{B} - \nabla{U}
    - \frac{1}{c^2}\bigg[ + \nabla{(2\Tilde{\omega}^{2}\Omega^{2}U)}
    -\nabla{\left(4\Tilde{\omega}\Omega U_{\phi}\right)} + \nabla{W} + \nabla{\left(2\Phi\right)} 
    \\ 
    + \nabla{\left(\frac{1}{2}{(H+\delta)}^2\right)} 
    + 4\Tilde{\omega}\left( U_{\phi} - 4U\Tilde{\omega}\Omega \right)\frac{d\Omega}{d\Tilde{\omega}}\hat{\Tilde{\omega}} + \Tilde{\omega}h^{2}(\Tilde{\omega},z)\hat{\Tilde{\omega}} \bigg] = 0.
\end{multline}
Finally, by using Eq.~\eqref{h^2}, the last two terms cancel each other, and the equation reads
\begin{equation}
    \nabla{\left(\Pi+\frac{P}{\varrho}\right)} = \nabla{\mathcal{U}},
\end{equation}
with
\begin{equation}
    \mathcal{U}= H+\frac{1}{c^2}\left[ 2\Phi+W+2\Tilde{\omega}^2\Omega^2U-4\Tilde{\omega}\Omega U_{\phi}+\frac{{(H+\delta)}^2}{2} \right]
\end{equation}
being the efficient potential.

\bibliographystyle{apalike}  
\bibliography{references} 

\begin{thebibliography}{}

\bibitem[{Chandrasekhar}, 1965a]{C1965b}
{Chandrasekhar}, S. (1965a).
\newblock {The Post-Newtonian Effects of General Relativity on the Equilibrium
  of Uniformly Rotating Bodies. I. The Maclaurin Spheroids and the Virial
  Theorem.}
\newblock {\em ApJ}, 142:1513.

\bibitem[{Chandrasekhar}, 1965b]{C1965a}
{Chandrasekhar}, S. (1965b).
\newblock {The Post-Newtonian Equations of Hydrodynamics in General
  Relativity.}
\newblock {\em ApJ}, 142:1488.

\bibitem[{Chandrasekhar}, 1965c]{C1965c}
{Chandrasekhar}, S. (1965c).
\newblock {The Stability of Gaseous Masses for Radial and Non-Radial
  Oscillations in the Post-Newtonian Approximation of General Relativity.}
\newblock {\em ApJ}, 142:1519.

\bibitem[{Chandrasekhar}, 1969]{C1969}
{Chandrasekhar}, S. (1969).
\newblock {Conservation Laws in General Relativity and in the Post-Newtonian
  Approximations}.
\newblock {\em ApJ}, 158:45.

\bibitem[{Chandrasekhar} and {Nutku}, 1969]{Chandra-Nutku-1969}
{Chandrasekhar}, S. and {Nutku}, Y. (1969).
\newblock {The Second Post-Newtonian Equations of Hydrodynamics in General
  Relativity}.
\newblock {\em ApJ}, 158:55.

\bibitem[{Cook} et~al., 1994]{CST94}
{Cook}, G.~B., {Shapiro}, S.~L., and {Teukolsky}, S.~A. (1994).
\newblock {Rapidly Rotating Polytropes in General Relativity}.
\newblock {\em ApJ}, 422:227.

\bibitem[{Fahlmann} and {Anand}, 1971]{FA1971}
{Fahlmann}, G.~G. and {Anand}, S.~P.~S. (1971).
\newblock {Rapidly Rotating Polytropes in the Post-Newtonian Approximation to
  General Relativity}.
\newblock {\em Astrophys. Space. Sci.}, 12(1):58--82.

\bibitem[{Font} et~al., 2000]{2000MNRAS.313..678F}
{Font}, J.~A., {Stergioulas}, N., and {Kokkotas}, K.~D. (2000).
\newblock {Non-linear hydrodynamical evolution of rotating relativistic stars:
  numerical methods and code tests}.
\newblock {\em MNRAS}, 313(4):678--688.

\bibitem[{Fowler}, 1966]{F1966}
{Fowler}, W.~A. (1966).
\newblock {The Stability of Supermassive Stars}.
\newblock {\em ApJ}, 144:180.

\bibitem[{Geroyannis}, 1990]{G1990ApJ...350..355G}
{Geroyannis}, V.~S. (1990).
\newblock {A Complex-Plane Strategy for Computing Rotating Polytropic Models:
  Numerical Results for Strong and Rapid Differential Rotation}.
\newblock {\em ApJ}, 350:355.

\bibitem[{Geroyannis}, 1991]{G1991}
{Geroyannis}, V.~S. (1991).
\newblock {An Iterative Technique for Computing Rotating Viscopolytropic
  Models}.
\newblock {\em Astrophys. Space. Sci.}, 186(1):27--56.

\bibitem[{Geroyannis} and {Karageorgopoulos}, 2014]{GK2014}
{Geroyannis}, V.~S. and {Karageorgopoulos}, V.~G. (2014).
\newblock {Computing rotating polytropic models in the post-Newtonian
  approximation: The problem revisited}.
\newblock {\em New Astronomy}, 28:9--16.

\bibitem[{Geroyannis} and {Karageorgopoulos}, 2015]{GK2015}
{Geroyannis}, V.~S. and {Karageorgopoulos}, V.~G. (2015).
\newblock {Critical rotation of general-relativistic polytropic models
  simulating neutron stars: A post-Newtonian hybrid approximative scheme}.
\newblock {\em New Astronomy}, 39:36--45.

\bibitem[{Geroyannis} and {Katelouzos}, 2008]{2008IJMPC..19.1863G}
{Geroyannis}, V.~S. and {Katelouzos}, A.~G. (2008).
\newblock {Numerical Treatment of Hartle's Perturbation Method for
  Differentially Rotating Neutron Stars Simulated by General-Relativistic
  Polytropic Models}.
\newblock {\em International Journal of Modern Physics C}, 19(12):1863--1908.

\bibitem[{Geroyannis} and {Sfaelos}, 2011]{GS11}
{Geroyannis}, V.~S. and {Sfaelos}, I.~E. (2011).
\newblock {Numerical Treatment of Rotating Neutron Stars Simulated by
  General-Relativistic Polytropic Models:. a Complex-Plane Strategy}.
\newblock {\em International Journal of Modern Physics C}, 22(3):219--248.

\bibitem[{Geroyannis} et~al., 1979]{GTV79}
{Geroyannis}, V.~S., {Tokis}, J.~N., and {Valvi}, F.~N. (1979).
\newblock {A Second-Order Perturbation Theory for Differentially Rotating
  Gaseous Polytropes}.
\newblock {\em Astrophys. Space. Sci.}, 64(2):359--389.

\bibitem[{Geroyannis} and {Valvi}, 2012]{GV2012}
{Geroyannis}, V.~S. and {Valvi}, F.~N. (2012).
\newblock {a Runge-Kutta-Fehlberg Code for the Complex Plane: Comparing with
  Similar Codes by Applying to Polytropic Models}.
\newblock {\em International Journal of Modern Physics C}, 23(5):1250038.

\bibitem[Haas et~al., 2022]{EinsteinToolkit:2022_11}
Haas, R., Cheng, C.-H., Diener, P., Etienne, Z., Ficarra, G., Ikeda, T.,
  Kalyanaraman, H., Kuo, N., Leung, L., Tian, C., Tsao, B.-J.~J., Wen, A.,
  Alcubierre, M., Alic, D., Allen, G., Ansorg, M., Armengol, F. G.~L.,
  Babiuc-Hamilton, M., Baiotti, L., Benger, W., Bentivegna, E., Bernuzzi, S.,
  Bode, T., Bozzola, G., Brandt, S.~R., Brendal, B., Bruegmann, B., Campanelli,
  M., Cipolletta, F., Corvino, G., Cupp, S., Pietri, R.~D., Dimmelmeier, H.,
  Dooley, R., Dorband, N., Elley, M., Khamra, Y.~E., Faber, J., Font, T.,
  Frieben, J., Giacomazzo, B., Goodale, T., Gundlach, C., Hawke, I., Hawley,
  S., Hinder, I., Huerta, E.~A., Husa, S., Iyer, S., Ji, L., Johnson, D.,
  Joshi, A.~V., Kastaun, W., Kellermann, T., Knapp, A., Koppitz, M., Laguna,
  P., Lanferman, G., L{\"o}ffler, F., Macpherson, H., Masso, J., Menger, L.,
  Merzky, A., Miller, J.~M., Miller, M., Moesta, P., Montero, P., Mundim, B.,
  Nelson, P., Nerozzi, A., Noble, S.~C., Ott, C., Paruchuri, R., Pollney, D.,
  Radice, D., Radke, T., Reisswig, C., Rezzolla, L., Rideout, D., Ripeanu, M.,
  Sala, L., Schewtschenko, J.~A., Schnetter, E., Schutz, B., Seidel, E.,
  Seidel, E., Shalf, J., Sible, K., Sperhake, U., Stergioulas, N., Suen, W.-M.,
  Szilagyi, B., Takahashi, R., Thomas, M., Thornburg, J., Tobias, M., Tonita,
  A., Walker, P., Wan, M.-B., Wardell, B., Werneck, L., Witek, H., Zilh{\~a}o,
  M., Zink, B., and Zlochower, Y. (2022).
\newblock {The Einstein Toolkit}.
\newblock To find out more, visit http://einsteintoolkit.org.

\bibitem[{Hachisu}, 1986]{H1986}
{Hachisu}, I. (1986).
\newblock {A Versatile Method for Obtaining Structures of Rapidly Rotating
  Stars}.
\newblock {\em ApJS}, 61:479.

\bibitem[{Hartle}, 1967]{Hartle_1_1967ApJ...150.1005H}
{Hartle}, J.~B. (1967).
\newblock {Slowly Rotating Relativistic Stars. I. Equations of Structure}.
\newblock {\em ApJ}, 150:1005.

\bibitem[{Hartle} and {Thorne}, 1968]{Hartle_2_1968ApJ...153..807H}
{Hartle}, J.~B. and {Thorne}, K.~S. (1968).
\newblock {Slowly Rotating Relativistic Stars. II. Models for Neutron Stars and
  Supermassive Stars}.
\newblock {\em ApJ}, 153:807.

\bibitem[{Horedt}, 2004]{2004Horedt}
{Horedt}, G.~P. (2004).
\newblock {\em {Polytropes - Applications in Astrophysics and Related Fields}},
  volume 306.
\newblock Springer.

\bibitem[{Krefetz}, 1966]{Kr1966}
{Krefetz}, E. (1966).
\newblock {A Variational Principle Governing the Equilibrium of a Uniformly
  Rotating Configuration in the Post-Newtonian Approximation}.
\newblock {\em ApJ}, 143:1004.

\bibitem[{Krefetz}, 1967a]{Kr1967b}
{Krefetz}, E. (1967a).
\newblock {The Appearance of a Rotating Configuration as Viewed from Infinity}.
\newblock {\em ApJ}, 148:613.

\bibitem[{Krefetz}, 1967b]{Kr1967a}
{Krefetz}, E. (1967b).
\newblock {The Equilibrium of Slowly Rotating Configurations in the
  Post-Newtonian Approximation: Corrections to Clairaut's Equation}.
\newblock {\em ApJ}, 148:589.

\bibitem[Liu, 2002]{liu2002postnewtonian}
Liu, Y.~T. (2002).
\newblock {Post-Newtonian Models of Differentially Rotating Neutron Stars.
  (arXiv:gr-qc/0207097v1 )}.

\bibitem[{L{\"o}ffler} et~al., 2015]{2015PhRvD..91f4057L}
{L{\"o}ffler}, F., {De Pietri}, R., {Feo}, A., {Maione}, F., and {Franci}, L.
  (2015).
\newblock {Stiffness effects on the dynamics of the bar-mode instability of
  neutron stars in full general relativity}.
\newblock {\em Phys. Rev. D}, 91(6):064057.

\bibitem[{Lyford} et~al., 2003]{LBS-2003ApJ...583..410L}
{Lyford}, N.~D., {Baumgarte}, T.~W., and {Shapiro}, S.~L. (2003).
\newblock {Effects of Differential Rotation on the Maximum Mass of Neutron
  Stars}.
\newblock {\em ApJ}, 583(1):410--415.

\bibitem[{Papasotiriou} and {Geroyannis}, 2002]{PG2002}
{Papasotiriou}, P.~J. and {Geroyannis}, V.~S. (2002).
\newblock {A SCILAB Program for Computing Rotating Magnetic Compact Objects}.
\newblock {\em International Journal of Modern Physics C}, 13(3):297--314.

\bibitem[{Seguin}, 1973]{S1973}
{Seguin}, F.~H. (1973).
\newblock {A Post-Newtonian Study of Differentially Rotating Polytropes}.
\newblock {\em ApJ}, 179:289--308.

\bibitem[{Stergioulas}, 1996]{1996PhDT.........4S}
{Stergioulas}, N. (1996).
\newblock {\em {The Structure and Stability of Rotating Relativistic Stars}}.
\newblock PhD thesis, University of Wisconsin, Milwaukee.

\bibitem[{Stergioulas}, 1998]{1998LRR.....1....8S}
{Stergioulas}, N. (1998).
\newblock {Rotating Stars in Relativity}.
\newblock {\em Living Reviews in Relativity}, 1(1):8.

\bibitem[{Stergioulas}, 2003]{sterg-2003}
{Stergioulas}, N. (2003).
\newblock {Rotating Stars in Relativity}.
\newblock {\em Living Reviews in Relativity}, 6(1):3.

\bibitem[{Stergioulas} et~al., 2004]{2004STERG_AP-FONT}
{Stergioulas}, N., {Apostolatos}, T.~A., and {Font}, J.~A. (2004).
\newblock {Non-linear pulsations in differentially rotating neutron stars:
  mass-shedding-induced damping and splitting of the fundamental mode}.
\newblock {\em MNRAS}, 352(4):1089--1101.

\bibitem[{Stergioulas} and {Friedman}, 1995]{Sterg-1995}
{Stergioulas}, N. and {Friedman}, J.~L. (1995).
\newblock {Comparing Models of Rapidly Rotating Relativistic Stars Constructed
  by Two Numerical Methods}.
\newblock {\em ApJ}, 444:306.

\end{thebibliography}

\end{document}